\documentclass[12pt]{article}
\usepackage{CJKutf8}
\usepackage{amsfonts}
\usepackage{mathrsfs}
\usepackage{amsthm}
\usepackage{amsmath,amssymb}
\usepackage{mathtools}
\usepackage{bbm}

\usepackage{bm}
\usepackage{latexsym}
\usepackage[utf8]{inputenc}
\usepackage[T1]{fontenc}
\usepackage{CJK}
\usepackage{float}
\usepackage{graphicx}
\usepackage{subfigure}
\usepackage{multirow}
\usepackage{multicol}
\usepackage{dcolumn}
\usepackage{indentfirst}
\usepackage{fancyhdr}
\usepackage{threeparttable}
\usepackage{booktabs}
\usepackage{tabularx}
\usepackage{epstopdf}
\usepackage{color}
\usepackage{rotating} 
\usepackage{threeparttable}
\usepackage{natbib}
\usepackage[ruled,linesnumbered,lined,commentsnumbered]{algorithm2e}

\allowdisplaybreaks

\usepackage{setspace}
\usepackage{enumitem}
\usepackage{lipsum}
\usepackage{microtype} 
\makeatletter
    
    \newcommand{\Rmnum}[1]{\expandafter\@slowromancap\romannumeral #1@}
  \makeatother                                                      
\allowdisplaybreaks

\theoremstyle{definition}

\newtheorem{theory}{Theorem} 
\newtheorem{remark}{Remark}

\usepackage[colorlinks,
            linkcolor=black,
            anchorcolor=black,
            citecolor=black,
            allcolors=black]{hyperref}
\usepackage{enumitem}
\usepackage{url}
\usepackage{caption}
\newcommand{\figref}[1]{\figurename~\ref{#1}}

\usepackage{tikz}
\usetikzlibrary{shapes.geometric, arrows}
\tikzstyle{node} = [rectangle, rounded corners, minimum width=0.5cm, minimum height=0.5cm,text centered, draw=black, fill=white!30]
\tikzstyle{arrow} = [thick,->,>=stealth]
\usepackage{nccmath}
\usepackage{xr}
\newcommand{\blind}{0}

\usepackage{authblk}

\begin{document}

\begin{CJK}{UTF8}{gbsn}

\def\spacingset#1{\renewcommand{\baselinestretch}%
{#1}\small\normalsize} \spacingset{1}

\date{}
\if0\blind
{
  \title{\bf 
   Convolution-Smoothing Based Locally Sparse Estimation for Functional 
   Quantile Regression
   }
  \author[a]{Hua Liu\footnote{These two authors contributed equally to this paper and shared the first authorship.}}
  \author[b]{Boyi Hu $^*$ }
  \author[c]{Jinhong You}
  \author[d]{Jiguo Cao \thanks{Corresponding Author. Email: jiguo\_cao@sfu.ca}}
  \affil[a]{\small School of Economics and Finance,
  Xi'an Jiaotong University}
  \affil[b]{\small Department of Neurology, College of Physicians and Surgeons, Columbia University}
  \affil[c]{\small School of Statistics and Data Science,
  Shanghai University of Finance and Economics}
  \affil[d]{\small Department of Statistics and Actuarial Science,
  Simon Fraser University}
\maketitle
} \fi

\if0\blind
{
  \bigskip
  \bigskip
  \bigskip
  \begin{center}
    {\LARGE\bf 
    Convolution-Smoothing based Locally Sparse Estimation for 
   Functional 
   Quantile Regression}
    \end{center}
  \medskip
} \fi

\bigskip
\begin{abstract}

Motivated by an application to study the impact of temperature, precipitation and irrigation on soybean yield, this article proposes a sparse semi-parametric functional quantile model. The model is called “sparse” because the functional coefficients are only nonzero in the local time region where the functional covariates have significant effects on the response under different quantile levels. To tackle the computational and theoretical challenges in optimizing the quantile loss function added with a concave penalty, we develop a novel Convolution-smoothing based Locally Sparse Estimation (CLoSE) method, to do three tasks in one step, including selecting significant functional covariates, identifying the nonzero region of functional coefficients to enhance the interpretability of the model and estimating the functional coefficients. We establish the functional oracle properties and simultaneous confidence bands for the estimated functional coefficients, along with the asymptotic normality for the estimated parameters. In addition, because it is difficult to estimate the conditional density function given the scalar and functional covariates, we propose the split wild bootstrap method to construct the confidence interval of the estimated parameters and simultaneous confidence band for the functional coefficients. We also establish the consistency of the split wild bootstrap method. The finite sample performance of the proposed CLoSE method is assessed with simulation studies. The proposed model and estimation procedure are also illustrated by identifying the active time regions when the daily temperature influences the soybean yield.

\end{abstract}

\noindent%
{\it Keywords:} 
Functional data analysis,
Functional oracle property,
Semi-parametric model

\spacingset{2}

\section{Introduction}
\begin{figure}[htbp]
  \centering
  \subfigure[]{\includegraphics[width=5cm]{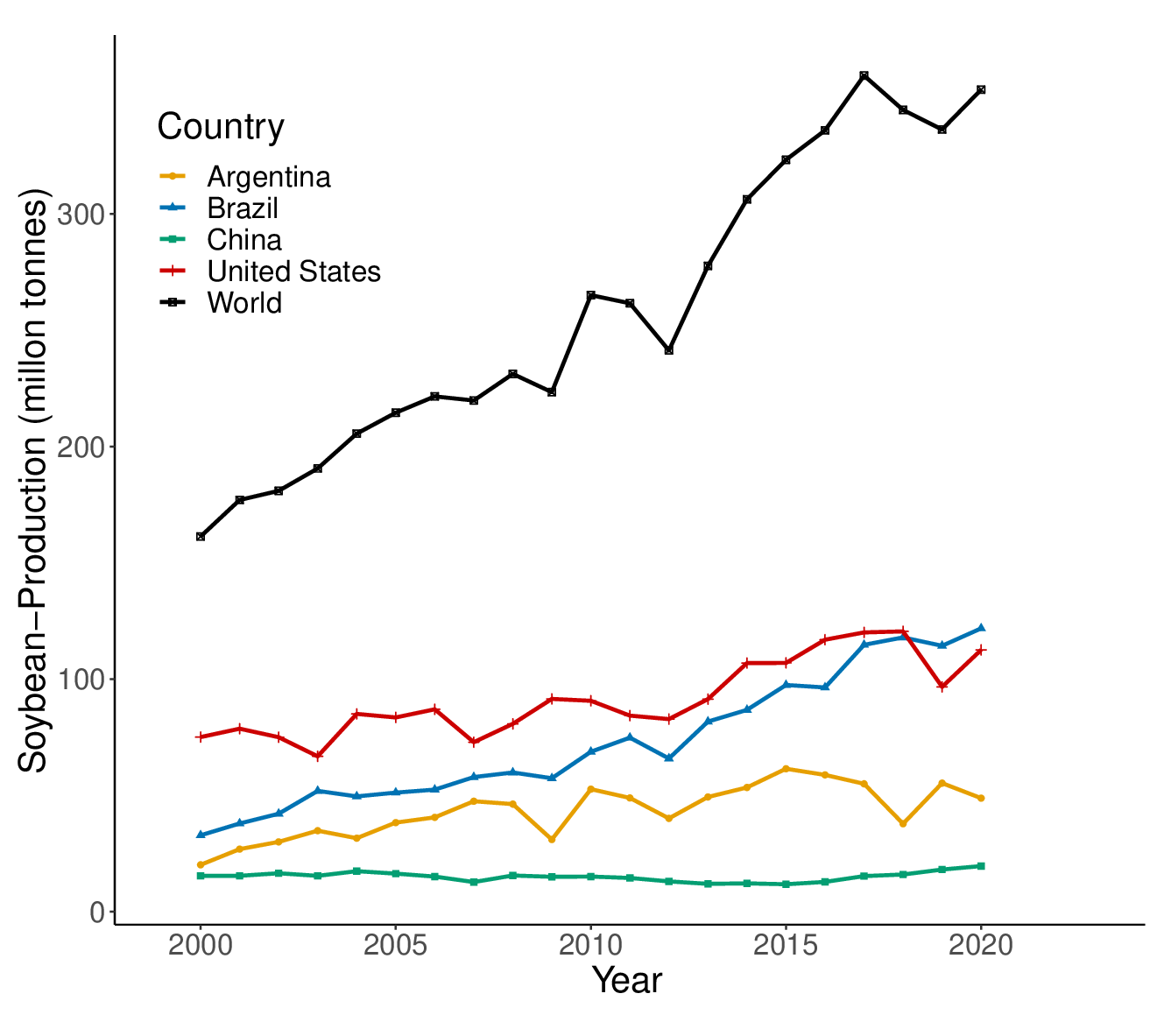}}
  \subfigure[]{\includegraphics[width=5cm]{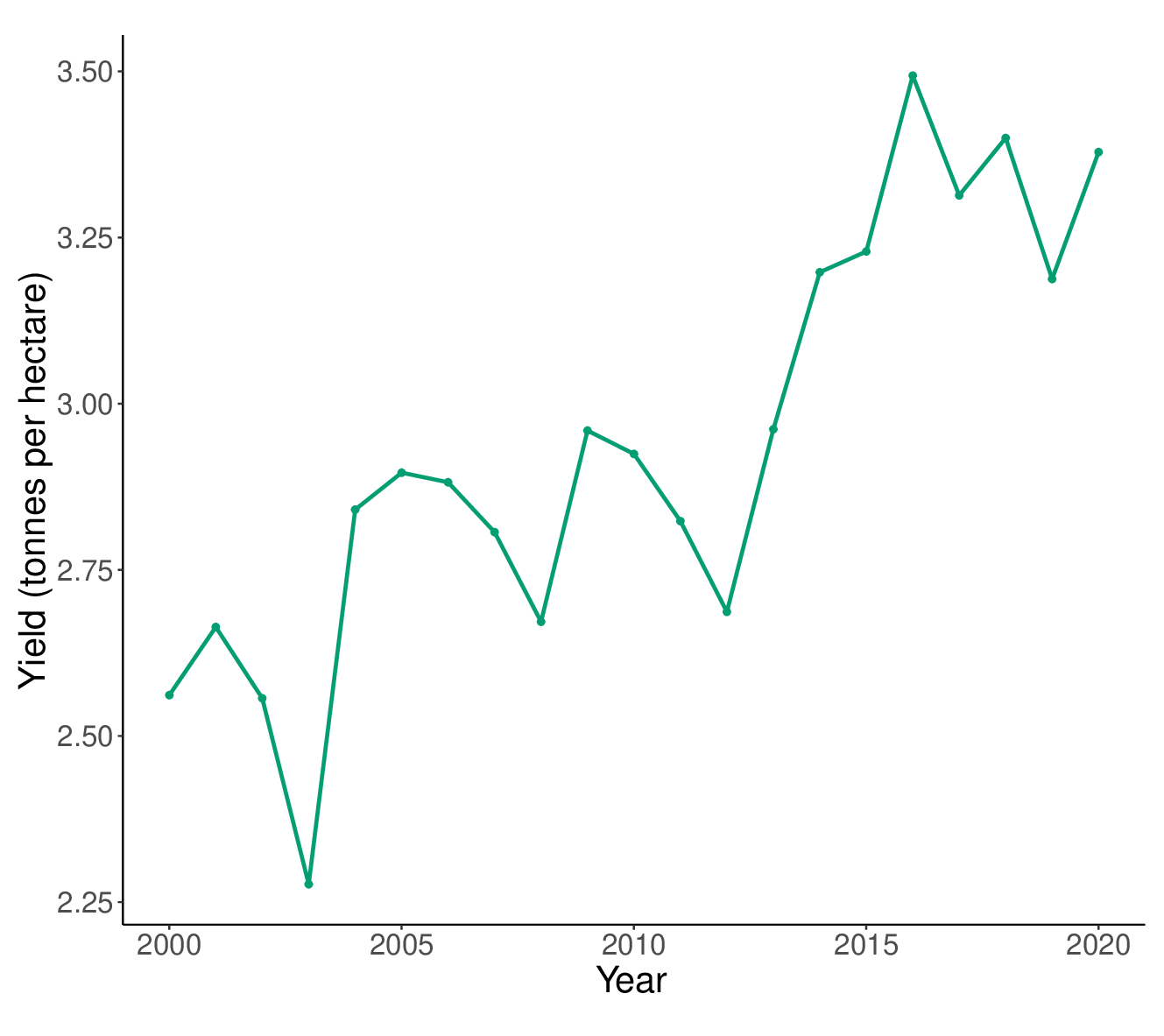}}
  \subfigure[]{\includegraphics[width=5cm]{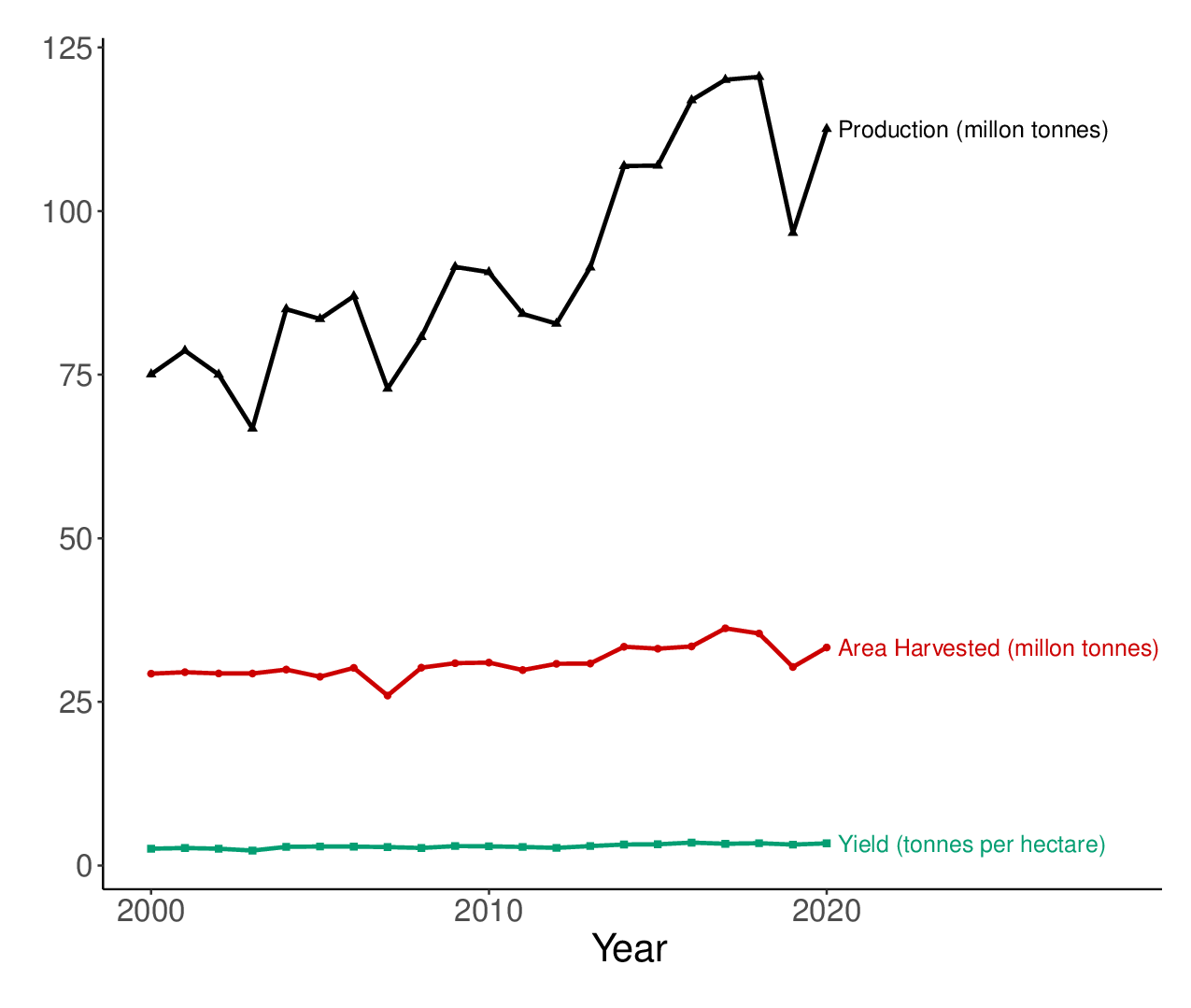}}
  \caption{(a) Soybean production data from 2000 to 2020 in the world and the four highest soybean-producing countries, namely the United States, Brazil, Argentina, and China.
  (b) Soybean yield in the United States from 2000 to 2020. (c) Comparison of soybean production and yield and area harvested in the United States from 2000 to 2020. These data are published by  \textit{the Food and Agriculture Organization of the United Nations}.}\label{yield}
\end{figure}

In agriculture, crop yield is a key focus worldwide because of its direct connection to the global needs for food, feed, and fuel. In addition, crops are highly liquid in the futures market, so crop price fluctuations can directly affect the stability of financial markets.
As one of the most important crops worldwide, more than three-quarters of soybeans are used to feed livestock, and only a small percentage (about 7\%) of global soybeans are used for typical soybean products such as tofu and soy milk. Meanwhile, the growing appetite for meat, dairy and soybean oil results in a rapidly increasing demand for soybeans as shown in Figure \ref{yield}-(a).
There are two main ways to increase production: to expand the amount of land to grow soybeans and to improve soybean yields (increasing per area harvest).
Taking data from the United States as an example, it is clear that the impressive improvement in soybean yields (Figure \ref{yield}-(b)) is not able to keep up with the increasing demand for soybean production (Figure \ref{yield}-(c)), which makes the government have to devote additional land to production. However, many scientists think increasing harvested area is a major underlying cause of deforestation. Therefore, it is urgent to improve soybean yields due to increasing product demand and environmental protection.

It is widely known that soybean is a crop with high demands on the natural environment and resources, especially in temperature and humidity. 
\textcolor{black}{For example, \cite{szczerba2021effect} investigated the effects of temperature on germination and seedling growth in four soybean cultivars, identifying an optimal temperature of approximately $25^{\circ}\mathrm{C}$ for both processes. Similarly, \cite{alsajri2022morpho} reported that the optimal temperature range for growth in two selected soybean cultivars was $24^{\circ}\mathrm{C}$–$27^{\circ}\mathrm{C}$, aligning with the findings of \cite{szczerba2021effect}.}
Hence, comprehending the intricate mechanisms of daily temperature, precipitation and irrigation to soybeans is crucial for enhancing both the growth and productivity of soybeans. The traditional approach for examining the correlation between a scalar response variable and combined (functional and scalar) covariates is the functional mean regression. 
This regression model primarily focuses on the conditional mean of the soybean yield distribution, thereby overlooking the impact of factors on the extremes of the response. However, in agricultural economics, investigations into the upper quantiles of yield hold greater interest and importance. In contrast, functional quantile regression
\citep{chen:2012,kato:2012} 
can unveil the influence of covariates on the extreme values of the response variable.

Furthermore, agricultural systems are dynamic, and seasonal variations can lead to sparse regions in the functional coefficient of the functional covariate (e.g. daily temperatures) on the scalar response (soybean yields). By identifying these sparse regions, a deeper understanding of the variability and dynamics of agricultural systems can be achieved.
Effective resource management can also be achieved, e.g., pinpointing these sparse regions can allow farmers to more accurately adjust interventions to reduce input costs such as water and fertilizer, thereby maximizing productivity.

This research is proposed to explore the impact of temperature, precipitation, and irrigation on soybean yield and identify the specific periods during which daily temperature exhibits a non-zero influence on annual soybean yield under different quantile levels. Motivated by this application problem but not limited to solving this one practical problem, we propose the following locally sparse semi-parametric functional quantile model,
\vspace{-0.5cm}
\begin{equation}\label{fqr}
    Q_{\tau}(Y|\bm{Z},\bm{X}(\cdot)) =
    \bm{Z}^T\bm{\alpha}_{\tau}+
    \int_{0}^{\mathcal{T}}\bm{X}^T(t)\bm{\beta}_{\tau}(t)d t,
\end{equation}

\vspace{-0.5cm}\noindent
where $Q_{\tau}(Y|\bm{Z},\bm{X}(\cdot))$ is the $\tau$-th conditional quantile of the scalar response $Y$ given $\bm{X}(\cdot) = (X_1(\cdot),\ldots,X_m(\cdot))^T$ and $\bm{Z}$ for a fixed quantile level $\tau\in(0,1)$. Without loss of generality, we assume that the domain for each $X_l(\cdot), l=1,...,m$ is the same, specifically characterized by $t\in [0,\mathcal{T}]$
and $\bm{\beta}_{\tau}(\cdot)=(\beta_{\tau,1}(\cdot),\ldots,
\beta_{\tau,m}(\cdot))^T$ is a vector of corresponding functional coefficeints. Meanwhile, some scalar covariates $\bm{Z}=(Z_1,\ldots,Z_d)^T$ are taken into account in this model and $\bm{\alpha}_{\tau}$ is a $d\times 1$ vector of coefficients, where we set $Z_1\equiv 1$ and $\alpha_{\tau,1}$ denotes the intercept throughout this article. 
In our soybean application, $Y$ is the annual soybean yield, $X_1(t)$ and $X_2(t)$ are the daily maximum and minimum temperature, respectively, and $Z_1$ and $Z_2$ are the annual precipitation and the ratio of irrigated area of each county in Kansas.

The functional coefficeint $\beta_{\tau,l}(\cdot)$ is assumed to be locally sparse, which means that $\beta_{\tau,l}(t)=0$ in some regions $\mathcal{N}$, where $\mathcal{N}$ is a subset of the whole time domain $[0,\mathcal{T}]$.
Then the local sparsity of $\bm{\beta}_{\tau}(\cdot)$ can depict the dynamic dependence of the $\tau$-th conditional quantile of the scalar response $Y$ on the functional covariates. 

The proposed sparse semi-parametric functional quantile model (\ref{fqr}) includes a variety of functional models as special cases. For example, when the identified sparse regions for all functional coefficient $\beta_{\tau,l}(t), l=1,\ldots,m$, equal to  $[0,\mathcal{T}]$, then model (\ref{fqr}) becomes the classic quantile regression \citep{koenker1978regression}. If $\alpha_{\tau,l}=0$ for all $l=1,\ldots,d$, and no functional coefficient $\beta_{\tau,l}$ has a locally sparse region, model (\ref{fqr}) is reduced to the functional quantile regression (FQR) with only functional covariates \citep{chen:2012,kato:2012}. 
Various partially functional quantile regression models are special cases of our proposed model. For instance, 
\cite{yao2017regularized} consider a partially functional quantile regression with a functional covariate and high-dimensional scalar covariates.
\cite{ma2019quantile} proposed a functional partially linear model with multiple functional covariates and ultrahigh-dimensional scalar covariates, and imposed two nonconvex penalties to select the significant functional and scalar covariates. To the best of our knowledge, no work has studied the local sparsity structure for functional quantile models, although it is common and important in various applications.

Most existing works considering local sparsity focus on the mean regression models \citep{wang2007group,james2009functional,zhou2013functional}. 
\textcolor{black}{A variety of regularization techniques have been developed to induce the local sparsity of functionas. One class of methods leverages the local support property of B-splines combined with group bridge penalties \citep{huang2009group}. 
For example, \cite{wang2015functional} investigated functional sparsity in nonparametric regression. Building on similar ideas, \cite{tu2020estimation} studied simultaneous domain selection for nonparametric varying-coefficient models with longitudinal data. In the context of functional linear regression, \cite{guan2020estimating} focused on identifying nonzero regions near the boundaries, while \cite{wang2023functional} aimed to detect locally sparse supports in function-on-scalar linear regression.
Another widely used approach is based on the concave penalty SCAD. \cite{lin2017locally} proposed a functional extension of the SCAD penalty \citep{fan2001variable}, known as functional SCAD (fSCAD), to achieve locally sparse estimation in scalar-on-function regression. This method has been further applied in multi-output settings, such as in 
\cite{li2022integrative}, for functional linear regression with multiple responses. Additionally, \cite{park2022sparse} developed a regularization method for functional linear discriminant analysis that similarly induces zero-effect regions.
While other penalty functions such as the group bridge can also be employed to induce local sparsity, in this paper we primarily adopt the second approach based on the fSCAD penalty.}

Our article makes several contributions.
First, we propose a locally sparse semi-parametric functional quantile regression model and develop a Convolution-smoothing based Locally Sparse Estimation (CLoSE) method. This unified framework simultaneously selects important functional covariates, identifies locally sparse regions to enhance interpretability, and estimates the corresponding functional coefficients. To alleviate the computational challenges associated with optimizing the piecewise linear quantile loss combined with a concave fSCAD penalty, we adopt a smoothed quantile loss function based on convolution techniques \citep{fernandes2021smoothing,he2021smoothed,tan2021high}.
Second, to provide a rigorous foundation for our approach, we establish several key theoretical results. Specifically, we prove the existence of a local minimizer for the associated optimization problem, ensuring the estimator is well-defined. We further demonstrate that the estimated functional coefficients possess functional oracle properties, including local sparsity and asymptotic normality. Consequently, we also derive simultaneous confidence bands (SCBs) for the estimated functional coefficients. Additionally, we establish the asymptotic normality of the model parameters, enabling valid statistical inference. These theoretical guarantees highlight the reliability and practical utility of the proposed method.
Finally, the analysis of soybean yield data from Kansas demonstrates that the time periods during which temperature significantly affects yield differ across quantile levels. In particular, different functional climatic factors emerge as significant at different parts of the yield distribution. These insights may inform targeted agricultural strategies, such as temperature regulation during critical growth stages, to optimize crop production.

The remainder of the paper is organized as follows. Section \ref{section2} proposes a method for identifying locally sparse regions and estimating functional coefficients on nonnull regions. Section \ref{section3} presents the asymptotic properties of the estimators. Section \ref{application} reports the real data analysis, and Section \ref{conclusion} concludes with discussions. Numerical studies, technical conditions, and proofs are provided in the supplementary material.


\section{CLoSE Method}\label{section2}
\subsection{Convolution-type Smoothing Approach}
Suppose that $\left\{(\bm{Z}_i,\bm{X}_i(t),Y_i, t\in[0,\mathcal{T}])\right\}_{i=1}^n$ is an independent and identically distributed random sample from $(\bm{Z},\bm{X}(t),Y,t\in[0,\mathcal{T}])$.
Then the locally sparse FQR estimator can be obtained by minimizing the following loss function 
\vspace{-0.5cm}
\begin{equation}\label{objective}
\medmath{
\begin{split}
    \mathcal{L}(\bm{\alpha}_{\tau},\bm{\beta}_{\tau},\gamma_l,\lambda_l) =&
    \frac{1}{n}\sum_{i=1}^n\rho_{\tau}\left(Y_i-\bm{Z}_i^T\bm{\alpha}_{\tau}-
        \int_{0}^{\mathcal{T}}\bm{X}_i^T(t)\bm{\beta}_{\tau}(t)d t\right) +\sum_{l=1}^m\gamma_l
    \left\|\bm{\mathcal{D}}^{q}\beta_{\tau,l}\right\|_2^2+\text{Locally sparse penalty},
\end{split}
}
\end{equation}
where $\rho_{\tau}(u) = u(\tau-I(u<0))$ is the quantile check function, and  $I(\cdot)$ is an indicator function. 
\textcolor{black}{However, $\rho_{\tau}(u)$ is not differentiable at $u = 0$, which poses challenges for optimization and theoretical analysis.}
In the loss function (\ref{objective}), the second term is a convex roughness penalty with the $q$th-order differential operator $\bm{\mathcal{D}}^{q}$ and the non-negative tuning parameters $\gamma_l,l=1,\ldots,m$, which control the smoothness of the 
estimated coefficient functions. 
For the third term, we choose the fSCAD penalty \citep{lin2017locally}, which is a concave penalty, to simultaneously identify the zero regions of the estimated functional coefficients and select the significant functional covariates.

The loss function (\ref{objective}) has no closed-form solution. Iterative procedures are often adopted to find the optimal solution.
However, in the iterative algorithm for minimizing the loss function (\ref{objective}), the combination of a {first}-order non-differentiable check function and a concave locally sparse penalty brings computation difficulty.  More specifically, local quadratic approximation (LQA) is a commonly used strategy for the optimization that involves fSCAD or SCAD penalty \citep{fan2001variable, lin2017locally}. Such an algorithm usually requires the calculation of gradient and Hessian matrix, which is not available for (\ref{objective}) because of the non-differentiable check function. To make the computation fast and stable, we consider an alternative convolution smoothed quantile loss function \citep{fernandes2021smoothing,he2021smoothed,tan2021high}. 

Let $e(\bm{\alpha}_{\tau},\bm{\beta}_{\tau})= Y-\bm{Z}^T\bm{\alpha}_{\tau}-        \int_{0}^{\mathcal{T}}\bm{X}^T(t)\bm{\beta}_{\tau}(t)d t$. Denote the conditional cumulative distribution function (CDF) and density function of $e(\bm{\alpha}_{\tau},\bm{\beta}_{\tau})$ given $\bm{Z}$ and $\bm{X}(t)$
as $F_{e|\bm{Z},\bm{X}}(\cdot)$ and $f_{e|\bm{Z},\bm{X}}(\cdot)$, respectively.
Then the population quantile loss can be expressed as 
\vspace{-0.5cm}
\begin{equation}\label{population}
\medmath{
    E[\rho_{\tau}(\bm{\alpha}_{\tau},\bm{\beta}_{\tau})] = E_{\bm{Z},\bm{X}}\left\{\int \rho_{\tau}(u) d F_{e|\bm{Z},\bm{X}}(u;\bm{\alpha}_{\tau},\bm{\beta}_{\tau})\right\}.
    }
\end{equation}

\vspace{-0.5cm}
\noindent When the unknown CDF $F_{e|\bm{Z},\bm{X}}(u;\bm{\alpha}_{\tau},\bm{\beta}_{\tau})$ in (\ref{population}) is replaced by the empirical distribution function $\widehat{F}(u;\bm{\alpha}_{\tau},\bm{\beta}_{\tau}) = 1/n\sum_{i=1}^n I (e_i(\bm{\alpha}_{\tau},\bm{\beta}_{\tau})\leq u\}$ of the residuals $e_i(\bm{\alpha}_{\tau},\bm{\beta}_{\tau})= Y_i-\bm{Z}_i^T\bm{\alpha}_{\tau}- \int_{0}^{\mathcal{T}}\bm{X}_i^T(t)\bm{\beta}_{\tau}(t)d t$, we can obtain the standard quantile loss, that is, the first term in (\ref{objective}).
However, the empirical distribution function $\widehat{F}(u;\bm{\alpha}_{\tau},\bm{\beta}_{\tau})$ is a discontinuous function. Thus, a kernel smoothing estimator of $F_{e|\bm{Z},\bm{X}}(u;\bm{\alpha}_{\tau},\bm{\beta}_{\tau})$ is a better choice \citep{fernandes2021smoothing,tan2021high}. The kernel density estimator is $\widehat{f}_h(u;\bm{\alpha}_{\tau},\bm{\beta}_{\tau})=1/n\sum \mathcal{K}_h(u-e_i(\bm{\alpha}_{\tau},\bm{\beta}_{\tau}))=(nh)^{-1}\sum \mathcal{K}((u-e_i(\bm{\alpha}_{\tau},\bm{\beta}_{\tau}))/h)$, where $\mathcal
K(\cdot)$ is a kernel function, $h$ is a bandwidth and $\mathcal{K}_h(u)=1/h\mathcal{K}(u/h)$.
The corresponding  kernel smoothing CDF estimator is $\widehat{F}_h(u;\bm{\alpha}_{\tau},\bm{\beta}_{\tau}) = n^{-1}\sum G_h(u-e_i(\bm{\alpha}_{\tau},\bm{\beta}_{\tau}))$, where $G_h(u)=G(u/h)$ and $G(u)= \int_{-\infty}^u\mathcal{K}(v)dv$.
When we replace the unknown CDF by the kernel smoothing estimator of the CDF $\widehat{F}_h(u;\bm{\alpha}_{\tau},\bm{\beta}_{\tau})$, we derive a new loss function with the convolution smoothed quantile loss:
\vspace{-0.5cm}
\begin{equation}\label{conquer}
\medmath{
\begin{split}
    \mathcal{L}^*(\bm{\alpha}_{\tau},\bm{\beta}_{\tau},\gamma_l,\lambda_l) =&
    \frac{1}{n}\sum_{i=1}^n\left(\rho_{\tau}*\mathcal{K}_h\right)\left(Y_i-\bm{Z}_i^T\bm{\alpha}_{\tau}- \int_{0}^{\mathcal{T}}\bm{X}_i^T(t)\bm{\beta}_{\tau}(t)d t\right)
    +\sum_{l=1}^m\gamma_l
    \left\|\bm{\mathcal{D}}^{q}\beta_{\tau,l}\right\|_2^2+ \text{fSCAD},
\end{split}
}
\end{equation}
where $\left(\rho_{\tau}*\mathcal{K}_h\right)(u)=\int_{-\infty}^{\infty}\rho_{\tau}(v)\mathcal{K}_h(v-u)dv$ and $*$ is the convolution operator. The smoothed quantile loss is a global convex function, which has first and second-order derivatives to make the computation faster. 

\subsection{Estimation Procedure}
We use B-spline basis functions to represent the functional coefficients $\beta_{\tau,l}(t),l=1,\ldots m$.  We set the order to be $p+1$ and place $K+1$ equally spaced knots $0=t_0<t_1<\cdots<t_{K-1}<t_K=\mathcal{T}$ in the domain $[0,\mathcal{T}]$ to define a set of B-spline basis functions. Then the functional coefficient
$\beta_{\tau,l}(t)$ can be approximated by the B-spline basis functions $\beta_{\tau,l}(t) \approx \bm{\mathcal{B}}^T(t)\bm{\theta}_{\tau,l}$,
where $\bm{\mathcal{B}}(t)=(B_1(t),\ldots,B_{K+p}(t))^T$ is the vector of B-spline basis functions and $\bm{\theta}_{\tau,l}=(\theta_{\tau,l,1},\ldots,
\theta_{\tau,l, K+p})^T$ is the corresponding coefficients. 
Denote $\bm{U}_i=\int_o^{\mathcal{T}} \bm{X}_i(t)\otimes \bm{\mathcal{B}}(t)dt$ and $\bm{\theta}_{\tau}^T = (\bm{\theta}_{\tau,1}^T,\cdots,\bm{\theta}_{\tau,m}^T)$,
then the first term of $\mathcal{L}^*(\bm{\alpha}_{\tau},\beta_{\tau},\gamma,\lambda)$ can be written as
$n^{-1}\sum_{i=1}^n\left(\rho_{\tau}*\mathcal{K}_h\right)(Y_i-\bm{Z}_i^T\bm{\alpha}_{\tau}-\bm{U}_i^T\bm{\theta}_{\tau})$.
The roughness penalty in \eqref{conquer} can be expressed as
\begin{equation}
    \label{roughness}
    \medmath{
    \sum_{l=1}^m\gamma_l
    \left\|\bigg(\bm{\mathcal{B}}^{(q)}(t)\bigg)^{T}\bm{\theta}_{\tau,l}\right\|_2^2=
    \sum_{l=1}^m\gamma_l \bm{\theta}_{\tau,l}^T \bm{V}\bm{\theta}_{\tau,l}=\bm{\theta}^T_{\tau}(\bm{\Gamma}\otimes\bm{V})\bm{\theta}_{\tau},
    }
\end{equation}
where $\bm{\Gamma}=\text{diag}(\gamma_1,\cdots,\gamma_m)$, $\bm{V}=\int_0^{\mathcal{T}} [\bm{\mathcal{B}}^{(q)}(t)][\bm{\mathcal{B}}^{(q)}(t)]^{T} dt$. 

\textcolor{black}{
Recall that $0=t_0<t_1<\cdots<t_{K-1}<t_K=\mathcal{T}$ is an equally spaced sequence in the domain $[0,\mathcal{T}]$.
According to Theorem 1 in \cite{lin2017locally}, as $K\rightarrow\infty$, the fSCAD penalty term of the $l$th function $\beta_{\tau,l}(t),t\in[0,\mathcal{T}]$ can be approximated by
\vspace{-0.5cm}
\begin{equation}\label{fSCAD1}
\begin{split}
   \frac{K}{\mathcal{T}}\int_0^\mathcal{T}p_{\lambda_l}(\left|\beta_{\tau,l}(t)\right|)dt
& \approx\sum_{j=1}^Kp_{\lambda_l}\left(\sqrt{\frac{K}{\mathcal{T}}\int_{t_{j-1}}^{t_j}\beta_{\tau,l}^2(t)dt}\right),
\end{split}
\end{equation}}

\vspace{-1.5cm}\noindent
\textcolor{black}{
here the SCAD penalty \citep{fan2001variable}
$p_{\lambda_l}(u)=\lambda_l u$ if $0\leq u\leq\lambda_l$, $p_{\lambda_l}(u)= -(u^2-2a\lambda_l u + \lambda_l^2)/(2(a-1))$ if $\lambda_l\leq u\leq a\lambda_l$, and $p_{\lambda_l}(u)=(a+1)\lambda_l^2/{2}$ if
$u\geq a\lambda_l$
with the non-negative tuning parameters $a$ and $\lambda_l$.}
{A recommended value for tuning parameter $a$ is 3.7 according to \cite{fan2001variable}.}

\textcolor{black}{The approximation (\ref{fSCAD1}) shows that the penalty of $\beta_{\tau,l}(t)$ on the whole domain $[0,\mathcal{T}]$ is equal to the avaraged SCAD penalty on each small subinterval $[t_{j-1},t_j]$. 
When $\beta_{\tau,l}(t)$ is significant on $[t_{j-1},t_j]$, then $p_{\lambda_l}\left(\sqrt{\frac{K}{\mathcal{T}}\int_{t_{j-1}}^{t_j}\beta_{\tau,l}^2(t)dt}\right)$ can not over shrink $\beta_{\tau,l}(t),t\in [t_{j-1},t_j]$.
Inversely, when $\beta(t)$ is vary small on the subinterval $[t_{j-1},t_j]$, the penalty $p_{\lambda_l}\left(\sqrt{\frac{K}{\mathcal{T}}\int_{t_{j-1}}^{t_j}\beta_{\tau,l}^2(t)dt}\right)$ can shrink $\beta_{\tau,l}(t)$ toward zero for every $t\in [t_{j-1},t_j]$, and then $\beta_{\tau,l}(t)$ is sparse on $[t_{j-1},t_j]$, which produces a locally sparse estimate.}
\textcolor{black}{In model (1), there are $m$-dimensional functional covariates $\bm{X}(t)$ and $m$ corresponding functional coefficients $\beta_{\tau,l},\ l=1,\ldots,m$.
Suppose different functional coefficients $\beta_{\tau,l}$ have different fSCAD tuning parameters $\lambda_l$, thus the 
fSCAD penalty on all functional coefficients is the summation of the fSCAD penalty on each functional coefficient:
\vspace{-0.5cm}
\begin{equation}\label{fSCAD}
\begin{split}
   \sum_{l=1}^m\frac{K}{\mathcal{T}}\int_0^\mathcal{T}p_{\lambda_l}(\left|\beta_{\tau,l}(t)\right|)dt
& \approx\sum_{l=1}^m\sum_{j=1}^Kp_{\lambda_l}\left(\sqrt{\frac{K}{\mathcal{T}}\int_{t_{j-1}}^{t_j}\beta_{\tau,l}^2(t)dt}\right).\\
\end{split}
\end{equation}
}

\vspace{-1.5cm}
Plugging the spline approximation into 
$\int_{t_{j-1}}^{t_j}\beta_{\tau,l}^2(t)dt$,
we can get the matrix representation, 
$
\int_{t_{j-1}}^{t_j}\beta_{\tau,l}^2(t)dt = \bm{\theta}_{\tau,l}^T \bm{W}_j\bm{\theta}_{\tau,l}
$
, where $\bm W_j$ is an $(K+p)$ by $(K+p)$ matrix with entries $w_{uv} = \int_{t_{j-1}}^{t_j}B_u(t)B_v(t)dt$ if $j \leq u, v\leq j+p$ and zero otherwise.
Thus the fSCAD penalty (\ref{fSCAD}) can be rewritten as 
\vspace{-0.5cm}
\begin{equation}\label{fsacdnew}
\medmath{
\sum_{l=1}^m\sum_{j=1}^Kp_{\lambda_l}\left(\sqrt{\frac{K}{\mathcal{T}}\int_{t_{j-1}}^{t_j}\beta_{\tau,l}^2(t)dt}\right)
=\sum_{l=1}^m\sum_{j=1}^Kp_{\lambda_l}\left(\sqrt{\frac{K}{\mathcal T}\bm{\theta}_{\tau,l}^T \bm{W}_j\bm{\theta}_{\tau,l}}\right).
}
\end{equation}

\vspace{-0.5cm}\noindent
Then, combined with the B-spline approximation,
(\ref{roughness}) and (\ref{fsacdnew}), we can rewrite the loss function (\ref{conquer}) as follows,
\vspace{-0.5cm}
\begin{equation}\label{obj2}
\medmath{
\begin{split}
    \mathcal{L}^*(\bm{\alpha}_{\tau},\bm{\theta}_{\tau},\gamma_l,\lambda_l)  \approx  
  & \frac{1}{n}\sum_{i=1}^n\left(\rho_{\tau}*\mathcal{K}_h\right)(Y_i-\bm{Z}_i^T\bm{\alpha}_{\tau}-\bm{U}_i^T\bm{\theta}_{\tau})
    +\bm{\theta}^T_{\tau}(\bm{\Gamma}\otimes\bm{V})\bm{\theta}_{\tau} +\sum_{l=1}^m\sum_{j=1}^Kp_{\lambda_l}\left(\sqrt{\frac{K}{\mathcal T}\bm{\theta}_{\tau,l}^T \bm{W}_j\bm{\theta}_{\tau,l}}\right).
\end{split}
}
\end{equation}

\textcolor{black}{
Overall, both the roughness penalty and the locally sparse penalty (fSCAD) serve as regularization mechanisms, each with a corresponding tuning parameter that governs its strength. These penalties play complementary roles in balancing model flexibility, smoothness, and interpretability.
The roughness penalty controls the global smoothness of the estimated functional coefficients  $\beta_{\tau,l}(t)$ with tuning parameters $\gamma_l$. A larger $\gamma_l$ enforces greater smoothness, reducing variance but potentially introducing bias by oversmoothing important features. A smaller $\gamma_l$ allows more flexibility but may lead to overfitting, especially in noisy data. 
The fSCAD penalty with the tuning parameters $\lambda_l$, on the other hand, promotes sparsity in the functional domain by shrinking the functional coefficients toward zero in specific subintervals the tuning parameter associated with sparsity. A larger $\lambda_l$ encourages more aggressive local shrinkage, potentially setting large portions of the functions to zero and improving interpretability. A smaller $\lambda_l$ allows for more detailed estimation but may retain irrelevant regions of the function, increasing model complexity. 
When used jointly, $\gamma_l,\ l=1,\ldots,m$ and $\lambda_l,\ l=1,\ldots,m$ provide a flexible framework for estimating functional coefficients that are both smooth and locally sparse.
}

However, the SCAD penalty $p_{\lambda_l}(\cdot)$ is not differentiable, which brings difficulty in the optimization. There are two ways to approximate $p_{\lambda_l}(\cdot)$, local quadratic approximation (LQA; \citealt{fan2001variable}) and local linear approximaation (LLA; \cite{zou2008one}). \cite{lin2017locally} found that LLA does not work well with $L^q$ norm of functions and they suggested using LQA in their paper. Therefore, we also choose the LQA to approximate the fSCAD penalty. When $u\approx u^{(0)}$, the LQA of the SCAD function $p_{\lambda_l}(u)$ is:
\vspace{-0.5cm}
\begin{equation}\label{scadd}
\medmath{
\begin{split}
     p_{\lambda_l}(|u|)
     & \approx p_{\lambda_l}(|u^{(0)}|)+\frac{1}{2}\frac{\dot{p}_{\lambda_l}(|u^{(0)}|)}{|u^{(0)}|}(u^2-u^{(0)2})
     = \frac{1}{2}\frac{\dot{p}_{\lambda_l}(|u^{(0)}|)}{|u^{(0)}|}u^2 
     + p_{\lambda_l}(|u^{(0)}|) 
     - \frac{1}{2}\frac{\dot{p}_{\lambda_l}(|u^{(0)}|)}{|u^{(0)}|}u^{(0)2}\\
     & \overset{\Delta}{=} \frac{1}{2}\frac{\dot{p}_{\lambda_l}(u^{(0)})}{|u^{(0)}|}u^2
     + R_1(u^{(0)}),
\end{split}
}
\end{equation}

\vspace{-0.5cm}\noindent
where $\dot{p}_{\lambda_l}(u)$ is the first order derivative of $p_{\lambda_l}(u)$ and $R_1(u^{(0)})$ is a constant of
$u^{(0)}$. 

Then given some initial estimator $\widehat{\bm{\theta}}_{\tau}^{(0)}=(\widehat{\bm{\theta}}_{\tau,1}^{(0)T},\cdots,\widehat{\bm{\theta}}_{\tau,m}^{(0)T})^T$, when $\bm{\theta}_{\tau}\approx \widehat{\bm{\theta}}_{\tau}^{(0)}$, we have
\vspace{-0.5cm}
\begin{equation}\label{lqa}
\medmath{
\begin{split}
\sum_{l=1}^m\sum_{j=1}^Kp_{\lambda_l}\left(\sqrt{\frac{K}{\mathcal T}\bm{\theta}_{\tau,l}^T \bm{W}_j\bm{\theta}_{\tau,l}}\right)
& \approx 
\frac{1}{2}\sum_{l=1}^m\sum_{j=1}^K
\frac{\dot{p}_{\lambda_l}\left(\sqrt{\frac{K}{\mathcal T}\widehat{\bm{\theta}}_{\tau,l}^{(0)T} \bm{W}_j\widehat{\bm{\theta}}_{\tau,l}^{(0)}}\right)}{\sqrt{\frac{K}{\mathcal T}\widehat{\bm{\theta}}_{\tau,l}^{(0)T} \bm{W}_j\widehat{\bm{\theta}}_{\tau,l}^{(0)}}}\frac{\bm{\theta}_{\tau,l}^T \bm{W}_j\bm{\theta}_{\tau,l}}{\mathcal{T}/K}+R_2(\widehat{\bm{\theta}}_{\tau}^{(0)})\\
& \approx 
\frac{1}{2}\sum_{l=1}^m\sum_{j=1}^K
\frac{\dot{p}_{\lambda_l}\left(\sqrt{\frac{K}{\mathcal T}\widehat{\bm{\theta}}_{\tau,l}^{(0)T} \bm{W}_j\widehat{\bm{\theta}}_{\tau,l}^{(0)}}\right)}{\sqrt{\frac{\mathcal T}{K}\widehat{\bm{\theta}}_{\tau,l}^{(0)T} \bm{W}_j\widehat{\bm{\theta}}_{\tau,l}^{(0)}}}\bm{\theta}_{\tau,l}^T \bm{W}_j\bm{\theta}_{\tau,l}+R_2(\widehat{\bm{\theta}}_{\tau}^{(0)}),
\end{split}
}
\end{equation}

\vspace{-0.5cm}\noindent
where 
\vspace{-0.5cm}
\begin{equation*}
\medmath{
    \begin{split}
        R_2(\widehat{\bm{\theta}}_{\tau}^{(0)})
        & = \sum_{l=1}^m\sum_{j=1}^Kp_{\lambda_l}\left(\sqrt{\frac{K}{\mathcal T}\widehat{\bm{\theta}}_{\tau,l}^{(0)T} \bm{W}_j\widehat{\bm{\theta}}_{\tau,l}^{(0)}}\right) -\frac{1}{2}\sum_{l=1}^m\sum_{j=1}^K
\dot{p}_{\lambda_l}\left(\sqrt{\frac{K}{\mathcal T}\widehat{\bm{\theta}}_{\tau,l}^{(0)T} \bm{W}_j\widehat{\bm{\theta}}_{\tau,l}^{(0)}}\right)\sqrt{\frac{K}{\mathcal T}\widehat{\bm{\theta}}_{\tau,l}^{(0)T} \bm{W}_j\widehat{\bm{\theta}}_{\tau,l}^{(0)}},
    \end{split}
    }
\end{equation*}

\vspace{-0.5cm}\noindent
only depends on the initial estimator $\widehat{\bm{\theta}}_{\tau}^{(0)}$.
Denote 
\vspace{-0.5cm}
\begin{equation*}
\medmath{
    \bm{W}_{\tau,l}^{(0)}=\frac{1}{2}\sum_{j=1}^K
\frac{\dot{p}_{\lambda_l}\left(\sqrt{\frac{K}{\mathcal T}\widehat{\bm{\theta}}_{\tau,l}^{(0)T} \bm{W}_j\widehat{\bm{\theta}}_{\tau,l}^{(0)}}\right)}{\sqrt{\frac{\mathcal T}{K}\widehat{\bm{\theta}}_{\tau,l}^{(0)T} \bm{W}_j\widehat{\bm{\theta}}_{\tau,l}^{(0)}}} \bm{W}_j,
\quad \text{and} \quad
\bm{W}_{\tau}^{(0)} = 
    \left(\begin{smallmatrix}
    \bm{W}_{\tau,1}^{(0)}& &\\
    &\ddots &\\
    && \bm{W}_{\tau,m}^{(0)}
    \end{smallmatrix}\right),
    }
\end{equation*}

\vspace{-0.5cm}\noindent
we can get $\sum_{l=1}^m\sum_{j=1}^Kp_{\lambda_l}\left(\sqrt{K/{\mathcal T}\bm{\theta}_{\tau,l}^T \bm{W}_j\bm{\theta}_{\tau,l}}\right)
 \approx \bm{\theta}_{\tau}^T\bm{W}_{\tau}^{(0)} \bm{\theta}_{\tau}+R_2(\widehat{\bm{\theta}}_{\tau}^{(0)})$.
Thus we can express the loss function (\ref{obj2}) as follows,
\vspace{-0.5cm}
\begin{equation}\label{lqaloss}
\medmath{
\begin{split}
    \mathcal{L}^*(\bm{\alpha}_{\tau},\bm{\theta}_{\tau},\gamma_l,\lambda_l\mid \widehat{\bm{\theta}}_{\tau}^{(0)})  \approx  
  & \frac{1}{n}\sum_{i=1}^n\left(\rho_{\tau}*\mathcal{K}_h\right)(Y_i-\bm{Z}_i^T\bm{\alpha}_{\tau}-\bm{U}_i^T\bm{\theta}_{\tau}) +\bm{\theta}^T_{\tau}(\bm{\Gamma}\otimes\bm{V})\bm{\theta}_{\tau} +\bm{\theta}_{\tau}^T\bm{W}_{\tau}^{(0)} \bm{\theta}_{\tau}+R_2(\widehat{\bm{\theta}}_{\tau}^{(0)}).
\end{split}
}
\end{equation}
Denote $\mathcal{L}_0^*(\bm{\alpha}_{\tau},\bm{\theta}_{\tau}) = 1/n\sum_{i=1}^n\left(\rho_{\tau}*\mathcal{K}_h\right)(Y_i-\bm{Z}_i^T\bm{\alpha}_{\tau}-\bm{U}_i^T\bm{\theta}_{\tau})$, we can get the first-order and second-order derivatives of $\mathcal{L}_0^*(\bm{\alpha}_{\tau},\bm{\theta}_{\tau})$ respectively, that is, $\dot{\mathcal{L}}^{*}_0(\bm{\alpha}_{\tau},\bm{\theta}_{\tau}) = 1/n\sum_{i=1}^n\{G_h(\bm{Z}_i^T\bm{\alpha}_{\tau}\\+
\bm{U}_i^T\bm{\theta}_{\tau}-Y_i)-\tau\}(\bm{Z}_i^T,\bm{U}_i^T)^T$,
$\ddot{\mathcal{L}}^{*}_0(\bm{\alpha}_{\tau},\bm{\theta}_{\tau})  = 1/n\sum_{i=1}^n\mathcal{K}_h(\bm{Z}_i^T\bm{\alpha}_{\tau}+\bm{U}_i^T\bm{\theta}_{\tau}-Y_i)(\bm{Z}_i^T,\bm{U}_i^T)^T(\bm{Z}_i^T,\bm{U}_i^T)$.

With the smoothed quantile loss and LQA on the fSCAD penalty, we are able to calculate its gradient and Hessian matrix. Then an iterative Newton–Raphson–type algorithm can be used to solve the optimization problem.
More specifically, for a fixed $\bm W_{\tau}^{(k)}$, we use a Newton–Raphson type algorithm to solve the minimization of $\mathcal{L}^*(\bm{\alpha}_{\tau},\bm{\theta}_{\tau},\gamma_l,\lambda_l\mid \widehat{\bm{\theta}}_{\tau}^{(k)})$ with respect to $(\bm{\alpha}_{\tau},\bm{\theta}_{\tau})$. After it converges, we update $\bm W_{\tau}^{(k)}$ to $\bm W_{\tau}^{(k+1)}$ and then minimize $\mathcal{L}^*(\bm{\alpha}_{\tau},\bm{\theta}_{\tau},\gamma_l,\lambda_l\mid \widehat{\bm{\theta}}_{\tau}^{(k+1)})$.
We repeat this procedure until the sequences of minimizers $(\bm{\alpha}_{\tau}^{(j)},\bm{\theta}_{\tau}^{(j)})$ converge.
Note that in practice, the term $\dot{p}_{\lambda_l}(u^{(0)})/|u^{(0)}|$ in \eqref{scadd} can go to infinity if $|u^{(0)}|$ is very small,
which can cause the algorithm to be unstable.
{To solve this issue, we adopt the modified LQA precedure from \cite{hunter2005variable} for $p_{\lambda_l}(\cdot)$.}
We summarize the computational details in Section S1 of the supplementary document.
\textcolor{black}{Let $S_1$ and $S_2$ denote the candidate sets for the tuning parameters $\lambda_l$ and $\gamma_l$, respectively. 
We tune these parameters based on the following strategy. 
For a given pair $(\lambda_l,\gamma_l)$, with $\lambda_l \in S_1$ and $\gamma_l \in S_2$, we first fit model \eqref{lqaloss} to identify the estimated null and non-null subregions, denoted by $\widehat{\mathcal{N}}(\beta_{\tau,l})$ and $\widehat{\mathcal{S}}(\beta_{\tau,l})$, respectively.
Next, we refit model \eqref{lqaloss} using only the data within the estimated signal region $\widehat{\mathcal{S}}(\beta_{\tau,l})$, i.e., $\bm{X}_i(t)$ for $t \in \widehat{\mathcal{S}}(\beta{\tau,l})$, and without applying the fSCAD penalty.
We then evaluate each candidate pair $(\lambda_l,\gamma_l)$ using the Bayesian Information Criterion (BIC) \citep{lee2014model}, and select the pair that minimizes the BIC.}

\vspace{-0.5cm}
\section{Large Sample Properties}\label{section3}
For any vector, we shall use $\|\cdot\|_{2}$ to denote the Euclidean norm. The null region and nonnull region of $\beta_{\tau,l}(t),l=1,\ldots,m$ are denoted by $\mathcal{N}(\beta_{\tau,l})$ and $\mathcal{S}(\beta_{\tau,l})$, respectively, where $\mathcal{N}(\beta_{\tau,l}) = \left\{t\in [0,\mathcal{T}]:\beta_{\tau,l}(t) = 0\right\}$ and $\mathcal{S}(\beta_{\tau,l}) = \left\{t\in [0,\mathcal{T}]:\beta_{\tau,l}(t) \neq 0\right\}$.
$\bm{\mathcal{B}}_{\tau,l,1}(t)$ denotes the $K_{\tau,l}^*$ dimensional sub-vector of $\bm{\mathcal{B}}(t)$ 
such that each $B_{j}(t)$ in $\bm{\mathcal{B}}_{\tau,l,1}(t)$ has a support inside $\mathcal{S}(\beta_{\tau,l})$.
Let $\bm{U}^*_{\tau}$ and $\bm{U}^*_{\tau,i}$ be a $K^*_{\tau} = \sum_{l=1}^m K_{\tau,l}^*$ dimensional vector of the corresponding elements of $\bm{U}$ and $\bm{U}_i$, respectively, associated with $\bm{\mathcal{B}}_{\tau,1,1},\ldots,\bm{\mathcal{B}}_{\tau,m,1}$, where $\bm{U}=\int_o^{\mathcal{T}} \bm{X}(t)\otimes \bm{\mathcal{B}}(t)dt$. We also define $\bm{Z}_{\tau}^* = (\bm{Z}^T,\bm{U}_{\tau}^{*T})^T$, $\bm{\Sigma}_{\tau,1}=E\left\{\bm{Z}^*_{\tau}\bm{Z}_{\tau}^{*T}\right\}$ and $\bm{\Sigma}_{\tau,2}= E\left\{f_{e|\bm{Z},\bm{X}}(0)\bm{Z}_{\tau}^*\bm{Z}_{\tau}^{*T}\right\}$. All the technical conditions needed for the theorems are listed in \textcolor{black}{Section S2 of the supplementary document.}

\vspace{-1cm}
\subsection{Functional Oracle Property and Asymptotic Normality}
\textcolor{black}{
}
\textcolor{black}{It is important to note that all theoretical results in this section rely on the assumption that the tuning parameters for the roughness penalty and the fSCAD penalty satisfy Conditions 5 and 9, respectively, and are assumed to be fixed and known.}

\begin{theory}
Under Conditions 1-9, there exists a local minimizer $(\widehat{\bm{\alpha}}_{\tau},\widehat{\bm{\beta}}_{\tau})$ of (\ref{conquer}) such that $\|\widehat{\bm{\alpha}}_{\tau}-\bm{\alpha}_{\tau}\|_2 = O_p(n^{-1/2})$ and $\|\widehat{\bm{\beta}}_{\tau}-\bm{\beta}_{\tau}\|_{2} = O_p(n^{-1/2}K^{1/2})$.
\end{theory}

From this theorem, it is clear that there exists a root-$n$ consistent estimator $\widehat{\bm{\alpha}}_{\tau}$ and a root-$n/K$ consistent estimator $\widehat{\bm{\beta}}_{\tau}(t)$.

\begin{theory}
[\textbf{Functional Oracle Property}]
\label{th2}
If Conditions 1-9 hold, as $n\rightarrow\infty$:
\begin{itemize} 
    \item [(i)] \textbf{Locally Sparsity}: 
    For every $t$ not in the support of $\beta_{\tau,l}(t)$, we have $\widehat{\beta}_{\tau,l}(t)=0$ with probability tending to one.
    \item [(ii)] \textbf{Asymptotic Normality}: 
    For $t$ such that $\beta_{\tau,l}(t)\neq 0$ we have
    $\sigma^{-1/2}_{\tau,l}(t)(\widehat{\beta}_{\tau,l}(t)-\beta_{\tau,l}(t))\overset{d}{\rightarrow} \mathbb{G}(t)$,
where $\sigma_{\tau,l}(t)=\tau(1-\tau)
        \bm{\Lambda}_{\tau,l}(t)\bm{\Sigma}_{\tau,2}^{-1}(\bm{\Sigma}_{\tau,1}/n)\bm{\Sigma}_{\tau,2}^{-1}\bm{\Lambda}^T_{\tau,l}(t)$, $\bm{\Lambda}_{\tau,l}(t)=\bm{\xi}_l^T\widetilde{\bm{\mathcal{B}}}_{\tau}(t)(\bm{0}_{{K^*_{\tau}}\times d},\bm{I}_{K^*_{\tau}})$, $\bm{\xi}_l$ is a $m\times 1$ unit vector in which the $l$-th element is 1, $\bm{I}_d$ be a $d\times d$ identity matrix, $\bm{0}_{d\times K_{\tau}^*}$ is a $d\times K_{\tau}^*$ zero matrix,
    $\widetilde{\bm{\mathcal{B}}}_{\tau,1}(t)=\text{Bdiag}
(\bm{\mathcal{B}}_{\tau,1,1}^T(t),  \cdots \bm{\mathcal{B}}_{\tau,m,1}^T(t))$,
and $\mathbb{G}(t)$ is a Gaussian random process with mean 0 defined on $S(\beta_{\tau,l})$ with
the covariance function
$\mathcal{C}(t,s)=\tau(1-\tau)\sigma^{-1/2}_{\tau,l}(t)\sigma^{-1/2}_{\tau,l}(s)
        \bm{\Lambda}_{\tau,l}(t)\bm{\Sigma}_{\tau,2}^{-1}(\bm{\Sigma}_{\tau,1}/n)\bm{\Sigma}_{\tau,2}^{-1}\bm{\Lambda}^T_{\tau,l}(s)$.
\end{itemize}
\end{theory}

\begin{theory}
[\textbf{Asymptotic Normality of the Estimators of Parameters}]
\label{th1}
If Conditions 1-9 hold, as $n\rightarrow\infty$, we have $\sqrt{n}(\widehat{\bm{\alpha}}_{\tau}-\bm{\alpha}_{\tau}^0)
        \overset{d}{\rightarrow}  \bm{\mathbb{N}}(\bm{0},\tau(1-\tau)\bm{\mathcal{I}}_{\tau}\bm{\Sigma}_{\tau,2}^{-1}\bm{\Sigma}_{\tau,1}\bm{\Sigma}_{\tau,2}^{-1}\mathcal{\bm{I}}_{\tau}^T)$,
    where $\bm{\alpha}_{\tau}^0$ is the true parameters and $\bm{\mathcal{I}}_{\tau}=(\bm{I}_d,\bm{0}_{d\times K^*_{\tau}})$.
\end{theory}

\begin{remark}
    Theorem 2 shows that the estimators $\widehat{\bm{\beta}}_{\tau}(t)$ possess the functional version oracle property, which makes the fitted model simpler and more interpretable.
Denote the left endpoint and right endpoint of $\mathcal{S}(\beta_{\tau,l})$ as $\mathcal{S}^L(\beta_{\tau,l})$ and  $\mathcal{S}^R(\beta_{\tau,l})$, respectively. Then we partition $\mathcal{S}(\beta_{\tau,l})$ into $\widetilde{K}_{\tau,l}+1$ equally spaced intervals with $\mathcal{S}^L(\beta_{\tau,l})=\nu_0<\nu_1<\cdots<\nu_{\widetilde{K}_{\tau,l}}<\nu_{\widetilde{K}_{\tau,l}+1}=\mathcal{S}^R(\beta_{\tau,l})$, where $\widetilde{K}_{\tau,l}\rightarrow \infty$.  
    \textcolor{black}{From Theorem 2, we have that for any $a \in (0,1)$,
$\lim_{n\to\infty}P\left\{\sup_{t\in \mathcal{S}_{\varepsilon}(\beta_{\tau,l})}\left|
\sigma_{\tau,l}^{-1/2}(t)\left\{\widehat{\beta}_{\tau,l}(t)-\beta_{\tau,l}(t)\right\}\right|\leq \mathcal{Q}_{\tau,l}(a) \right\} = 1-a,$
where $\mathcal{S}_{\varepsilon}(\beta_{\tau,l}) = \left\{\nu_0,\ldots,\nu_{\mathcal{N}_{\tau,l}+1}\right\}$ as a subset of $\mathcal{S}(\beta_{\tau,l})$ becomes denser as $n\to\infty$
, $\mathcal{Q}_{\tau,l}(a) = (2\log |\mathcal{S}_{\varepsilon}(\beta_{\tau,l})|)^{1/2} - (2\log |\mathcal{S}_{\varepsilon}(\beta_{\tau,l})|)^{-1/2}\left\{\log(-0.5\log(1-a))
+0.5\left[\log(\log |\mathcal{S}_{\varepsilon}(\beta_{\tau,l})|) + \log (4\pi)\right]\right\}$, and $|\mathcal{S}_{\varepsilon}(\beta_{\tau,l})|$ denote the cardinality of the set
$|\mathcal{S}_{\varepsilon}(\beta_{\tau,l})|$. 
Then an asymptotic $100(1-a)\%$ simultaneous confidence band (SCB)
for $\beta_{\tau,l}(t)$ over $\mathcal{S}_{\varepsilon}(\beta_{\tau,l})$ is given by $\widehat{\beta}_{\tau,l}(t)\pm\sigma_{\tau,l}^{1/2}(t)\mathcal{Q}_{\tau,l}(a)$.}
In addition, for any given fixed point $t$ such that $\beta_{\tau,l}(t)\neq 0$, we have $\sigma_{\tau,l}^{-1/2}(t)(\widehat{\beta}_{\tau,l}(t)-\beta_{\tau,l}(t))\overset{d}{\rightarrow}  \mathbb{N}\left(\bm{0},1\right)$, then the asymptotic $100(1-a)\%$ point-wise confidence band (PCB) of $\beta_{\tau,l}$ is $\widehat{\beta}_{\tau,l}(t)\pm \sigma_{\tau,l}^{1/2}(t)z_{a/2}$ and the width of SCB is inflated by $\mathcal{Q}_{\tau,l}(a)/z_{a/2}$, where $z_{a/2}$ is the upper $a/2$-quantile of the standard normal distribution. 

Theorem 3 establishes the asymptotic normality of the parameters $\alpha_{\tau,l}$ and its corresponding $100(1-a)\%$ point-wise confidence interval is $\widehat{\alpha}_{\tau,l}\pm n^{-1/2}\sqrt{\tau(1-\tau)\widetilde{\bm{\xi}}_l^T\bm{\mathcal{I}}_{\tau}\bm{\Sigma}_{\tau,2}^{-1}\bm{\Sigma}_{\tau,1}\bm{\Sigma}_{\tau,2}^{-1}\mathcal{\bm{I}}_{\tau}^T\widetilde{\bm{\xi}}_l}z_{a/2}$, where
$\widetilde{\bm{\xi}}_l$ is a $d\times 1$ unit vector in which the $l$-th element is 1 and $l=1,\ldots,d$.
\end{remark}

\vspace{-1cm}
\subsection{Wild Bootstrap}
\textcolor{black}{In Theorems \ref{th2} and \ref{th1}, the covariance function of $\widehat{\beta}_{\tau,l}(t)$: $\mathcal{C}(t,s)=\tau(1-\tau)\sigma^{-1/2}_{\tau,l}(t)\sigma^{-1/2}_{\tau,l}(s)\bm{\Lambda}_{\tau,l}(t)\\
\bm{\Sigma}_{\tau,2}^{-1}(\bm{\Sigma}_{\tau,1}/n)
\bm{\Sigma}_{\tau,2}^{-1}\bm{\Lambda}^T_{\tau,l}(s)$ 
and the covariance matrix of $\widehat{\bm{\alpha}}_{\tau}$:
$n^{-1}\tau(1-\tau)\bm{\mathcal{I}}_{\tau}\bm{\Sigma}_{\tau,2}^{-1}\bm{\Sigma}_{\tau,1}\bm{\Sigma}_{\tau,2}^{-1}\mathcal{\bm{I}}_{\tau}^T$ are both unknown because that
$\bm{\Sigma}_{\tau,1}=E\left\{\bm{Z}^*_{\tau}\bm{Z}_{\tau}^{*T}\right\}$ and $\bm{\Sigma}_{\tau,2}= E\left\{f_{e|\bm{Z},\bm{X}}(0)\bm{Z}_{\tau}^*\bm{Z}_{\tau}^{*T}\right\}$
are both unknown.}
Estimating the unknown conditional density function of the error given both scalar and functional covariates is particularly challenging, making accurate estimation of the covariance function and matrix essential before constructing SCBs and pointwise confidence intervals. Motivated by the wild bootstrap for classical quantile regression \citep{feng2011wild} and the refitted wild bootstrap for high-dimensional quantile regression \citep{cheng2022regularized}, we propose a modified wild bootstrap procedure for sparse semiparametric functional quantile regression to simultaneously estimate the covariance function of $\widehat{\beta}_{\tau,l}(t)$ and the covariance matrix of $\widehat{\bm{\alpha}}_{\tau}$. Detailed steps are provided in \textcolor{black}{Section S3 of the supplementary document}.

\begin{theory}\label{wild}
    Under the Conditions 1-9,
    using the wild bootstrap procedure
    , we have 
    $\sup_{u\in{\mathbb{R}}}\left|P\left\{\sup_{t\in \mathcal{S}_{\varepsilon}(\beta_{\tau,l})}\sqrt{n/K}(\widehat{\beta}_{\tau,l}^{(b)}(t)-\widehat{\beta}_{\tau,l}(t))\leq u \right\}-
    P\left\{\sup_{t\in \mathcal{S}_{\varepsilon}(\beta_{\tau,l})}\sqrt{n/K}(\widehat{\beta}_{\tau,l}(t)-\right.\right.\\
    \left.\left.\beta_{\tau,l}(t))\leq u \right\}\right|\to 0$,
    and 
    $\sup_{v\in{\mathbb{R}}}\left|P(n^{1/2}(\widehat{\alpha}_{\tau,l}^{(b)}-\widehat{\alpha}_{\tau,l})\leq v)-P(n^{1/2}(\widehat{\alpha}_{\tau,l}-\alpha_{\tau,l}^0)\leq v)\right|\to 0$.
\end{theory}
Theorem \ref{wild} provides a theoretical justification for using the wild bootstrap method to estimate the sample covariance function and covariance matrix for conducting statistical inference.

\vspace{-0.5cm}
\section{Soybean Yield Data Analysis}\label{application}
 \textcolor{black}{In this section, we apply the proposed locally sparse functional quantile regression model to study the relationship between soybean yield and daily temperature profiles, including both maximum and minimum temperatures. Functional quantile regression is particularly suitable in this context as it allows us to investigate how different parts of the temperature curve influence the conditional quantiles of crop yield, capturing both central tendencies and extreme responses. The locally sparse penalty plays a key role by enabling automatic identification of critical time intervals during the growing season when temperature has a significant effect on yield. Moreover, we further utilize the simultaneous confidence bands derived in the theoretical section to perform inference analysis in this real-world application, demonstrating the practical utility of our theoretical developments.
}

Climate factors such as temperature and rainfall have significant effects on soybean germination and growth. In North America, these climate factors can account for 15\% variation of the soybean yield \citep{vogel2019effects}. For the stability of soybean production, it is important to keep tracking these factors over the growing season. If we can figure out when and how the temperatures combined with other environmental and non-environmental factors influence the soybean yield, then we may obtain a better soybean planting and harvesting strategy. We can also make some interventions when the temperature is too low or too high. In addition, to have a more thorough understanding of this relationship, analyzing the conditional quantiles of soybean yield should be more meaningful than analyzing the conditional mean only. For these reasons, in this analysis, we want to identify the impact of daily minimum and daily maximum temperatures on the soybean yield for different quantiles.

Kansas in the United States has 4.7 million acres of soybean planted, producing 200 million bushels, and ranks 10th in the United States in terms of soybean yield. The data on soybean yield and other related non-environmental variables in Kansas state between 1991 and 2006 are collected and organized from the United States Department of Agriculture (USDA) website (\url{https://quickstats.nass.usda.gov/}).
The corresponding measurements on climate variables are collected from the National Oceanic and Atmospheric Administration (NOAA) website (\url{https://www.ncei.noaa.gov/products}) for Kansas within the same time range.

\begin{figure}[htbp]
\centering
    \includegraphics[width=10cm]{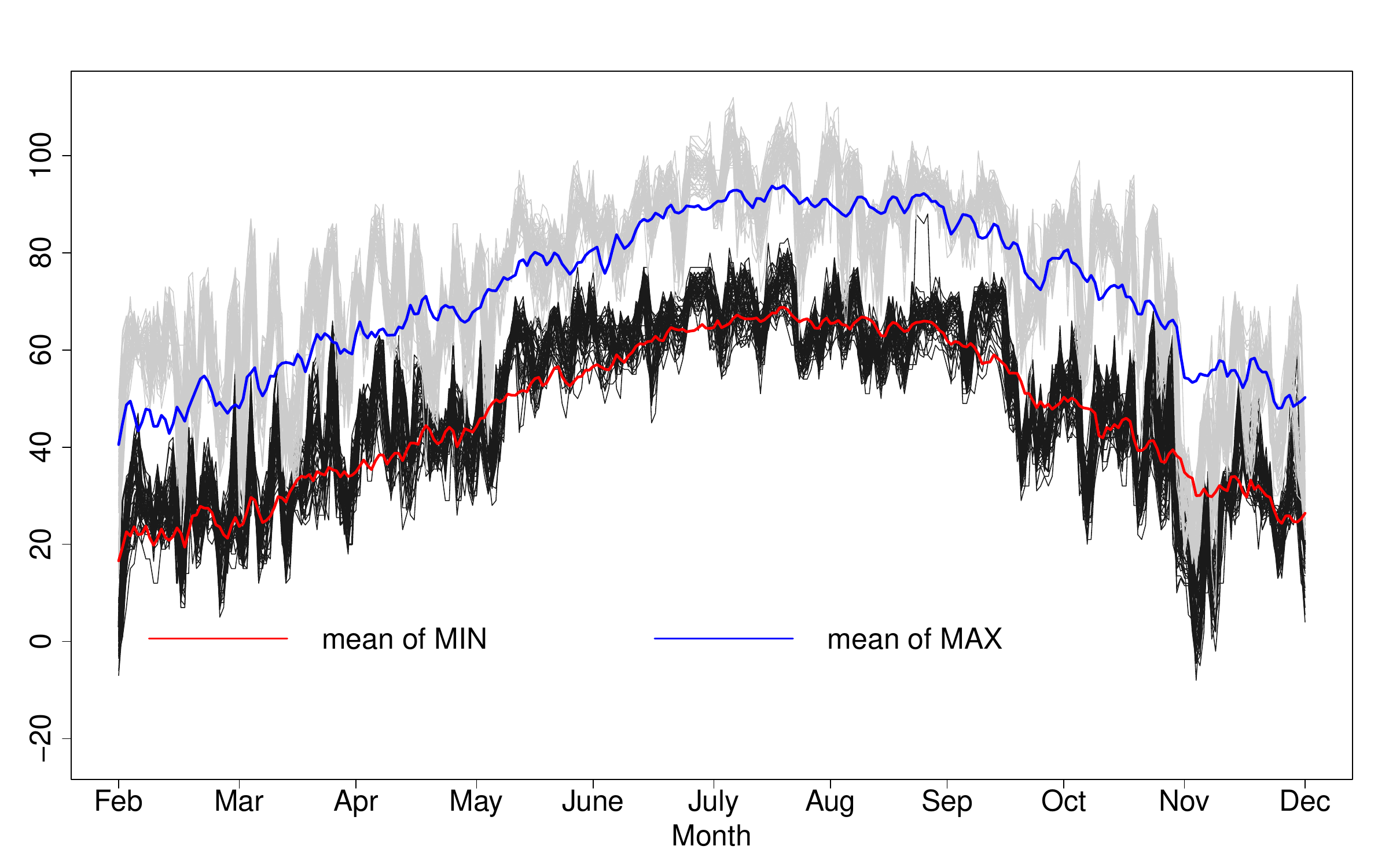}
\caption{A sample of daily minimum and maximum temperature curves of counties in Kansas. The unit of the y-axis is the Fahrenheit temperature scale.}
\label{raw_temp_data}
\end{figure}

To be more specific, the soybean yield-related data collected from the USDA website contains the county-level annual soybean yield (measured in bushels per acre), the size of harvested land and the size of irrigated area among each harvested land. The climate data collected from the NOAA website contains daily minimum temperatures, daily maximum temperatures and annual precipitations at the climate station level.
To link the observations from different websites together, we first identify the latitudes and longitudes of each climate station and the center of each county of Kansas. Next, we label the location of each climate station by comparing its distance to all the county centers. More specifically, for each climate station, the closest county center is its label of location. To obtain the county-level daily minimum and maximum temperature and annual precipitation observations, we average the corresponding climate station level observations over all the climate stations within each county. In this way, we integrate all the observations at the county level. {Also note that the data set from the USDA website only includes observation of a number of counties for each year in Kansas but not all of them.} In the following analysis, we treat the annual precipitation, and the ratio of the size of the irrigated area over the size of harvested land of each county as two scalar predictors and we treat daily minimum temperature and daily maximum temperature curves as two functional predictors.
\figref{raw_temp_data} displays a sample of the daily minimum temperature and daily maximum temperature curves of Kansas.

Let $X_1(t)$ and $X_2(t)$ denote the daily maximum temperature and daily minimum temperature between February and November respectively.
Let $Z_1$ and $Z_2$ denote the annual precipitation and the ratio of irrigated area of each county in Kansas.
Let $Y$ denote the annual soybean yield of each county in Kansas.
The model we want to investigate is 
$Q_{\tau}(Y|X_1(\cdot),X_2(\cdot),Z_1,Z_2) = c(\tau) + 
    \alpha_{1,\tau}Z_1+ \alpha_{2,\tau}Z_2 + 
    \int_{\mathcal{T}}X_1(t)\bm{\beta}_{1,\tau}(t)dt + \int_{\mathcal{T}}X_2(t)\bm{\beta}_{2,\tau}(t)dt$.
We fit the model at three different quantiles $\tau = 0.25, 0.5, 0.75$, which represent three different scenarios of the soybean yield: the ``worst'' case, the normal case and the ``best'' case.
We apply the bootstrap procedure mentioned above to compute the simultaneous confident bands for $\beta_{1,\tau}$ and $\beta_{2,\tau}$, and the 90\%-confidence intervals for $\alpha_{1,\tau}$ and $\alpha_{2,\tau}$.


\begin{figure}[htbp]
  \centering
  \subfigure[$\beta_{1,0.25}(\cdot)$]{\includegraphics[width=5cm]{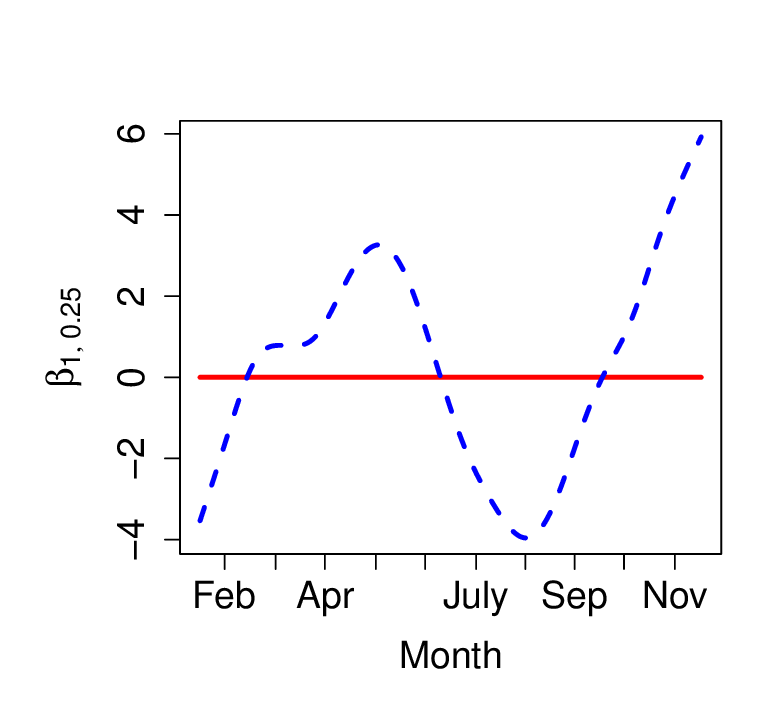}}
  \subfigure[$\beta_{1,0.5}(\cdot)$]{\includegraphics[width=5cm]{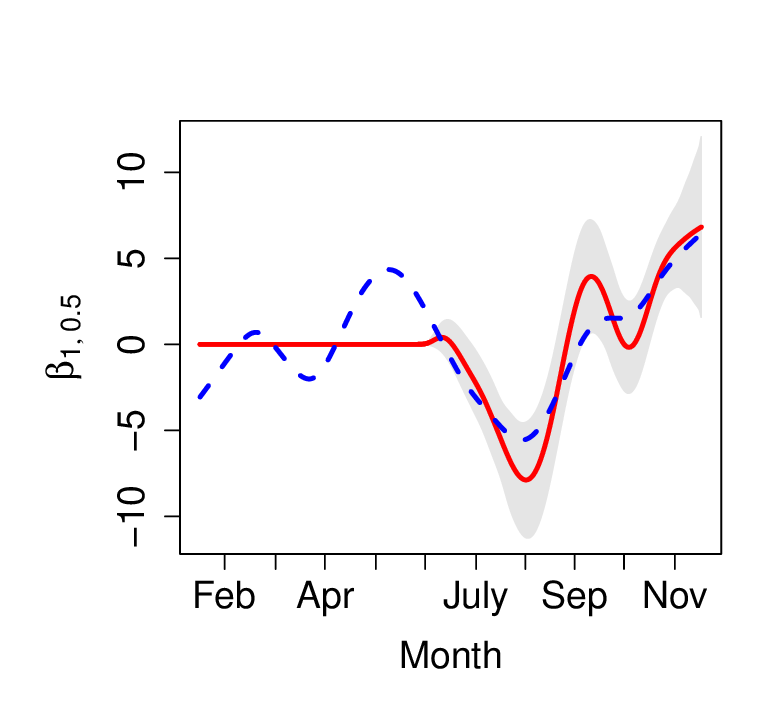}}
  \subfigure[$\beta_{1,0.75}(\cdot)$]{\includegraphics[width=5cm]{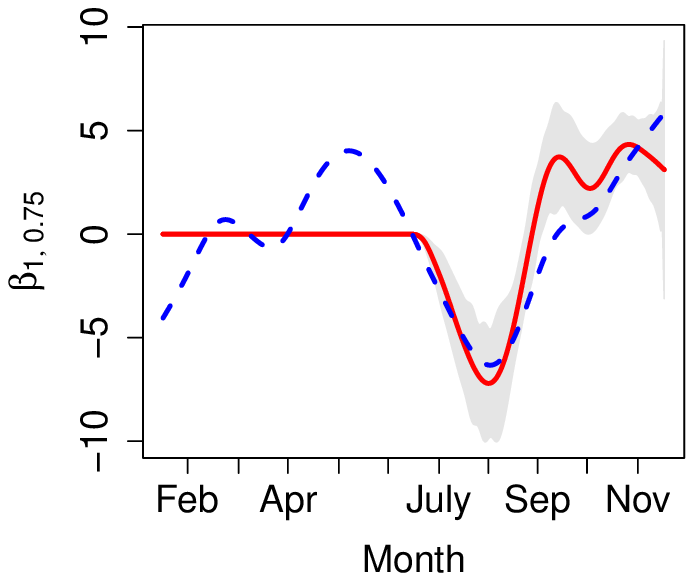}}
  \\\vspace{-0.4cm}
  \subfigure[$\beta_{2,0.25}(\cdot)$]{\includegraphics[width=5cm]{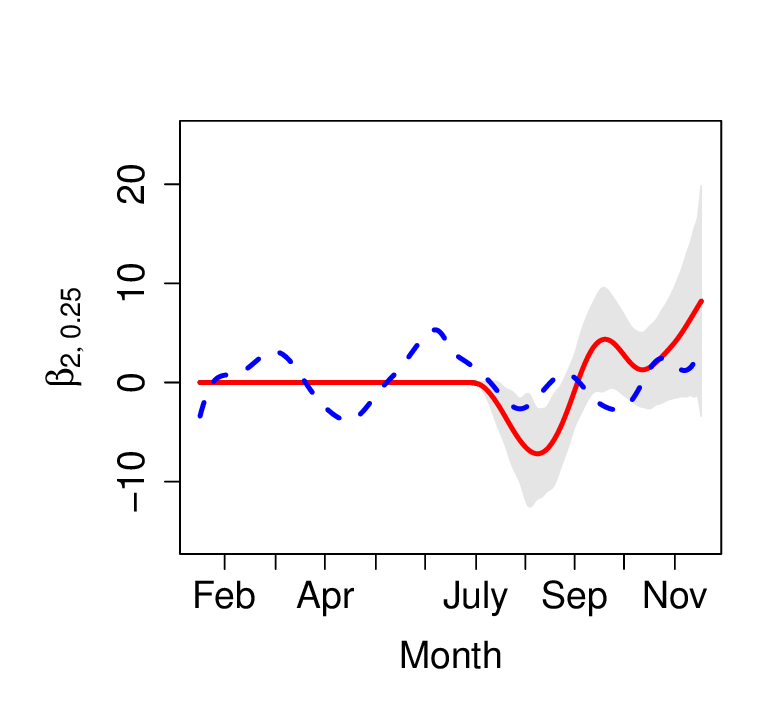}}
  \subfigure[$\beta_{2,0.5}(\cdot)$]{\includegraphics[width=5cm]{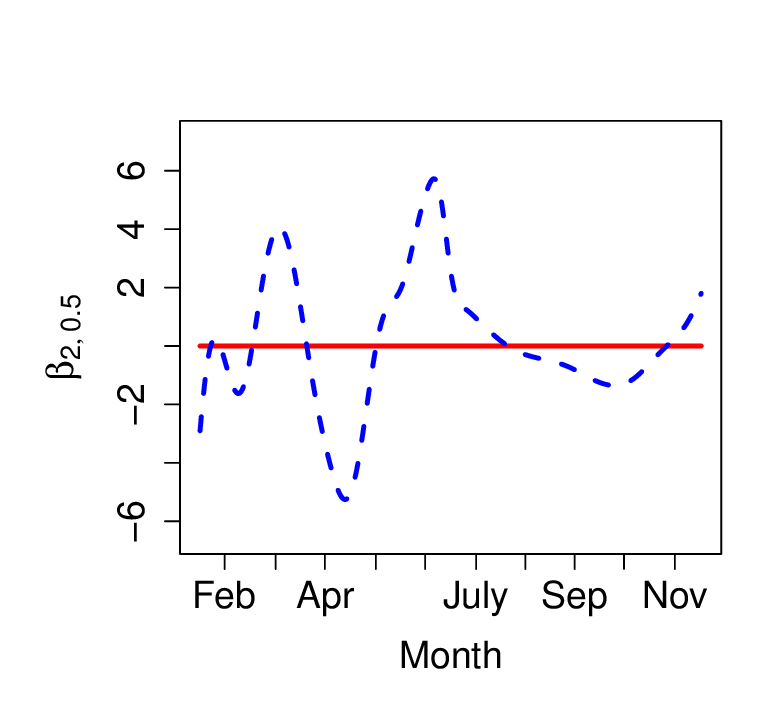}}
  \subfigure[$\beta_{2,0.75}(\cdot)$]{\includegraphics[width=5cm]{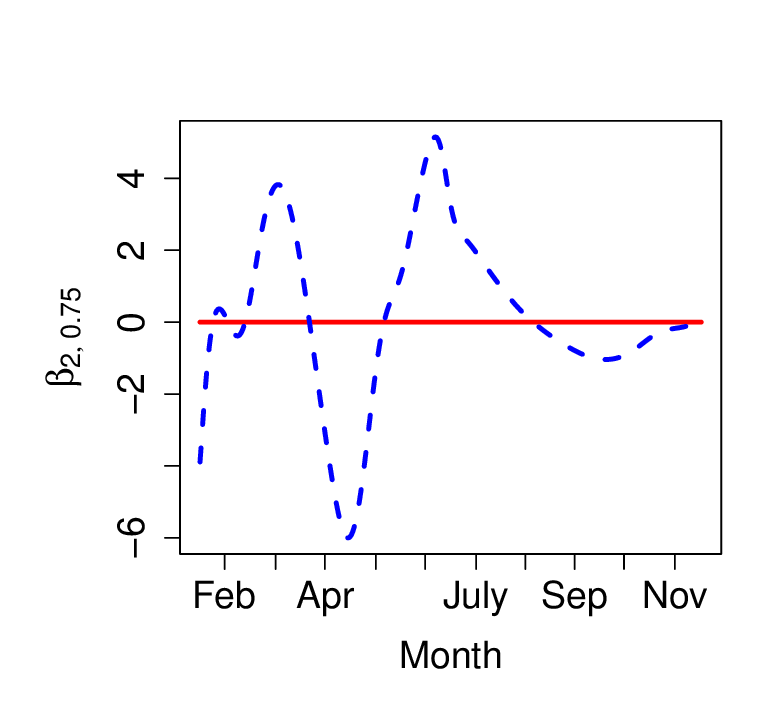}}
  \vspace{-0.35cm}
  \caption{The estimated slope functions using the convolution-smoothing based locally sparse estimation (CLoSE) method (red solid line) and the smoothed quantile loss (SQL) method (blue dashed line) at different quantile levels, $\tau=0.25, 0.50$ and $0.75$, from the soybean dataset of Kansas. The gray areas are the corresponding 95\% simultaneous confidence bands.}
\label{fig:scb}
\end{figure}

\figref{fig:scb} displays the simultaneous confident bands for $\beta_{1,\tau}$ and $\beta_{2,\tau}$ with $\tau=0.25,0.5$ and $0.75$. 
Specifically, 
\figref{fig:scb}-(a) and \figref{fig:scb}-(d) show that the
daily maximum temperature has no influence on the $25\%$ quantile of the soybean yield. Only the daily minimum temperature after late July matters for this ``worst'' scenario of the soybean yield. From \figref{fig:scb}-(e) and \figref{fig:scb}-(f), we observe that the daily minimum temperature has no effect on the soybean yield for the 50\% and 75\% quantiles of the soybean yield. 
On the other hand, \figref{fig:scb}-(b) and \figref{fig:scb}-(c) show that the daily maximum temperatures play an important role in these two quantiles of the soybean yield. More specifically, from \figref{fig:scb}-(b) and \figref{fig:scb}-(c), we can observe that regarding the 50\% and 75\% quantiles of the soybean yield, the maximum temperature during the hot summer has a negative effect and the maximum temperature during the fall and winter have a positive influence.
This may be due to the fact that before reaching $30^\circ C$ $(86^\circ F)$, the increasing air temperature can lead to an increase in soybean yield. But after reaching $30^\circ C$ $(86^\circ F)$, a higher temperature is negatively related to the soybean yield \citep{schlenker2009nonlinear}.
{According to the information from the website \url{https://webapp.agron.ksu.edu/agr_social/m_eu_article.throck?article_id=1152}, the most active soybean planting dates of different counties of Kansas range from mid-May to late June.} This can explain why we observe the sparsity of $\hat\beta_{1,\tau}(t)$ and $\hat\beta_{2,\tau}(t)$ between February and late June.
{In addition, the proposed CLoSE method tends to use only one of the two functional predictors: daily minimum and daily maximum temperature curves.
This may be due to the high correlation of these two functional predictors.
The two predictors may contain repetitive information and therefore using only one of them in the model should be enough.
While the SQL method uses whichever predictors input into the model, regardless of correlation.}

\begin{table}[htbp]
\centering
 \begin{threeparttable}
\caption{The estimate and 95\% Confidence Intervals (CI) for $\alpha_{1,\tau}$ and $\alpha_{2,\tau}$ at different quantile levels, $\tau = 0.25, 0.50$ and $0.75$, using the convolution-smoothing based locally sparse estimation (CLoSE) method from the soybean dataset of Kansas.}
\label{table:soybean_thetahat}
\renewcommand\arraystretch{1}
\setlength{\tabcolsep}{6mm}{
\begin{tabular}{crccc}
 \toprule
 \multirow{2}{*}{$\tau$} & \multicolumn{2}{c}{$\alpha_{1,\tau}$} &  \multicolumn{2}{c}{$\alpha_{2,\tau}$} \cr
  \cmidrule(rr){2-3} \cmidrule(lr){4-5} 
  & Estimate &  95\% CI  & Estimate & 95\% CI \cr
   \midrule
    0.25 & 19.26  & [-94.23, 84.60] & 15.19 & [13.03, 23.69]\cr
    0.50 & 5.36  & [-56.55, 14.34] & 16.69 & [14.33, 19.42]\cr
    0.75 & 7.23  & [-32.79, 28.42] & 15.17 & [13.22, 18.34]\cr
 \bottomrule
 \end{tabular}}
 \end{threeparttable}
\end{table}

Table \ref{table:soybean_thetahat} summarizes 95\%-confidence intervals for $\alpha_{1,\tau}$ and $\alpha_{2,\tau}$. We observe that for all three quantiles, the proportion of irrigated areas of each county is significant to the annual soybean yield, while the annual precipitation is not significant for any quantiles.

\vspace{-0.5cm}
\section{Conclusions and Discussion}\label{conclusion}

In this paper, {we propose a locally sparse semi-parametric functional quantile model to study the dynamic dependence of scalar response on functional covariates.} A convolution-smoothing based locally sparse estimation (CLoSE) method is developed to identify the locally sparse regions of the functional coefficients and estimate the parameters and the functional coefficients on the non-null regions. We also establish the functional version of the oracle property of the functional coefficients and the asymptotic properties of the estimated parameters. 
The CLoSE method addresses 
the difficulty in computation 
by first convolving the quantile loss function with a smoothing kernel. This step smooths out the function and makes it easier to optimize. The CLoSE method has been shown to be effective in estimating the locally sparse semi-parametric functional quantile model in our simulation studies and real applications.

When the data size is large-scale, the computation burden brought by the fSCAD penalty and the selection of several tuning parameters is heavy even if we use the convolution-smoothed quantile loss. Improving the computation efficiency when analyzing large-scale functional data is indeed an important topic worth investigating in future research. One possible approach is to develop new algorithms that can handle large-scale data more efficiently. For example, one could consider using parallel computing techniques or developing distributed algorithms that can run on clusters of computers.


\vspace{-0.5cm}
\section*{Supplementary Materials}
\par
The supplementary document includes the summarized algorithm of the proposed Convolution-smoothing based Locally Sparse Estimation (CLoSE) method, and 
the required conditions, the wild bootstrap method, simulation studies and 
technical proofs of the theorems. The computing scripts for replicating the simulation studies and the real data application are also provided.

\vspace{-1cm}
\section*{Declarations}
\textbf{Conflict of interest} The authors have no conflicts of interest to declare.

\vspace{-0.5cm}
\bibliographystyle{apalike}      
\bibliography{bibfile,FDA,revise}   

\CJKindent
\end{CJK}

\end{document}


\def\spacingset#1{\renewcommand{\baselinestretch}%
{#1}\normalsize} \spacingset{1.5}

\bigskip

\title{\large\bf Supplementary Document for the Manuscript entitled\\
``Convolution-Smoothing Based Locally Sparse Estimation for Functional 
   Quantile Regression
''}
\author{Hua Liu, Boyi Hu, Jinhong You, Jiguo Cao}
\author{}
\date{}
\maketitle


The supplementary document includes the summarized algorithm of the proposed Convolution-smoothing based Locally Sparse Estimation (CLoSE) method, 
the required conditions,
the wild bootstrap method, simulation studies and 
technical proofs of the theorems in the manuscript.

\appendix
\renewcommand{\appendixname}{Proofs of main results}
\setcounter{equation}{0}
\numberwithin{equation}{section}
\renewcommand{\thetable}{S.\arabic{table}}
\renewcommand{\thefigure}{S.\arabic{figure}}
\def\theequation{S\arabic{section}.\arabic{equation}}
\def\thesection{S\arabic{section}}

\section{The Convolution-smoothing based Locally Sparse
Estimation (CLoSE) Method}
We summarize the computational details of the proposed Convolution-smoothing based Locally
Sparse Estimation (CLoSE) method in Algorithm \ref{al1}.
Note that for each iteration, $\eta_{j,k}$ is chosen such that the objective function decreases after the update.
\IncMargin{1em}
\begin{algorithm}[htbp]
\SetKwData{Left}{left}\SetKwData{This}{this}\SetKwData{Up}{up}
\SetKwFunction{Union}{Union}\SetKwFunction{FindCompress}{FindCompress}
\SetKwInOut{Input}{Input}\SetKwInOut{Output}{Output}
\Input{Data, quantile level $\tau$, bandwidth $h$}

\BlankLine
\textbf{Initialization}: 
\begin{equation*}
\begin{pmatrix}
\widehat{\bm{\alpha}}_{\tau}^{(0)}\\
\widehat{\bm{\theta}}_{\tau}^{(0)}
\end{pmatrix}
    =\argmin_{\bm{\alpha}_{\tau},\bm{\theta}_{\tau}}\left\{
    \mathcal{L}_0^*(\bm{\alpha}_{\tau},\bm{\theta}_{\tau})
+\bm{\theta}^T_{\tau}(\bm{\Gamma}\otimes\bm{V})\bm{\theta}_{\tau}\right\},
\end{equation*}
where the tuning parameters $\gamma_l$ in the matrix $\bm{\Gamma}$ are selected by cross validation\;
\BlankLine

{Let $k_2=1$}

\While{not converged}{
{Let $k_1=1$, $\widehat{\bm{\alpha}}_{\tau}^{(0,k_2)}=\widehat{\bm{\alpha}}_{\tau}^{(k_2-1)}$ and 
$\widehat{\bm{\theta}}_{\tau}^{(0,k_2)}=\widehat{\bm{\theta}}_{\tau}^{(k_2-1)}$};\

\While{not converged}{

Update $\widehat{\bm{\alpha}}_{\tau}^{(k_1,k_2)}$ and $\widehat{\bm{\theta}}_{\tau}^{(k_1,k_2)}$ as follows:\
\begin{equation*}
    \begin{pmatrix}
    \widehat{\bm{\alpha}}_{\tau}^{(k_1,k_2)}\\
    \widehat{\bm{\theta}}_{\tau}^{(k_1,k_2)}
    \end{pmatrix}=
    \begin{pmatrix}
    \widehat{\bm{\alpha}}_{\tau}^{(k_1-1,k_2)}\\
    \widehat{\bm{\theta}}_{\tau}^{(k_1-1,k_2)}
    \end{pmatrix}-\eta_{k_1,k_2}\left\{\bm{D}_2^{(k_1-1,k_2)}\right\}^{-1}\bm{D}_1^{(k_1-1,k_2)},
\end{equation*}
where 
\begin{equation*}
    \bm{D}_1^{(k_1-1,k_2)} = \dot{\mathcal{L}}^{*}_0\left(\widehat{\bm{\alpha}}_{\tau}^{(k_1-1,k_2)},
    \widehat{\bm{\theta}}_{\tau}^{(k_1-1,k_2)}\right) 
    +2 \begin{pmatrix}
    \bm{0}_{d\times d} &\\
    & \bm{W}_{\tau}^{(k_2-1)}+\bm{\Gamma}\otimes \bm{V}
    \end{pmatrix}\begin{pmatrix}
    \widehat{\bm{\alpha}}_{\tau}^{(k_1-1,k_2)}\\
    \widehat{\bm{\theta}}_{\tau}^{(k_1-1,k_2)}
    \end{pmatrix},
\end{equation*}
\begin{equation*}
    \bm{D}_2^{(k_1-1,k_2)} = \ddot{\mathcal{L}}^{*}_0\left(\widehat{\bm{\alpha}}_{\tau}^{(k_1-1,k_2)},
    \widehat{\bm{\theta}}_{\tau}^{(k_1-1,k_2)}\right) 
    +2 \begin{pmatrix}
    \bm{0}_{d\times d} &\\
    & \bm{W}_{\tau}^{(k_2-1)}+\bm{\Gamma}\otimes \bm{V}
    \end{pmatrix}.
\end{equation*}

\
Let $k_1:=k_1+1$.}\
The limit is denoted as 
$\widehat{\bm{\alpha}}_{\tau}^{(k_2)}$ and 
$\widehat{\bm{\theta}}_{\tau}^{(k_2)}$

Update $\bm{W}_{\tau}^{(k_2)}$;

Let $k_2:=k_2+1$;
}
\BlankLine
\Output{The final estimators $\widehat{\bm{\alpha}}_{\tau}$ and $\widehat{\bm{\theta}}_{\tau}$}

\caption{
Algorithm for CLoSE Method
}\label{al1}
\end{algorithm}
\DecMargin{1em}

\newpage
\section{Conditions}\label{Conditions}
To establish the asymptotic results of the estimated parameters, and the oracle properties of the estimated functional coefficients, we first need to give some Conditions.
\begin{definition}\label{XZ-moment}
    (i) For the functional predictor $X_l(t),l=1,\ldots,m$, it holds that $\|X_l\|_2$ is almost surely bounded, where $\|X_l\|_2^2 = \int_0^{\mathcal{T}}X_l^2(t)dt$. Moreover, $\rho_{min}(\bm{U}\bm{U}^T)*K$ and $\rho_{max}(\bm{U}\bm{U}^T)*K$ are bounded away from $0$ and $\infty$ as $n\to \infty$, where $\rho_{\min}(\bm{A})$ and $\rho_{\max}(\bm{A})$
    denote the minimal and maximal eigenvalues of the matrix $\bm{A}$, respectively.
    (ii) For the scalar predictors, the components of $\bm{Z}$ have bounded support. $E(\bm{Z}\bm{Z}^T)$ is positive definite and has eigenvalues bounded
away from zero.
    
\end{definition}

\begin{definition}\label{beta-smooth}
Let $\upsilon$ be a nonnegative integer, and $\kappa\in(0,1]$ such that $r = \upsilon+\kappa\geq p+1$. We assume the unknown functional coefficient $\beta_{\tau,l}(\cdot)\in $  $\mathcal{H}^{(r)}(\mathcal{S}(\beta_{\tau,l}))$, which is the class of function $f$ on $\mathcal{S}(\beta_{\tau,l})$ whose $\upsilon$th deriative exists and satisfies a Lipschitz condition of order $\kappa$: $|f^{(\upsilon)}(t)-f^{(\upsilon)}(s)|\leq C_{\upsilon}|s-t|^{\kappa}$, for $s,t\in \mathcal{S}(\beta_{\tau,l})$ and some constant $C_{\upsilon}>0$.
\end{definition}

\begin{definition}\label{density}
The conditional density function $f_{e|\bm{Z},\bm{X}}(\cdot)$ is bounded away from zero, and second times continuously differentiable.
\end{definition}

\begin{definition}
    \label{kernel}
    Suppose $\mathcal{K}(\cdot)$ is a symmetric, bounded, continuous and non-negative function integrating to one, which means that $\mathcal{K}(u)=\mathcal{K}(-u)$, $\mathcal{K}\geq 0$ and $\int_{-\infty}^{\infty}\mathcal{K}(u)=1$ for all $u\in \mathbb{R}$.
    Furthermore, $\mathcal{K}(\cdot)$ is second-order continuously differentiable and bounded.
\end{definition}

\begin{definition}\label{knots}
   Let $\delta_j=t_{j+1}-t_j$ and $\delta = \max_{0\leq j\leq K} (t_{j+1}-t_j)$. There exists a constant $M > 0$, such that
    \begin{equation}\label{eq5}
    \delta/ \min_{0\leq j\leq K} (t_{j+1}-t_j)\leq M,\ \ \max_{0\leq j\leq K-1}|\delta_{j+1}-\delta_j|= o(K^{-1}).
    \end{equation}
\end{definition}

\begin{definition}\label{basis-number}
    The number of knots $K=o(\sqrt{n})$ and $K = \omega(n^{1/(2r+1)})$, where $K = \omega(n^{1/(2r+1)})$ means $K/n^{1/(2r+1)}\rightarrow\infty$ as $n\rightarrow\infty$.
\end{definition}

\begin{definition}\label{roughness}
    For the roughness penalty, we assume tuning parameter $\gamma_l,l=1\ldots,m$ satisfies that  $\gamma_l=o(n^{1/2}K^{1/2-2q}$), where $q\leq p$.
\end{definition}

\begin{definition}
    \label{bandwidth}
    The positive bandwidth $h$ satisfies $nh^4\to 0$ and $hK^r\to 0$.
\end{definition}

\begin{definition}
    \label{fscad}
    For the fSCAD penalty term, $\lambda_l\to 0$ as $n\to \infty$, 
    (i) $\max_{l}\sqrt{\int_{\mathcal{S}(\beta_{\tau,l}^0)}
        \dot{p}_{\lambda_l}(|\bm{\mathcal{B}}^T(t)\bm{\theta}_{\tau,l}^0|)^2dt} \\= o(n^{-1/2}K^{-1})$, and 
    $\max_{l}\sqrt{\int_{\mathcal{S}(\beta_{\tau,l}^0)}
        \ddot{p}_{\lambda_l}(|\bm{\mathcal{B}}^T(t)\bm{\theta}_{\tau,l}^0|)^2dt} =o(1)$.
    (ii) $K^{1/2}n^{-1/2}\lambda_l^{-1}\rightarrow 0$ and $\liminf_{n\rightarrow \infty}
    \liminf_{u\rightarrow 0^+}\dot{p}_{\lambda_l}(u)\lambda_l^{-1}>0$.
\end{definition}

\begin{remark}
    Condition \ref{XZ-moment}-(i) gives some conditions for functional predictors and is essential to obtain the oracle property of the estimators \citep{lin2017locally}.
    For the scalar predictors, Condition \ref{XZ-moment}-(ii) imposes some moment conditions. Condition \ref{beta-smooth} is about the smoothness of the functional coefficients $\beta_{\tau,l}(t)$, which has been widely used in the literature of nonparametric estimation. The common conditions on the conditional density function of $e(\bm{\alpha}_{\tau},\bm{\beta}_{\tau})$ in the quantile regression context is given in Condition \ref{density}.
    Condition \ref{kernel} is a necessary condition on the kernel function, which is also required in \cite{fernandes2021smoothing}, \cite{he2021smoothed} and \cite{tan2021high}. Condition \ref{knots} gives the assumption about the knots used in B-spline approximation, which implies that $\delta \sim K^{-1}$, i.e., $\delta$ and $K^{-1}$ are rate-wise equivalent from (\ref{eq5}). Condition \ref{basis-number} is imposed to make the spline approximation bias asymptotically negligible. Condition \ref{roughness} can make the shrinkage bias negligible brought by roughness penalty. Condition \ref{bandwidth} ensures that the convolution smoothing has an asymptotically negligible bias on the estimator of $\bm{\alpha}_{\tau}^0$ and $\beta_{\tau,l}$. Condition \ref{fscad}-(i) is the regularity condition to ensure the bias brought by the sparsity penalty is asymptotically negligible \citep{fan2004nonconcave}, and Condition \ref{fscad}-(ii) is used to obtain the functional oracle property of $\hat{\beta}_{\tau,l}(t)$.
\end{remark}

\clearpage
\section{Wild Bootstrap Method}
We randomly split the original dataset into two even parts and carry out this wild bootstrapping using 
the following steps:

\begin{enumerate}
[itemindent=2em]
\item [\bf{\textit{Step} 1}.] 
We first minimize the objective function (\ref{obj2}) based on the first part dataset to 
obtain the estimators $\widetilde{\bm\alpha}_{\tau,I}$ and $\widetilde{\bm\theta}_{\tau,I}$.
\item [\bf{\textit{Step} 2}.] 
Based on the non-zero coefficients identified by the estimator \(\widetilde{\bm{\theta}}_{\tau,I}\), we use the second part of the dataset to obtain the estimators using the convolution loss function. Specifically, we apply the roughness penalty only to the non-sparse regions identified in \textbf{\textit{Step 1}, without incorporating the fSCAD penalty. 
Define the set $\widehat{S}_{\bm{\theta},I} = \{j: \widetilde{\theta}_{\tau,I,j}\neq 0, 1\leq j\leq m(K+p)\}$ and $\widehat{\bm{T}}_{\bm{\theta},I} = \{\bm{v}\in \mathbb{R}^{m(K+p)}: v_j= 0, \forall j\in \widehat{S}_{\bm{\theta},I}^c\}$, where 
$\widehat{S}_{\bm{\theta},I}^c$ is the complement of the set $\widehat{S}_{\bm{\theta},I}$. To make the notation consistent, we denote the estimators based on the $Data_{II}$ as $\widehat{\bm{\alpha}}_{\tau,II}$ and $\widetilde{\bm{\theta}}_{\tau,II}$, which is from
\begin{equation}\label{dataI}
\begin{split}
     (\widehat{\bm{\alpha}}_{\tau,II},
    \widetilde{\bm{\theta}}_{\tau,II}) 
=  &  \arg\min_{\bm{\alpha}_{\tau},\bm{\theta}_{\tau}\in \widehat{\bm{T}}_{\bm{\theta},I} }
  \frac{1}{n_1}\sum_{i=1}^{n_1}\left(\rho_{\tau}*\mathcal{K}_h\right)(Y_i-\bm{Z}_i^T\bm{\alpha}_{\tau}-\bm{U}_i^T\bm{\theta}_{\tau})\\
   & +\bm{\theta}^T_{\tau}(\bm{\Gamma}\otimes\bm{V})\bm{\theta}_{\tau},
\end{split}
\end{equation}
then $\widetilde{\bm{\theta}}_{\tau,II}$ includes those zero terms obtained from \textbf{\textit{Step 1}}.
}

\item [\bf{\textit{Step} 3}.]
Independently generate weights $w_i$ such that 
\begin{itemize}
    \item [(1)] there are two positive constants $c_1$ and $c_2$ satisfying $-c_1=\sup\{w\in\mathbb{W}:w\leq 0\}$ and 
    $c_2=\inf\{w\in\mathbb{W}:w\geq 0\}$, where $\mathbb{W}$ is the support of $w$;
    \item [(2)] $E\left[w_i^{-1}I(w_i>0)\right]=-E\left[w_i^{-1}I(w_i<0)\right]=1/2$ and $E\left[|w_i|\right]<\infty$;
    \item [(3)] the $\tau$th quantile of $w$ is zero.
\end{itemize}
\item [\bf{\textit{Step} 4}.]
Calculate the residuals based on the second dataset: $\hat e_i = Y_i -\bm{Z}_i^T\hat{\bm\alpha}_{\tau,II}-\bm{U}_i^T\tilde{\bm\theta}_{\tau,II}$.
Then used the second part dataset to obtain the bootstrapped samples 
denoted by $Y_i^{(b)} = \bm{Z}_i^T\hat{\bm\alpha}_{\tau,II}+\bm{U}_i^T\tilde{\bm\theta}_{\tau,II} + e_i^{(b)}$, where $e_i^{(b)}=w_i|\hat e_i|$.
\item [\bf{\textit{Step} 5}.]
Resolve the objective function (\ref{dataI})
based on the bootstrapped samples
and denote the estimate by $\hat{\bm\alpha}_{\tau}^{(b)}$ and $\hat{\bm\theta}_{\tau}^{(b)}$.
\item [\bf{\textit{Step} 6}.]
Repeat \textbf{\textit{Step} 2 - \textit{Step} 5} $B$ times and then estimate the variance function 
$\sigma_{\tau,l}(t)$
of $\hat\beta_{\tau,l}(t)$ from the $B$ estimations at each $t$.  Meanwhile, we can also obtain the estimated covariance matrix of $\bm\alpha_{\tau}$.
\end{enumerate}
We repeat the \textbf{\textit{Step} 1 - \textit{Step} 6} by switching the two parts of the datasets. Subsequently, we compute the average of the estimated variance function and covariance matrix derived from two parts of the dataset to obtain the final estimators of the variance function and covariance matrix.




\clearpage
\section{Simulation Studies}\label{simulation}
In this section, we conduct simulation studies to evaluate the finite sample performance of our proposed CLoSE method in comparison with the smoothed quantile loss (SQL) method. The difference between the SQL method and the CLoSE method is that the SQL method only uses the combination of the smoothed quantile loss and the roughness penalty to estimate $\beta_{\tau}(t)$, while the CLoSE method contains the additional local sparse penalty in the objective function to estimate $\beta_{\tau}(t)$. 

\subsection{Simulations with One Functional Covariate}
We first use the following semi-parametric functional quantile model with one functional covariate to generate data:
\begin{equation}
    Y_i = 
    \bm{Z}_i^T\bm{\alpha}_{\tau}+
    \int_{-1}^{1}X_i(t)\beta_{\tau}(t)d t + e_i(\tau), i =1,\ldots,n,
    \label{eq:simulation_one_covariate}
\end{equation}
where $X_i(t)$ is a Wiener process, $Z_{i1}, Z_{i2}\overset{iid}{\sim }N(0,\sigma_z^2)$ with $\sigma_z=0.1$, $\bm{Z}_i = (1,Z_{i1},Z_{i2})^T$, $\bm{\alpha}_{\tau}=(0,1,1)^T$, $\beta_{\tau}(t) = \sin(2\pi t)\mathbbm{1}\{-0.5\leq t\leq 0.5\}$, and $e_i(\tau) = e_i - F_e^{-1}(\tau)$, where $F_e(\cdot)$ is the CDF of $e_i$. For data generating model \eqref{eq:simulation_one_covariate}, we use the following two scenarios to generate $e_{i}$.
\begin{itemize}
    \item \textbf{Scenario I:} $e_{i}$ are i.i.d from the normal distribution, namely, $e_{i}\overset{iid}\sim N(0,\sigma^2)$ with $\sigma = 0.02$.
    Under this scenario, the distribution of the random error is symmetric.
    \item \textbf{Scenario II:} $e_{i}$ are i.i.d from a heavy-tailed distribution, Cauchy distribution, namely, $e_{i}\overset{iid}\sim C(0,0.01)$.
\end{itemize}

In the simulations, we use cubic B-spline basis functions. In the convolution smoothing loss function, we use the Gaussian kernel $\mathcal{K}(u)= (2\pi)^{-1/2}e^{-u^2/2}$, and bandwidth $h=((K+p+d)/n)^{2/5}$ \citep{he2021smoothed}. Then the corresponding smoothed loss function is
$
\left(\rho_{\tau}*\mathcal{K}_h\right)(u) = (h/2)l_G(u/h) + (\tau-1/2)u
$
where $l_G(u) = (2/\pi)^{1/2}e^{-u^2/2} + u(1-2\Phi(-u))$ and $\Phi(\cdot)$ is the cumulative distribution function of the standard normal distribution. The simulation is repeated 100 times.

Regarding the estimation for $\beta_{\tau}(t)$, we consider two aspects of the performance of the proposed estimator $\hat \beta_{\tau}(t)$: (i) identification of null and nonnull regions and (ii) the difference between $\hat \beta_{\tau}(t)$ and $\beta_{\tau}(t)$.
Let $T$ denote the set of time points where $X(t)$ is evaluated.
Define the null region $\mathcal{N}(\beta_{\tau}) = \{t\in T:\beta_{\tau}(t)=0\}$ and the nonnull region $\mathcal{S}(\beta_{\tau}) = \{t\in T:\beta_{\tau}(t)\not=0\}$.
Define the true discovery rate (TDR) for $\mathcal{N}(\beta_{\tau})$ and the false discovery rate (FDR) for $\mathcal{S}(\beta_{\tau})$ as 
$\text{TDR} = |\mathcal{N}(\hat\beta_{\tau})\cap \mathcal{N}(\beta_{\tau})|/|\mathcal{N}(\beta_{\tau})|$ and $\text{FDR} = |\mathcal{N}(\hat\beta_{\tau})\cap \mathcal{S}(\beta_{\tau})|/|\mathcal{S}(\beta_{\tau})|$.
We use 
them to evaluate the performance of the proposed method on identifying the null and nonnull regions of $\beta_{\tau}(t)$ for $\tau = 0.1, \ldots, 0.8$. Ideally, we want a larger TDR and a smaller FDR. A larger TDR means more null regions of functional predictors are correctly detected. A smaller FDR means a less nonnull region of functional predictors is mistakenly identified as the null region.  

\begin{table}[htbp]
\centering
\caption{True discovery rate (TDR) for the null region $\mathcal{N}(\beta_{\tau})$ and false discovery rate (FDR) for the nonnull region $\mathcal{S}(\beta_{\tau})$ using the convolution -smoothing based locally sparse estimation (CLoSE) method when the errors are simulated from the Normal or Cauchy distribution. Here $n$ denotes the sample size and $\tau$ denotes the quantile.}
\renewcommand\arraystretch{1}
\setlength{\tabcolsep}{8mm}{
\begin{tabular}{cc|cc|cc}
\toprule
& & \multicolumn{2}{c|}{Normal} &\multicolumn{2}{c}{Cauchy} \\
\hline
 n & $\tau$ & TDR & FDR & TDR & FDR\\
 \hline
 & 0.2 & 98.5\% & 4.2\% & 89.3\% & 17.2\%\\
 & 0.3 & 99.2\% & 4.3\% & 95.1\% & 13.9\%\\
 & 0.4 & 98.9\% & 3.3\% & 88.6\% & 17.5\%\\
 500 & 0.5 & 98.7\% & 2.8\% & 98.4\% & 10.7\%\\
 & 0.6 & 98.7\% & 3.1\% & 92.5\% & 17.6\%\\
 & 0.7 & 99.0\% & 3.5\% & 96.1\% & 17.2\%\\
 & 0.8 & 98.6\% & 3.8\% & 86.9\% & 16.3\%\\
 \hline
 & 0.2 & 99.3\% & 2.5\% & 95.8\% & 12.2\%\\
 & 0.3 & 98.9\% & 2.7\% & 98.6\% & 11.6\%\\
 & 0.4 & 95.9\% & 1.9\% & 99.9\% & 11.5\%\\
1000 & 0.5 & 99.5\% & 2.1\% & 99.6\% & 7.8\%\\
 & 0.6 & 96.1\% & 1.5\% & 99.7\% & 10.6\%\\
 & 0.7 & 98.6\% & 2.0\% & 97.5\% & 12.6\%\\
 & 0.8 & 98.5\% & 2.4\% & 93.4\% & 13.6\%\\
\bottomrule
\end{tabular}}\label{table:TDR}
\end{table}

The simulation results on TDR and FDR are summarized in Table \ref{table:TDR}. Table \ref{table:TDR} shows that the TDR is overall satisfying for all chosen quantiles under both scenarios. The FDR of the Normal scenario is much better than that of the Cauchy scenario. This may be due to the fact that the Cauchy distribution is heavy-tailed and the variance of the error cannot be controlled. However, under both scenarios, FDR decreases as the sample sizes increase for all chosen quantiles, which is desirable.

\begin{table}[htbp]
\centering
\caption{The $L_2$-norm of the difference between the estimator $\hat{\beta}_{\tau}(t)$ and the true function $\beta_{\tau}(t)$ $\|\hat{\beta}_{\tau}(t,u)-\beta_{\tau}(t)\|_2$ using the convolution-smoothing based locally sparse estimation (CLoSE) method and the smoothed quantile loss (SQL) method when the errors are simulated from the normal or Cauchy distribution. Here $n$ denotes the sample size and $\tau$ denotes the quantile.}
\renewcommand\arraystretch{1}
\setlength{\tabcolsep}{8mm}{
\begin{tabular}{cc|cc|ccc}
 \toprule
 & & \multicolumn{2}{c|}{Normal} &\multicolumn{2}{c}{Cauchy}\\
 \hline
 n & $\tau$  & SQL & CLoSE & SQL & CLoSE \\
  \hline
    & 0.2 & 0.058 & 0.028 & 0.110 & 0.097 \\
    & 0.3 & 0.037 & 0.027 & 0.090 & 0.086\\
    & 0.4 & 0.039 & 0.026 & 0.102 & 0.091\\
    500 & 0.5 & 0.036 & 0.023 & 0.076 & 0.046\\
    & 0.6 & 0.043 & 0.028 & 0.103 & 0.092\\
    & 0.7 & 0.039 & 0.024 & 0.093 & 0.088\\
    & 0.8 & 0.064 & 0.026 & 0.105 & 0.088\\
  \hline
    & 0.2 & 0.043  & 0.026 & 0.075 & 0.070\\
    & 0.3 & 0.041  & 0.026 & 0.069 & 0.057\\
    & 0.4 & 0.036  & 0.028 & 0.065 & 0.062\\
 1000 & 0.5 & 0.029  & 0.022 & 0.057 & 0.035\\
    & 0.6 & 0.037  & 0.027 & 0.065 & 0.058\\
    & 0.7 & 0.044  & 0.025 & 0.071 & 0.059\\
    & 0.8 & 0.044  & 0.026 & 0.082 & 0.075\\   
 \bottomrule
 \end{tabular}}\label{table:L2diff}
\end{table}

\begin{figure}[htbp]
\centering
\includegraphics[width=10cm]{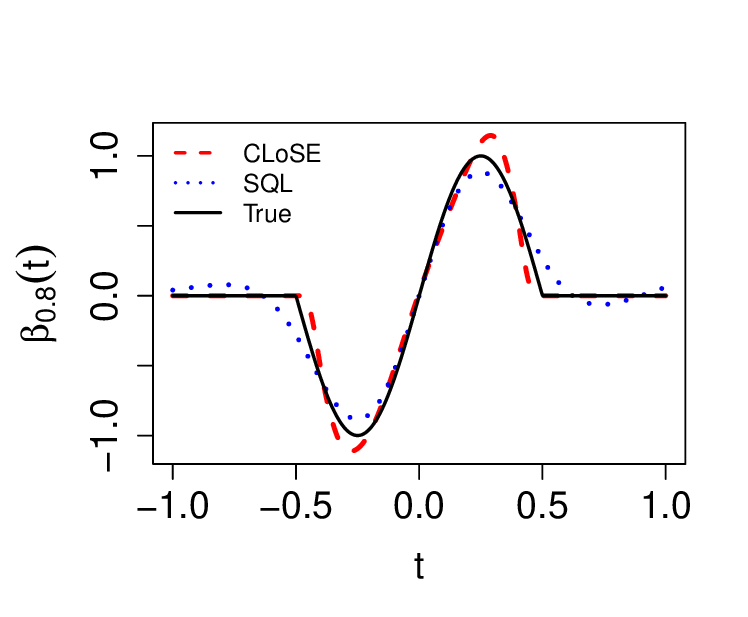}
\caption{The estimator $\hat\beta_{0.8}(t)$ in one simulation replicate using the convolution-smoothing based locally sparse estimation (CLoSE) method (red dashed line) and the smoothed quantile loss (SQL) method (blue dotted line) when the errors are simulated from the normal distribution when the sample size $n=500$. The true $\beta_{0.8}(t)$ is displayed as the black solid line.}
\label{fig:betahat_0.8}
\end{figure}

Table \ref{table:L2diff} shows the $L_2$-norm of the difference between the estimator $\hat{\beta}_{\tau}(t)$ and the true function $\beta_{\tau}(t)$ from two different methods. It shows that the performance of both methods is improved as the sample size increases while the proposed method always outperforms the SQL method without the locally sparse regularization. The advantage of the proposed method is especially significant when the sample size is small. Figure \ref{fig:betahat_0.8} displays the estimator $\hat\beta_{0.8}(t)$ 
under the Normal scenario using the CLoSE method and the SQL method. It shows that the CLoSE method can obtain a strictly zero estimate for the functional coefficient $\beta_{0.8}(t)$ in the null regions, while $\hat{\beta}_{0.8}(t)$ estimated from the SQL method is always nonzero in the null regions.

Table \ref{table:par_est} shows the performance of the estimator for the parametric components $\bm\alpha_{\tau}$ using the CLoSE method. {The mean squared errors of the estimators for $\alpha_{1,\tau}$ and $\alpha_{2,\tau}$ both decrease as the sample size increases under two scenarios for all chosen quantiles.}
\begin{table}[htbp]
\centering
\caption{Biases and standard errors (SEs) of the estimator for $\alpha_{1,\tau}$ and $\alpha_{2,\tau}$ using the convolution-smoothing based locally sparse estimation (CLoSE) method when the errors are simulated from the normal or Cauchy distribution. Here $n$ denotes the sample size, and $\tau$ denotes the quantile.}
\renewcommand\arraystretch{1}
\setlength{\tabcolsep}{1.5mm}{
\begin{tabular}{cc|cccc|cccc}
\toprule
\multirow{3}{*}{$n$}& \multirow{3}{*}{$\tau$}& \multicolumn{4}{c|}{Normal} &\multicolumn{4}{c}{Cauchy}\\
\cmidrule(lr){3-6} \cmidrule(lr){7-10}
 &  & Bias($\alpha_{1,\tau}$)& SE$(\alpha_{1,\tau})$&  Bias($\alpha_{2,\tau}$) & SE$(\alpha_{2,\tau})$ 
& Bias($\alpha_{1,\tau}$)& SE$(\alpha_{1,\tau})$ & Bias($\alpha_{2,\tau}$) & SE$(\alpha_{2,\tau})$\\
& & $\times 10^2$ &  $\times 10^2$&  $\times 10^2$& $\times 10^2$
& $\times 10^2$ &$\times 10^2$ & $\times 10^2$ &$\times 10^2$\\
 \midrule
 & 0.2 & 0.05 & 0.90  & 0.11 & 0.99  & 0.09 & 4.40 & -0.92 & 4.36\\
 & 0.3 & 0.04 & 0.87 & 0.12 & 0.99  & 0.24 & 3.65 & -0.88 & 3.62\\
 & 0.4 & 0.04 & 0.90 & 0.12 & 0.97  & 0.31 & 3.37 & -0.80 & 3.19\\
 500 & 0.5 & 0.05 & 0.89 & 0.11 & 0.97 &  0.34 & 3.34 & -0.85 & 3.09\\
 & 0.6 & 0.04 & 0.87  & 0.11 & 0.99  & 0.50 & 3.52 & -0.92 & 3.15\\
 & 0.7 & 0.05 & 0.88 & 0.11 & 0.98 & 0.49 & 3.85 & -0.99 & 3.43\\
 & 0.8 & 0.04 & 0.88  & 0.12 & 0.98  & 0.75 & 4.66 & -1.20 & 4.07\\
 \midrule
 & 0.2 & -0.07 & 0.63 & -0.14 & 0.68  &-0.09 & 2.34 & 0.30 & 2.80\\
 & 0.3 & -0.06 & 0.62  & -0.15 & 0.66 & -0.03 & 1.97 & 0.16 & 2.37\\
 & 0.4 & -0.05 & 0.65  & -0.14 & 0.68  & 0.04 & 2.03 & 0.10 & 2.34\\
1000 & 0.5 & -0.06 & 0.61 & -0.15 & 0.66 & 0.02 & 1.82 & -0.02 & 2.19\\
 & 0.6 & -0.08 & 0.63 & -0.15 & 0.65  & 0.03 & 2.01 & -0.10 & 2.36\\
 & 0.7 & -0.06 & 0.62  & -0.15 & 0.63 & 0.01 & 2.14 & -0.29 & 2.50\\
 & 0.8 & -0.05 & 0.64  & -0.15 & 0.66 & 0.03 & 2.66 & -0.51 & 3.05\\
\bottomrule
\end{tabular}}\label{table:par_est}
\end{table}

\newpage

\subsection{Simulations with Two Functional Covariates}
To demonstrate the flexibility of the proposed method, we further conduct simulation using a data generating model with two functional covariates:
\begin{equation}
    Y_i = 
    \bm{Z}_i^T\bm{\alpha}_{\tau}+
    \int_{-1}^{1}X_{i,1}(t)\beta_{\tau, 1}(t)d t +  \int_{-1}^{1}X_{i,2}(t)\beta_{\tau, 2}(t)d t + e_i(\tau), i =1,\ldots,n,
    \label{eq:simulation_two_covariates}
\end{equation}
where both $X_{i,1}(t)$ and $X_{i,2}(t)$ are Wiener processes, $Z_{i1}, Z_{i2}\overset{iid}{\sim }N(0,\sigma_z^2)$ with $\sigma_z=0.1$, $\bm{Z}_i = (1,Z_{i1},Z_{i2})^T$, $\bm{\alpha}_{\tau}=(0,1,1)^T$, $\beta_{\tau, 1}(t) = \sin(2\pi t)\mathbbm{1}\{t\leq 0\}$, $\beta_{\tau, 2}(t) = \sin(2\pi t)\mathbbm{1}\{-0.5\leq t\leq 0.5\}$,  and $e_i(\tau) = e_i - F_e^{-1}(\tau)$, where $F_e(\cdot)$ is the CDF of $e_i$ and $e_{i}$ are i.i.d from the normal distribution, namely, $e_{i}\overset{iid}\sim N(0,\sigma^2)$ with $\sigma = 0.02$.


\begin{table}[htbp]
\centering
\caption{True discovery rate (TDR) for the null regions $\mathcal{N}(\beta_{\tau,1})$ and $\mathcal{N}(\beta_{\tau,2})$, and false discovery rate (FDR) for the non-null regions $\mathcal{S}(\beta_{\tau,1})$ and $\mathcal{S}(\beta_{\tau,2})$, as well as the integrated mean squared error (IMSE) of the estimators $\hat\beta_{\tau,1}$ and $\hat\beta_{\tau,2}$ obtained using the convolution-smoothing based locally sparse estimation (CLoSE) method. The average coverage probabilities of $\beta_{\tau,1}$ and $\beta_{\tau,2}$ are also reported using the proposed wild bootstrap procedure. Results are shown for scenarios with two functional predictors, where $n$ denotes the sample size and $\tau$ denotes the quantile level.}
\renewcommand\arraystretch{1}
\setlength{\tabcolsep}{3mm}{\begin{tabular}{cc|cccc|cccc}
\toprule
& & \multicolumn{4}{c|}{$\beta_{\tau,1}$} &\multicolumn{4}{c}{$\beta_{\tau,2}$} \\
\hline
 n & $\tau$ & IMSE $\times 10^2$ & FDR & TDR  & CP & IMSE $\times 10^2$ & FDR & TDR  & CP \\
 \hline
 & 0.1 & 0.75 & 0.0\% & 83.8\%  & 100\% & 1.39 & 0.8\% & 67.7\%   & 100\%\\
 500 & 0.5 & 0.21 & 0.0\% & 75.9\%  & 92\% & 0.26 & 0.0\% & 57.0\% & 98\%\\
 & 0.9 & 0.79 & 0.0\% & 83.3\% & 100\% & 1.21 & 0.1\% & 69.1\% & 100\%\\
\bottomrule
\end{tabular}}\label{table:two_covariates}
\end{table}

The simulation results on TDR, FDR, IMSE and average coverage probability under the setting of two functional covariates are summarized in Table \ref{table:two_covariates}. 
We examine the two extreme quantiles 0.1 and 0.9 as well as the median. The TDR and FDR are overall satisfying for all chosen quantiles for both $\beta_{\tau,1}(t)$ and $\beta_{\tau,2}(t)$. 
The average coverage probabilities of both $\beta_{\tau,1}(t)$ and $\beta_{\tau,2}(t)$ for median are close to the nominal 95\% while the coverage probabilities for extreme quantiles, 0.1 and 0.9, are much larger than the nominal level indicating the estimated estimator variance is larger than the true variance.

\clearpage
\section{Some Lemmas}
To prove our theorems in the manuscripts, we start by proving the following lemmas. 
\begin{lemma}\label{lemma1}
		Under Condition \ref{beta-smooth}, $\beta_{\tau,l}^0(t)-\beta_{\tau,l}(t)=b_a(t)+o(K^{-r})$,
  where $\beta_{\tau,l}^0(t)=\mathcal{B}^T(t)\bm{\theta}_{\tau,l}^0$ is the best B-spline function in approximating $\beta_{\tau,l}$, and $b_a(t)=O(k^{-r})$ is the spline approximation bias.
  
\end{lemma}

\begin{proof1}
		The proof of this lemma can be found in \cite{barrow1978asymptotic}.
\end{proof1}

\begin{lemma}\label{lemma2}
		Under Condition \ref{XZ-moment} and \ref{knots},
  \begin{itemize}
      \item 
  [(i)] there exists constants $C_{G}>c_{G}>0$ such that 
                \begin{equation*}               c_{G}K^{-1}\leq\rho_{min}\left(\bm{U}\bm{U}^T\right)\leq\rho_{max}\left(\bm{U}\bm{U}^T\right)\leq C_{G}K^{-1},
			\end{equation*}
			where $\rho_{min}$ and $\rho_{max}$ denote the
			smallest and largest eigenvalues of a matrix, respectively.
		\item [(ii)] we can have $\|\bm{U}\|_{\infty}=O(K^{-1})$.
\end{itemize}	
\end{lemma}

\begin{lemma}\label{lemma3}
		Under Condition \ref{knots}, we can get 
  \begin{itemize}
      \item  [(i)] $\|\bm{V}\|_{\infty}=O(K^{2q-1})$; 
      \item  [(ii)] for any non-zero vector $\bm{u}$, 
      there are two positive constants $c_{D}<C_{D}$ such that $c_{D} K^{2q-1} \|\bm{\mu}\|_2^2\leq \bm{\mu}^{T}\bm{V}\bm{\mu}\leq C_{D} K^{2q-1}\|\bm{\mu}\|_2^2$.
  \end{itemize}
	\end{lemma}
 

\begin{lemma}\label{lemma4}
If Conditions \ref{XZ-moment}, \ref{beta-smooth}, \ref{density}, \ref{kernel} and \ref{bandwidth} hold,
for any $\delta>0$, there exists a constant $C$ such that 
   \begin{equation}
       Pr(\|\dot{\mathcal{L}}_0^*(\bm{\alpha}_{\tau}^0,\bm{\theta}_{\tau}^0)\|_{2}<C\sqrt{1/n})>1-\delta,
   \end{equation}
   holds for all sufficiently large $K$ and $n$.
\end{lemma}
\begin{proof1}
    Recall that 
\begin{equation*}
\begin{split}
\dot{\mathcal{L}}^{*}_0(\bm{\alpha}_{\tau}^0,\bm{\theta}_{\tau}^0) &= \frac{1}{n}\sum_{i=1}^n\left\{G_h(\bm{Z}_i^T\bm{\alpha}_{\tau}^0+
\bm{U}_i^T\bm{\theta}_{\tau}^0-Y_i)-\tau\right\}(\bm{Z}_i^T,\bm{U}_i^T)^T.\\
\end{split}
\end{equation*}
Note that each B-spline basis function $B_k(t)$ is non-negative, nonzero over no more than $p+1$ consecutive subintervals and $\sum_{k=1}^{K+p}B_k(t)=1$ for all $t\in[0,\mathcal{T}]$, we can get $\int_0^{\mathcal{T}}B^2_k(t)=O(K^{-1})$. Then, with Condition \ref{XZ-moment}, we have
\begin{equation*}
    \|\bm{U}_i\|_{\infty} = O(K^{-1/2}), \qquad and \qquad
    \|\bm{U}_i\|_{2} = O(1).
\end{equation*}
For $G_h(\bm{Z}_i^T\bm{\alpha}_{\tau}^0+
\bm{U}_i^T\bm{\theta}_{\tau}^0-Y_i)$, we have
\begin{equation*}
    \begin{split}
& G_h(\bm{Z}_i^T\bm{\alpha}_{\tau}^0+\bm{U}_i^T\bm{\theta}_{\tau}^0-Y_i)\\
= & G_h\left(\bm{Z}_i^T\bm{\alpha}_{\tau}^0+
\bm{U}_i^T\bm{\theta}_{\tau}^0-Y_i+\int_{0}^{\mathcal{T}}\bm{X}_i^T(t)\bm{\beta}_{\tau}(t)dt
-\int_{0}^{\mathcal{T}}\bm{X}_i^T(t)\bm{\beta}_{\tau}(t)dt\right)\\
= & G_h\left(
-e_i+\bm{U}_i^T\bm{\theta}_{\tau}^0-\int_{0}^{\mathcal{T}}\bm{X}_i^T(t)\bm{\beta}_{\tau}(t)dt
\right),
    \end{split}
\end{equation*}
by Lagrange mean value theorem, we can get
\begin{equation*}
    \begin{split}
      & G_h\left(
-e_i+\bm{U}_i^T\bm{\theta}_{\tau}^0-\int_{0}^{\mathcal{T}}\bm{X}_i^T(t)\bm{\beta}_{\tau}(t)dt
\right)  \\
=& G_h(-e_i)+\dot{G}_h\left(
-e_i+\int_{0}^{\mathcal{T}}\bm{X}_i^T(t)\bm{\beta}^*_{\tau}(t)dt
\right) \left(\bm{U}_i^T\bm{\theta}_{\tau}^0-\int_{0}^{\mathcal{T}}\bm{X}_i^T(t)\bm{\beta}_{\tau}(t)dt\right) \\
=& G_h(-e_i)+\mathcal{K}_h\left(
-e_i+\int_{0}^{\mathcal{T}}\bm{X}_i^T(t)\bm{\beta}^*_{\tau}(t)dt
\right) \left(\bm{U}_i^T\bm{\theta}_{\tau}^0-\int_{0}^{\mathcal{T}}\bm{X}_i^T(t)\bm{\beta}_{\tau}(t)dt\right), 
    \end{split}
\end{equation*}
where $\bm{\beta}^*_{\tau}(t)$ lies between $\bm{\beta}_{\tau}(t)$ and $\bm{U}_{i}^T\bm{\theta}_{\tau}^0$.

Then $\dot{\mathcal{L}}_0^*(\bm{\alpha}_{\tau}^0,\bm{\theta}_{\tau}^0) $ can be expressed as:
\begin{equation*}
    \begin{split}
         \dot{\mathcal{L}}_0^*(\bm{\alpha}_{\tau}^0,\bm{\theta}_{\tau}^0)      
= & 
 \frac{1}{n}\sum_{i=1}^n\left\{\mathcal{K}_h\left(
-e_i+\int_{0}^{\mathcal{T}}\bm{X}_i^T(t)\bm{\beta}^*_{\tau}(t)dt
\right) \left(\bm{U}_i^T\bm{\theta}_{\tau}^0-\int_{0}^{\mathcal{T}}\bm{X}_i^T(t)\bm{\beta}_{\tau}(t)dt\right)\right\}(\bm{Z}_i^T,\bm{U}_i^T)^T\\
& \qquad+ \frac{1}{n}\sum_{i=1}^n\left\{G_h(-e_i)-\tau\right\}(\bm{Z}_i^T,\bm{U}_i^T)^T\\
 & \overset{\Delta}{=} \bm{I}_1+\bm{I}_2.
    \end{split}
\end{equation*}

First, we will prove that $\|\bm{I}_1\|_{2}=O_p(K^{-r}(h^2+(nh)^{-1/2}))$. By Lemma A.\ref{lemma1} and Condition \ref{XZ-moment}, we have $\left|\bm{U}_i^T\bm{\theta}_{\tau}^0-\int_{0}^{\mathcal{T}}\bm{X}_i^T(t)\bm{\beta}_{\tau}(t)dt\right|=O_p(K^{-r})$. Then, by the weak law of large numbers and Condition \ref{XZ-moment}, we can get $\|\bm{I}_1\|_{2}=O_p(K^{-r}(h^2+(nh)^{-1/2}))$.

Moreover, we want to prove the bound of the term $\bm{I}_2$. It follows from Condition \ref{density} that
\begin{equation*}
    \begin{split}
        E\left(\frac{1}{n}\sum_{i=1}^n G_h(-e_i)\mid\bm{Z}_i,\bm{X}_i(t)\right)
        & = E( G_h(-e_i)\mid\bm{Z}_i,\bm{X}_i(t))\\
        & = \int G(-\varepsilon/h)f_{e\mid \bm{Z},\bm{X}}(\varepsilon)d\varepsilon\\
        & = \frac{1}{h}\int \mathcal{K}(-\varepsilon/h)F_{e\mid \bm{Z},\bm{X}}(\varepsilon)du\\
        & =  \int \mathcal{K}(-u)F_{e\mid \bm{Z},\bm{X}}(hu)d\varepsilon\\
        & =  \int \mathcal{K}(u)\left\{F_{e\mid \bm{Z},\bm{X}}(0)+huf_{e\mid \bm{Z},\bm{X}}(0)+O(h^2)\right\}du\\
        & = \tau +O(h^2),
    \end{split}
\end{equation*}
then, using Condition \ref{XZ-moment}, it is easy to have
\begin{equation}\label{Gbound}
    \begin{split}
E\left\{\frac{1}{n}\sum_{i=1}^n\left\{G_h(-e_i)-\tau\right\}(\bm{Z}_i^T,\bm{U}_i^T)^T\right\} & = E\left\{E\left\{\frac{1}{n}\sum_{i=1}^n\left\{G_h(-e_i)-\tau\right\}\mid\bm{Z}_i,\bm{X}_i(t)\right\}(\bm{Z}_i^T,\bm{U}_i^T)^T\right\}\\
& = O(h^2).
    \end{split}
\end{equation}

For any matrix $\bm{A}$, let $\bm{A}^{\otimes 2} = \bm{A}\bm{A}^T$. Given that
\begin{equation*}
    \begin{split}
        Cov\left\{\frac{1}{n}\sum_{i=1}^n\left\{G_h(-e_i)-\tau\right\}(\bm{Z}_i^T,\bm{U}_i^T)^T\right\}
        &= \frac{1}{n}Cov\left\{\left\{G_h(-e_i)-\tau\right\}(\bm{Z}_i^T,\bm{U}_i^T)^T\right\}\\
        & = \frac{1}{n} E\left\{\left\{G_h(-e_i)-\tau\right\}^2\left((\bm{Z}_i^T,\bm{U}_i^T)^T\right)^{\otimes 2}
        \right\}\\
        & = \frac{1}{n}E\left\{G_h^2(-e_i)\left((\bm{Z}_i^T,\bm{U}_i^T)^T\right)^{\otimes 2}
        \right\}\\
        &\quad - 2\tau\times \frac{1}{n}E\left\{G_h(-e_i)\left((\bm{Z}_i^T,\bm{U}_i^T)^T\right)^{\otimes 2}
        \right\}\\
        & \quad + \tau^2 \frac{1}{n}E\left\{\left((\bm{Z}_i^T,\bm{U}_i^T)^T\right)^{\otimes 2}
        \right\}.
    \end{split}
\end{equation*}
Denote $\widetilde{\mathcal{K}}(u)=2G(u)\mathcal{K}(u)$, 
$c_K = \int u \widetilde{\mathcal{K}}(u)du $
and note that $\int \widetilde{\mathcal{K}}(u)du = 1 $.
Similarly, using integration by parts and a change of variable, we have
\begin{equation*}
    \begin{split}
        E(G_h^2(-\varepsilon_i)\mid \bm{Z}_i,\bm{X}_i(t)) & = \int G_h^2(-\varepsilon)  f_{e\mid \bm{Z},\bm{X}}(\varepsilon)d\varepsilon\\
        & = \frac{1}{h} \int \widetilde{\mathcal{K}}(-\varepsilon/h) F_{e\mid \bm{Z},\bm{X}}(\varepsilon)d\varepsilon\\
        & = \tau - h f_{e\mid \bm{Z},\bm{X}}(0) \int u \widetilde{\mathcal{K}}(u)du + O(h^2)\\
        & = \tau+O(h).
    \end{split}
\end{equation*}
According to the bounded condition of $\bm{Z}_i$ and $\bm{X}_i(t)$ (Condition \ref{XZ-moment}), we can get
\begin{equation}\label{G2bound}
    \begin{split}
        E\left\{G_h^2(-e_i)\left\|\left((\bm{Z}_i^T,\bm{U}_i^T)^T\right)^{\otimes 2 }\right\|_{\infty}
        \right\} & = E\left\{\left\|\left((\bm{Z}_i^T,\bm{U}_i^T)^T\right)^{\otimes 2}\right\|_{\infty} E(G_h^2(-e_i)\mid \bm{Z}_i,\bm{X}_i(t))
        \right\} \\
        & = \tau+O(h).\\
    \end{split}
\end{equation}
Combining (\ref{Gbound}) and (\ref{G2bound}), we cant get 
$\left\|Cov\left\{n^{-1}\sum_{i=1}^n\left\{G_h(-e_i)-\tau\right\}(\bm{Z}_i^T,\bm{U}_i^T)^T\right\}\right\|_{2}= 1/n(\tau(1-\tau)+O(h))=O(1/n)$. Then $\|\bm{I}_2\|_{2}=O_p(h^2+\sqrt{1/n})$ and $\|\dot{\mathcal{L}}_0^*(\bm{\alpha}_{\tau}^0,\bm{\theta}_{\tau}^0)\|_{2} = O_p(K^{-r}(h^2+(nh)^{-1/2})+h^2+\sqrt{1/n})=O_p(\sqrt{1/n})$.
\end{proof1}

\begin{lemma}\label{lemma6}
    If Conditions \ref{XZ-moment}, \ref{beta-smooth}, \ref{density}, \ref{kernel} and \ref{bandwidth} hold hold, we have 
    \begin{equation}\label{normal}
        \frac{1}{\sqrt{n}}\sum_{i=1}^n\left\{E\left((\bm{Z}^T,\bm{U}^T)^T\right)^{\otimes2}\right\}^{-1/2}\left\{G_h(-e_i)-\tau\right\}(\bm{Z}_i^T,\bm{U}_i^T)^T \overset{d}{\rightarrow}  \bm{\mathbb{N}}(\bm{0},\tau(1-\tau)\bm{I}_{d+m\times (K+p)}).
    \end{equation}
\end{lemma}

\begin{proof1}
    Denote $\bm{\mathcal{M}}_i = (\bm{Z}_i^T,\bm{U}_i^T)^T $, $\bm{\mathcal{M}} = E\left\{\left((\bm{Z}^T,\bm{U}^T)^T\right)^{\otimes2}\right\}$, and 
    \begin{equation*}
       \bm{G}^*_{\tau,ni} = \frac{1}{\sqrt{n}}\bm{\mathcal{M}}^{-1/2}\left\{G_h(-e_i)-\tau\right\}\bm{\mathcal{M}}_i,
    \end{equation*}
then 
\begin{equation*}
        \sum_{i=1}^n \bm{G}^*_{\tau,ni} = \frac{1}{\sqrt{n}}\sum_{i=1}^n\bm{\mathcal{M}}^{-1/2}\left\{G_h(-e_i)-\tau\right\}\bm{\mathcal{M}}_i.
    \end{equation*}
From the proof of Lemma A.\ref{lemma4}, we can get $E(G_h(-e_i)-\tau) = O(h^2)$ and 
$Var(G_h(-e_i)-\tau) = \tau(1-\tau)+O(h)$, then 
\begin{equation*}
    E(\bm{G}^*_{\tau,ni}) = O(h^2n^{-1/2}), \qquad Cov(\bm{G}^*_{\tau,ni}) = \frac{1}{n}\tau(1-\tau)\bm{\mathcal{M}}^{-1/2}E\left(\bm{\mathcal{M}}_i\bm{\mathcal{M}}_i^T\right)\bm{\mathcal{M}}^{-1/2}+O(h/n).
\end{equation*}
Correspondingly, 
\begin{equation*}
    E\left(\sum_{i=1}^n \bm{G}^*_{\tau,ni}\right) = 
    O(h^2n^{-1/2}), \qquad Cov\left(\sum_{i=1}^n\bm{G}^*_{\tau,ni}\right) = \tau(1-\tau)\bm{I}+O(h).
\end{equation*}
By Cram{\'e}r–Wold Theorem, to establish the asymptotic normality of the vector $\sum_{i=1}^n \bm{G}^*_{\tau,ni}$, we first need to prove the asymptotic normality of $\sum_{i=1}^n \bm{\mu}^T \bm{G}^*_{\tau,ni}$ for any unit vector $\bm{\mu}$, where $\mu_j,j=1,\ldots,d+m\times (K+p)$ is the $j$-th column of the identity matrix $I_{d+m\times (K+p)}$.

Denote $\Phi_i  = n^{-1/2}\bm{\mu}^T\bm{\mathcal{M}}^{-1/2}\bm{\mathcal{M}}_i$, then 
$\bm{\mu}^T \bm{G}^*_{\tau,ni} = \Phi_i G_h(-e_i)-\tau$. Next, we will check the following condition, which ensures Lindeberg's condition is both sufficient and necessary:
\begin{equation}\label{linder}
    \max_{1\leq i\leq n} \frac{Var(\Phi_i G_h(-e_i)-\tau))}{\sum_{i=1}^n Var(\Phi_i G_h(-e_i)-\tau))} = \max_{1\leq i\leq n} \frac{E(\Phi_i^2) }{\sum_{i=1}^n E(\Phi_i^2) }\rightarrow 0,
\end{equation}
as $n\rightarrow \infty$. By the definition of $\Phi_i$, we have 
\begin{equation*}
    \begin{split}
        E(\Phi_i^2 )& = \frac{1}{n}\bm{\mu}^T\bm{\mathcal{M}}^{-1/2}E(\bm{\mathcal{M}}_i\bm{\mathcal{M}}_i^T)\bm{\mathcal{M}}^{-1/2}\bm{\mu}\\
        & = \frac{1}{n}tr\left\{\bm{\mu}^T\bm{\mathcal{M}}^{-1/2}E(\bm{\mathcal{M}}_i\bm{\mathcal{M}}_i^T)\bm{\mathcal{M}}^{-1/2}\bm{\mu}\right\}\\
        & = \frac{1}{n}tr\left\{\bm{\mu}\bm{\mu}^T\bm{\mathcal{M}}^{-1/2}E(\bm{\mathcal{M}}_i\bm{\mathcal{M}}_i^T)\bm{\mathcal{M}}^{-1/2}\right\}\\
        & = \frac{1}{n}tr\left\{\bm{\mathcal{M}}^{-1}E(\bm{\mathcal{M}}_i\bm{\mathcal{M}}_i^T)\right\}\\
        & \leq \frac{1}{n}tr\left\{\bm{\mathcal{M}}^{-1}\right\}tr\left\{E(\bm{\mathcal{M}}_i\bm{\mathcal{M}}_i^T)\right\}\\
        & \leq C_{\Phi}/n,
    \end{split}
\end{equation*}
then we can obtain
\begin{equation*}
    \max_{1\leq i\leq n} \frac{E(\Phi_i^2) }{\sum_{i=1}^n E(\Phi_i^2) } = O\left(\frac{1}{n}\right)\rightarrow 0.
\end{equation*}
Thus condition (\ref{linder}) holds, which means that $\sum_{i=1}^n \bm{\mu}^T \bm{G}^*_{\tau,ni}$ has a asymptotic normal distribution. By Cram{\'e}r–Wold Theorem and the above results, it is straightforward to prove that 
$\sum_{i=1}^n \bm{G}^*_{\tau,ni} = n^{-1/2}\sum_{i=1}^n\bm{\mathcal{M}}^{-1/2}\left\{G_h(-e_i)-\tau\right\}\bm{\mathcal{M}}_i$ has an asymptotic multivariate normal distribution
\begin{equation*}
    \frac{1}{\sqrt{n}}\sum_{i=1}^n\bm{\mathcal{M}}^{-1/2}\left\{G_h(-e_i)-\tau\right\}\bm{\mathcal{M}}_i \overset{d}{\rightarrow}  \bm{\mathbb{N}}(\bm{0},\tau(1-\tau)\bm{I}_{d+m\times (K+p)}),
\end{equation*}
thus (\ref{normal}) holds and the proof is completed.
\end{proof1}

\begin{lemma}\label{lemma5}
Under Conditions \ref{XZ-moment}, \ref{beta-smooth}, \ref{density}, \ref{kernel} and \ref{bandwidth}, we have 
\begin{equation}\label{L2}
\ddot{\mathcal{L}}^{*}_0(\bm{\alpha}_{\tau}^0,\bm{\theta}_{\tau}^0)  = E\left\{f_{e\mid \bm{Z},\bm{X}}(0)
 \begin{pmatrix}
     \bm{Z}\\
     \bm{U}\\
 \end{pmatrix}^{\otimes 2}\right\}+O_p\left(\frac{1}{\sqrt{nh}}\right),
\end{equation}
\end{lemma}
\begin{proof1}
Recall that 
    \begin{equation*}
\begin{split}
\ddot{\mathcal{L}}^{*}_0(\bm{\alpha}_{\tau}^0,\bm{\theta}_{\tau}^0) & = \frac{1}{n}\sum_{i=1}^n\mathcal{K}_h(\bm{Z}_i^T\bm{\alpha}_{\tau}^0+\bm{U}_i^T\bm{\theta}_{\tau}^0-Y_i)\begin{pmatrix}
     \bm{Z}_{i}\\
     \bm{U}_{i}\\
 \end{pmatrix}^{\otimes 2}.
\end{split}
\end{equation*}
For $\mathcal{K}_h(\bm{Z}_i^T\bm{\alpha}_{\tau}^0+
\bm{U}_i^T\bm{\theta}_{\tau}^0-Y_i)$, we have
\begin{equation*}
    \begin{split}
& \mathcal{K}_h(\bm{Z}_i^T\bm{\alpha}_{\tau}^0+\bm{U}_i^T\bm{\theta}_{\tau}^0-Y_i)\\
= & \mathcal{K}_h\left(\bm{Z}_i^T\bm{\alpha}_{\tau}^0+
\bm{U}_i^T\bm{\theta}_{\tau}^0-Y_i+\int_{0}^{\mathcal{T}}\bm{X}_i^T(t)\bm{\beta}_{\tau}(t)dt
-\int_{0}^{\mathcal{T}}\bm{X}_i^T(t)\bm{\beta}_{\tau}(t)dt\right)\\
= & \mathcal{K}_h\left(
-e_i+\bm{U}_i^T\bm{\theta}_{\tau}^0-\int_{0}^{\mathcal{T}}\bm{X}_i^T(t)\bm{\beta}_{\tau}(t)dt
\right),
    \end{split}
\end{equation*}
by Lagrange mean value theorem, we can get
\begin{equation*}
    \begin{split}
      & \mathcal{K}_h\left(
-e_i+\bm{U}_i^T\bm{\theta}_{\tau}^0-\int_{0}^{\mathcal{T}}\bm{X}_i^T(t)\bm{\beta}_{\tau}(t)dt
\right)  \\
=& \mathcal{K}_h(-e_i)+\dot{\mathcal{K}}_h\left(
-e_i+\int_{0}^{\mathcal{T}}\bm{X}_i^T(t)\bm{\beta}^*_{\tau}(t)dt
\right) \left(\bm{U}_i^T\bm{\theta}_{\tau}^0-\int_{0}^{\mathcal{T}}\bm{X}_i^T(t)\bm{\beta}_{\tau}(t)dt\right) \\
=& \mathcal{K}_h(-e_i)+\frac{1}{h^2}\dot{\mathcal{K}}_h\left(
-e_i+\int_{0}^{\mathcal{T}}\bm{X}_i^T(t)\bm{\beta}^*_{\tau}(t)dt
\right) \left(\bm{U}_i^T\bm{\theta}_{\tau}^0-\int_{0}^{\mathcal{T}}\bm{X}_i^T(t)\bm{\beta}_{\tau}(t)dt\right), 
    \end{split}
\end{equation*}
where $\dot{\mathcal{K}}_h(\cdot) = \dot{\mathcal{K}}(\cdot/h)$, $\bm{\beta}^*_{\tau}(t)$ lies between $\bm{\beta}_{\tau}(t)$ and $\bm{U}_{i}^T\bm{\theta}_{\tau}^0$.

Then $\ddot{\mathcal{L}}_0^*(\bm{\alpha}_{\tau}^0,\bm{\theta}_{\tau}^0) $ can be expressed as:
\begin{equation*}
    \begin{split}
         \ddot{\mathcal{L}}_0^*(\bm{\alpha}_{\tau}^0,\bm{\theta}_{\tau}^0)      
= & 
 \frac{1}{nh^2}\sum_{i=1}^n\left\{\dot{\mathcal{K}}_h\left(
-e_i+\int_{0}^{\mathcal{T}}\bm{X}_i^T(t)\bm{\beta}^*_{\tau}(t)dt
\right) \left(\bm{U}_i^T\bm{\theta}_{\tau}^0-\int_{0}^{\mathcal{T}}\bm{X}_i^T(t)\bm{\beta}_{\tau}(t)dt\right)\right\}\begin{pmatrix}
     \bm{Z}_{i}\\
     \bm{U}_{i}\\
 \end{pmatrix}^{\otimes 2}\\
& \qquad+ \frac{1}{n}\sum_{i=1}^n\mathcal{K}_h(-e_i)\begin{pmatrix}
     \bm{Z}_{i}\\
     \bm{U}_{i}\\
 \end{pmatrix}^{\otimes 2}\\
 & \overset{\Delta}{=} \bm{I}_1+\bm{I}_2.
    \end{split}
\end{equation*}

\textbf{(1) First, we consider the first term $\bm{I}_1$.} $\left|\bm{U}_i^T\bm{\theta}_{\tau}^0-\int_{0}^{\mathcal{T}}\bm{X}_i^T(t)\bm{\beta}_{\tau}(t)dt\right|=O(K^{-r})$, and 
\begin{equation*}
   \frac{1}{nh^2}\sum_{i=1}^n\dot{\mathcal{K}}_h\left(-e_i+\int_{0}^{\mathcal{T}}\bm{X}_i^T(t)\bm{\beta}^*_{\tau}(t)dt
\right) = O_p\left(\frac{1}{h}+\frac{1}{\sqrt{nh^3}}\right),
\end{equation*}
then 
\begin{equation}\label{L2_1}
    \bm{I}_1 = o_p\left(\frac{1}{\sqrt{nh}}\right),
\end{equation}
with the Condition $hK^r\rightarrow\infty$.

\textbf{(2) We consider the second term $\bm{I}_2$.} For $n^{-1}\sum_{i=1}^n\mathcal{K}_h(-e_i)$, using change of variable, we have 
\begin{equation*}
    \begin{split}
        E\left\{\frac{1}{n}\sum_{i=1}^n\mathcal{K}_h(-e_i)\mid\bm{Z}_i,\bm{X}_i(t)\right\}
        & = E\left\{\mathcal{K}_h(-e_i)\mid\bm{Z}_i,\bm{X}_i(t)\right\}\\
        & = \int \mathcal{K}_h(-\varepsilon)f_{e|\bm{Z},\bm{X}}(\varepsilon)d\varepsilon\\
        & = \int \mathcal{K}_h(-u)f_{e|\bm{Z},\bm{X}}(hu)du\\
        & = \int \mathcal{K}_h(u)\left[f_{e|\bm{Z},\bm{X}}(0)+O(h)\right]du\\
        & = f_{e|\bm{Z},\bm{X}}(0)+O(h^2), 
    \end{split}
\end{equation*}

\begin{equation*}
    \begin{split}
        E\left\{\mathcal{K}^2(-e_i/h)\mid\bm{Z}_i,\bm{X}_i(t)\right\}
        & = \int \mathcal{K}^2(-\varepsilon/h)f_{e|\bm{Z},\bm{X}}(\varepsilon)d\varepsilon\\
        & = h\int \mathcal{K}^2(-u)f_{e|\bm{Z},\bm{X}}(hu)du\\
        & = O(h), 
    \end{split}
\end{equation*}
moreover,
\begin{equation*}
    \begin{split}
        Var\left\{\frac{1}{n}\sum_{i=1}^n\mathcal{K}_h(-e_i)\mid\bm{Z}_i,\bm{X}_i(t)\right\}
        & = \frac{1}{nh^2}Var\left\{\mathcal{K}(-e_i/h)\mid\bm{Z}_i,\bm{X}_i(t)\right\}\\
        & = \frac{1}{nh^2}E\left\{\mathcal{K}^2(-e_i/h)\mid\bm{Z}_i,\bm{X}_i(t)\right\}\\
        & \quad
        - \frac{1}{nh^2}\left\{E\left\{\mathcal{K}(-e_i/h)\mid\bm{Z}_i,\bm{X}_i(t)\right\}\right\}^2\\
        & = O(n^{-1}h^{-1}).
    \end{split}
\end{equation*}
Then, we have 
\begin{equation}\label{L2_2}
    \bm{I}_2=\frac{1}{n}\sum_{i=1}^n\mathcal{K}_h(-e_i)\begin{pmatrix}
     \bm{Z}_{i}\\
     \bm{U}_{i}\\
 \end{pmatrix}^{\otimes 2} = E\left\{f_{e|\bm{Z},\bm{X}}(0)
 \begin{pmatrix}
     \bm{Z}\\
     \bm{U}\\
 \end{pmatrix}^{\otimes 2}\right\}+O_p\left(h^2+\frac{1}{\sqrt{nh}}\right).
\end{equation}
Combining (\ref{L2_1}) and (\ref{L2_2}), we can get that 
\begin{equation*}
\ddot{\mathcal{L}}^{*}_0(\bm{\alpha}_{\tau}^0,\bm{\theta}_{\tau}^0)  = E\left\{f_{e|\bm{Z},\bm{X}}(0)
 \begin{pmatrix}
     \bm{Z}\\
     \bm{U}\\
 \end{pmatrix}^{\otimes 2}\right\}+O_p\left(\frac{1}{\sqrt{nh}}\right),
\end{equation*}
with Condition $nh^4\rightarrow 0$. Then the proof is completed.
\end{proof1}

\clearpage
\section{Proofs of Theorems}
\subsection{Proof of Theorem 1}
We want to prove that for any $\delta>0$, there exists a sufficiently large constant $\Delta$ such that for sufficiently large $n$, 
\begin{equation}\label{local}
    Pr\left\{\inf_{\|\bm{\alpha}_{\tau}-\bm{\alpha}_{\tau}^0\|_2 \leq \Delta /\sqrt{n}
    \atop
    \|\bm{\theta}_{\tau}-\bm{\theta}_{\tau}^0\|_2\leq \Delta K/\sqrt{n}
    }  \mathcal{L}^*(\bm{\alpha}_{\tau},\bm{\theta}_{\tau})-
    \mathcal{L}^*(\bm{\alpha}_{\tau}^0,\bm{\theta}_{\tau}^0)>0\right\}\geq 1-\delta.
\end{equation}
(\ref{local}) implies with probability at least $1-\delta$ that
there is a local minimizer $(\widehat{\bm{\alpha}}_{\tau},\widehat{\bm{\theta}}_{\tau})$ such that $\|\widehat{\bm{\alpha}}_{\tau}-\bm{\alpha}^0_{\tau}\|_2=O_p(n^{-1/2})$ and 
$\|\widehat{\bm{\theta}}_{\tau}-\bm{\theta}^0_{\tau}\|_2=O_p(n^{-1/2}K)$.

By Taylor expansion, we have
\begin{equation}\label{diff}
\begin{split}
   & \mathcal{L}^*(\bm{\alpha}_{\tau},\bm{\theta}_{\tau})-
    \mathcal{L}^*(\bm{\alpha}_{\tau}^0,\bm{\theta}_{\tau}^0)\\
   =& \mathcal{L}_1^*(\bm{\alpha}_{\tau},\bm{\theta}_{\tau})-
    \mathcal{L}_1^*(\bm{\alpha}_{\tau}^0,\bm{\theta}_{\tau}^0)
    +\sum_{l=1}^m \frac{K}{\mathcal{T}}
    \left[\int_0^\mathcal{T}p_{\lambda_l}(\left|\bm{\mathcal{B}}^T(t)\bm{\theta}_{\tau,l}\right|)dt-
    \int_0^\mathcal{T}p_{\lambda_l}(\left|\bm{\mathcal{B}}^T(t)\bm{\theta}_{\tau,l}^0\right|)dt
    \right]
    \\
  \geq & \mathcal{L}_1^*(\bm{\alpha}_{\tau},\bm{\theta}_{\tau})-
    \mathcal{L}_1^*(\bm{\alpha}_{\tau}^0,\bm{\theta}_{\tau}^0)
    +\sum_{l=1}^m \frac{K}{\mathcal{T}}
    \left[\int_{\mathcal{S}(\beta_{\tau,l}^0)}
    p_{\lambda_l}(\left|\bm{\mathcal{B}}^T(t)\bm{\theta}_{\tau,l}\right|)dt-
    \int_{\mathcal{S}(\beta_{\tau,l}^0)}p_{\lambda_l}(\left|\bm{\mathcal{B}}^T(t)\bm{\theta}_{\tau,l}^0\right|)dt
    \right]
    \\
 \approx & \left\{\dot{\mathcal{L}}_1^*(\bm{\alpha}_{\tau}^0,\bm{\theta}_{\tau}^0)\right\}^T 
 \begin{pmatrix}
     \bm{\alpha}_{\tau}-\bm{\alpha}_{\tau}^0\\
     \bm{\theta}_{\tau}-\bm{\theta}_{\tau}^0\\
 \end{pmatrix}+\frac{1}{2}\begin{pmatrix}
     \bm{\alpha}_{\tau}-\bm{\alpha}_{\tau}^0\\
     \bm{\theta}_{\tau}-\bm{\theta}_{\tau}^0\\
 \end{pmatrix}^T \ddot{\mathcal{L}}_1^*(\bm{\alpha}_{\tau}^0,\bm{\theta}_{\tau}^0)
 \begin{pmatrix}
     \bm{\alpha}_{\tau}-\bm{\alpha}_{\tau}^0\\
     \bm{\theta}_{\tau}-\bm{\theta}_{\tau}^0\\
 \end{pmatrix}
 \\
 & \qquad +\sum_{l=1}^m \frac{K}{\mathcal{T}}
    \left[\int_{\mathcal{S}(\beta_{\tau,l}^0)}
    p_{\lambda_l}(\left|\bm{\mathcal{B}}^T(t)\bm{\theta}_{\tau,l}\right|)dt-
    \int_{\mathcal{S}(\beta_{\tau,l}^0)}p_{\lambda_l}(\left|\bm{\mathcal{B}}^T(t)\bm{\theta}_{\tau,l}^0\right|)dt
    \right]\\
 =& I+II+III.
\end{split}
\end{equation}
\textbf{(1) We consider the first term $I$ in (\ref{diff}).} 
Recall 
\begin{equation*}
    \dot{\mathcal{L}}_1^*(\bm{\alpha}_{\tau}^0,\bm{\theta}_{\tau}^0) = \dot{\mathcal{L}}_0^*(\bm{\alpha}_{\tau}^0,\bm{\theta}_{\tau}^0) +2 
    \begin{pmatrix}
        \bm{0}_{d\times d} & \\
        & \bm{\Gamma}\otimes\bm{V}
    \end{pmatrix}
   \begin{pmatrix}
       \bm{\alpha}_{\tau}^0\\
       \bm{\theta}_{\tau}^0
   \end{pmatrix},
\end{equation*}

For this term, with Lemma A.\ref{lemma4} and the Condition \ref{roughness}, then
\newline
\begin{equation*}
    \begin{split}
       & |I| = \left|\left\{\dot{\mathcal{L}}_1^*(\bm{\alpha}_{\tau}^0,\bm{\theta}_{\tau}^0)\right\}^T 
 \begin{pmatrix}
     \bm{\alpha}_{\tau}-\bm{\alpha}_{\tau}^0\\
     \bm{\theta}_{\tau}-\bm{\theta}_{\tau}^0\\
 \end{pmatrix}\right|\\
  \leq   & \left|\left\{\dot{\mathcal{L}}_0^*(\bm{\alpha}_{\tau}^0,\bm{\theta}_{\tau}^0)\right\}^T 
 \begin{pmatrix}
     \bm{\alpha}_{\tau}-\bm{\alpha}_{\tau}^0\\
     \bm{\theta}_{\tau}-\bm{\theta}_{\tau}^0\\
 \end{pmatrix}\right|
 + 2\left|\left\{ 
    \begin{pmatrix}
        \bm{0}_{d\times d} & \\
        & \bm{\Gamma}\otimes\bm{V}
    \end{pmatrix}
   \begin{pmatrix}
       \bm{\alpha}_{\tau}^0\\
       \bm{\theta}_{\tau}^0
   \end{pmatrix}\right\}^T 
 \begin{pmatrix}
     \bm{\alpha}_{\tau}-\bm{\alpha}_{\tau}^0\\
     \bm{\theta}_{\tau}-\bm{\theta}_{\tau}^0\\
 \end{pmatrix}\right|\\
   = & \left|\left\{\dot{\mathcal{L}}_0^*(\bm{\alpha}_{\tau}^0,\bm{\theta}_{\tau}^0)\right\}^T 
 \begin{pmatrix}
     \bm{\alpha}_{\tau}-\bm{\alpha}_{\tau}^0\\
     \bm{\theta}_{\tau}-\bm{\theta}_{\tau}^0\\
 \end{pmatrix}\right|
 + 2\left|\bm{\theta}_{\tau}^{0T} (\bm{\Gamma}\otimes\bm{V})^T (\bm{\theta}_{\tau}-\bm{\theta}_{\tau}^0)\right|\\
    = & \left|\left\{\dot{\mathcal{L}}_0^*(\bm{\alpha}_{\tau}^0,\bm{\theta}_{\tau}^0)\right\}^T 
 \begin{pmatrix}
     \bm{\alpha}_{\tau}-\bm{\alpha}_{\tau}^0\\
     \bm{\theta}_{\tau}-\bm{\theta}_{\tau}^0\\
 \end{pmatrix}\right|
 + 2\left\|\bm{\theta}_{\tau}^{0T} (\bm{\Gamma}\otimes\bm{V})^T\right\|_2 \left\|\bm{\theta}_{\tau}-\bm{\theta}_{\tau}^0\right\|_2\\
  \leq & \left|\frac{1}{n}\sum_{i=1}^n\left\{G_h(\bm{Z}_i^T\bm{\alpha}_{\tau}^0+
\bm{U}_i^T\bm{\theta}_{\tau}^0-Y_i)-\tau\right\}
\left\{\bm{Z}_i^T(\bm{\alpha}_{\tau}-\bm{\alpha}_{\tau}^0)+\bm{U}_i^T(\bm{\theta}_{\tau}-\bm{\theta}_{\tau}^0)\right\}\right|\\
& \quad+ 2\left\|\bm{\theta}_{\tau}^{0T} (\bm{\Gamma}\otimes\bm{V})^T\right\|_2 \left\|\bm{\theta}_{\tau}-\bm{\theta}_{\tau}^0\right\|_2\\
  \leq & \left|\frac{1}{n}\sum_{i=1}^n\left\{G_h(\bm{Z}_i^T\bm{\alpha}_{\tau}^0+
\bm{U}_i^T\bm{\theta}_{\tau}^0-Y_i)-\tau\right\}
\left\{\bm{Z}_i^T(\bm{\alpha}_{\tau}-\bm{\alpha}_{\tau}^0)\right\}\right|\\
& \quad + \left|\frac{1}{n}\sum_{i=1}^n\left\{G_h(\bm{Z}_i^T\bm{\alpha}_{\tau}^0+
\bm{U}_i^T\bm{\theta}_{\tau}^0-Y_i)-\tau\right\}
\left\{\bm{U}_i^T(\bm{\theta}_{\tau}-\bm{\theta}_{\tau}^0)\right\}\right|\\
& \quad+ 2\left\|\bm{\theta}_{\tau}^{0T} (\bm{\Gamma}\otimes\bm{V})^T\right\|_2 \left\|\bm{\theta}_{\tau}-\bm{\theta}_{\tau}^0\right\|_2\\
\leq & \left\|\frac{1}{n}\sum_{i=1}^n\left\{G_h(\bm{Z}_i^T\bm{\alpha}_{\tau}^0+
\bm{U}_i^T\bm{\theta}_{\tau}^0-Y_i)-\tau\right\}
\bm{Z}_i^T\right\|_2\left\|\bm{\alpha}_{\tau}-\bm{\alpha}_{\tau}^0\right\|_2\\
& \quad + \left\|\frac{1}{n}\sum_{i=1}^n\left\{G_h(\bm{Z}_i^T\bm{\alpha}_{\tau}^0+
\bm{U}_i^T\bm{\theta}_{\tau}^0-Y_i)-\tau\right\}
\bm{U}_i^T\right\|_2\left\|\bm{\theta}_{\tau}-\bm{\theta}_{\tau}^0\right\|_2\\
& \quad+ 2\left\|\bm{\theta}_{\tau}^{0T} (\bm{\Gamma}\otimes\bm{V})^T\right\|_2 \left\|\bm{\theta}_{\tau}-\bm{\theta}_{\tau}^0\right\|_2\\
    \end{split}
\end{equation*}
With the bounded Condition \ref{XZ-moment} and Lemma A.\ref{lemma4}, we have 
\begin{equation*}
    \begin{split}
         \left\|\frac{1}{n}\sum_{i=1}^n\left\{G_h(\bm{Z}_i^T\bm{\alpha}_{\tau}^0+
\bm{U}_i^T\bm{\theta}_{\tau}^0-Y_i)-\tau\right\}
\bm{Z}_i^T\right\|_2 & = C\left|\frac{1}{n}\sum_{i=1}^n\left\{G_h(\bm{Z}_i^T\bm{\alpha}_{\tau}^0+
\bm{U}_i^T\bm{\theta}_{\tau}^0-Y_i)-\tau\right\}\right|\\
& = O (n^{-1/2}).
    \end{split}
\end{equation*}
Thus, we have 
\begin{equation}\label{I_1}
    \begin{split}
        \left\|\frac{1}{n}\sum_{i=1}^n\left\{G_h(\bm{Z}_i^T\bm{\alpha}_{\tau}^0+
\bm{U}_i^T\bm{\theta}_{\tau}^0-Y_i)-\tau\right\}
\bm{Z}_i^T\right\|_2\left\|\bm{\alpha}_{\tau}-\bm{\alpha}_{\tau}^0\right\|_2 & = O_p(n^{-1})=o_p(K/n),
    \end{split}
\end{equation}
as the Condition $\left\|\bm{\alpha}_{\tau}-\bm{\alpha}_{\tau}^0\right\|_2 = O_p(n^{-1/2})$.

Note that $\|\bm{U}_i\|_2=O(1)$ and Lemma A.\ref{lemma4}, it has
\begin{equation}\label{I_2}
    \left\|\frac{1}{n}\sum_{i=1}^n\left\{G_h(\bm{Z}_i^T\bm{\alpha}_{\tau}^0+
\bm{U}_i^T\bm{\theta}_{\tau}^0-Y_i)-\tau\right\}
\bm{U}_i^T\right\|_2\left\|\bm{\theta}_{\tau}-\bm{\theta}_{\tau}^0\right\|_2
= O_p(K/n),
\end{equation}
with the Condition $\left\|\bm{\theta}_{\tau}-\bm{\theta}_{\tau}^0\right\|_2=O_p(K/n^{1/2})$.

As $\sup_{l}\gamma_l = o(n^{-1/2}K^{1/2-2d})$ with Condition \ref{roughness} , $\|\bm{\theta}_{\tau}^0\|_2 = O(K^{1/2})$ and Lemma A.\ref{lemma3}, we can obtain
\begin{equation}\label{I_3}
    \begin{split}
        \left\|\bm{\theta}_{\tau}^{0T} (\bm{\Gamma}\otimes\bm{V})^T\right\|_2
         \leq \sup_{l} (\gamma_l) \lambda_{\max}(\bm{V})\|\bm{\theta}_{\tau}^{0}\|_2
         = o(K/n).
    \end{split}
\end{equation}
Combining (\ref{I_1}), (\ref{I_2}) and (\ref{I_3}), we have
\begin{equation}\label{I_com}
    I = O_p(K/n).
\end{equation}

\textbf{(2) For the second term $II$.}
\begin{equation*}
    \begin{split}
       &  \frac{1}{2}\begin{pmatrix}
     \bm{\alpha}_{\tau}-\bm{\alpha}_{\tau}^0\\
     \bm{\theta}_{\tau}-\bm{\theta}_{\tau}^0\\
 \end{pmatrix}^T \ddot{\mathcal{L}}_1^*(\bm{\alpha}_{\tau}^0,\bm{\theta}_{\tau}^0)
 \begin{pmatrix}
     \bm{\alpha}_{\tau}-\bm{\alpha}_{\tau}^0\\
     \bm{\theta}_{\tau}-\bm{\theta}_{\tau}^0\\
 \end{pmatrix}\\
 = & \frac{1}{2}\begin{pmatrix}
     \bm{\alpha}_{\tau}-\bm{\alpha}_{\tau}^0\\
     \bm{\theta}_{\tau}-\bm{\theta}_{\tau}^0\\
 \end{pmatrix}^T\left\{\frac{1}{n}\sum_{i=1}^n\mathcal{K}_h(\bm{Z}_i^T\bm{\alpha}_{\tau}^0+\bm{U}_i^T\bm{\theta}_{\tau}^0-Y_i)\begin{pmatrix}
     \bm{Z}_{i}\\
     \bm{U}_{i}\\
 \end{pmatrix}^{\otimes 2}
 \right\}\begin{pmatrix}
     \bm{\alpha}_{\tau}-\bm{\alpha}_{\tau}^0\\
     \bm{\theta}_{\tau}-\bm{\theta}_{\tau}^0\\
 \end{pmatrix}\\
  & \quad +
  (\bm{\theta}_{\tau}-\bm{\theta}_{\tau}^0)^T(\bm{\Gamma}\otimes \bm{V})(\bm{\theta}_{\tau}-\bm{\theta}_{\tau}^0)
    \end{split}
\end{equation*}
From Lemma A.\ref{lemma5}, we have
\begin{equation*}
    \begin{split}
        \frac{1}{n}\sum_{i=1}^n\mathcal{K}_h(\bm{Z}_i^T\bm{\alpha}_{\tau}^0+\bm{U}_i^T\bm{\theta}_{\tau}^0-Y_i)\begin{pmatrix}
     \bm{Z}_{i}\\
     \bm{U}_{i}\\
 \end{pmatrix}^{\otimes 2} = E\left\{f_{e|\bm{Z},\bm{X}}(0)\begin{pmatrix}
     \bm{Z}\\
     \bm{U}\\
 \end{pmatrix}^{\otimes 2}\right\} +O_p(\frac{1}{\sqrt{nh}}).
    \end{split}
\end{equation*}
Based on Conditions \ref{density} and \ref{kernel}, we can also have
\begin{equation}\label{II_1}
\begin{split}
   & \frac{1}{2}\begin{pmatrix}
     \bm{\alpha}_{\tau}-\bm{\alpha}_{\tau}^0\\
     \bm{\theta}_{\tau}-\bm{\theta}_{\tau}^0\\
 \end{pmatrix}^T\left\{\frac{1}{n}\sum_{i=1}^n\mathcal{K}_h(\bm{Z}_i^T\bm{\alpha}_{\tau}^0+\bm{U}_i^T\bm{\theta}_{\tau}^0-Y_i)\begin{pmatrix}
     \bm{Z}_{i}\\
     \bm{U}_{i}\\
 \end{pmatrix}^{\otimes 2}
 \right\}\begin{pmatrix}
     \bm{\alpha}_{\tau}-\bm{\alpha}_{\tau}^0\\
     \bm{\theta}_{\tau}-\bm{\theta}_{\tau}^0\\
 \end{pmatrix}\\
 = & \frac{1}{2}\begin{pmatrix}
     \bm{\alpha}_{\tau}-\bm{\alpha}_{\tau}^0\\
     \bm{\theta}_{\tau}-\bm{\theta}_{\tau}^0\\
 \end{pmatrix}^T E\left\{f_{e|\bm{Z},\bm{X}}(0)\begin{pmatrix}
     \bm{Z}\\
     \bm{U}\\
 \end{pmatrix}^{\otimes 2}
 \right\}\begin{pmatrix}
     \bm{\alpha}_{\tau}-\bm{\alpha}_{\tau}^0\\
     \bm{\theta}_{\tau}-\bm{\theta}_{\tau}^0\\
 \end{pmatrix}+O_p(\frac{1}{\sqrt{nh}}\frac{K^2}{n})\\
  = & \frac{1}{2}\begin{pmatrix}
     \bm{\alpha}_{\tau}-\bm{\alpha}_{\tau}^0\\
     \bm{\theta}_{\tau}-\bm{\theta}_{\tau}^0\\
 \end{pmatrix}^TE\left\{f_{e|\bm{Z},\bm{X}}(0)
 \begin{pmatrix}
     \bm{Z}\\
     \bm{U}\\
 \end{pmatrix}^{\otimes 2}
 \right\}\begin{pmatrix}
     \bm{\alpha}_{\tau}-\bm{\alpha}_{\tau}^0\\
     \bm{\theta}_{\tau}-\bm{\theta}_{\tau}^0\\
 \end{pmatrix}+o_p\left(\frac{K}{n}\right).\\
 \end{split}
\end{equation}
And, we have
\begin{equation}\label{II_2}
    \begin{split}
        (\bm{\theta}_{\tau}-\bm{\theta}_{\tau}^0)^T(\bm{\Gamma}\otimes \bm{V})(\bm{\theta}_{\tau}-\bm{\theta}_{\tau}^0) \leq \sup_l(\gamma_l)\rho_{\max}(\bm{V})\left\|\bm{\theta}_{\tau}-\bm{\theta}_{\tau}^0\right\|_2^2 = o_p\left(\frac{K}{n}\right).
    \end{split}
\end{equation}
Then, (\ref{II_1}) and (\ref{II_2}) lead to 
\begin{equation}\label{II_com}
    II = \frac{1}{2}\begin{pmatrix}
     \bm{\alpha}_{\tau}-\bm{\alpha}_{\tau}^0\\
     \bm{\theta}_{\tau}-\bm{\theta}_{\tau}^0\\
 \end{pmatrix}^TE\left\{f_{e|\bm{Z},\bm{X}}(0)\begin{pmatrix}
     \bm{Z}\\
     \bm{U}\\
 \end{pmatrix}^{\otimes 2}
 \right\}\begin{pmatrix}
     \bm{\alpha}_{\tau}-\bm{\alpha}_{\tau}^0\\
     \bm{\theta}_{\tau}-\bm{\theta}_{\tau}^0\\
 \end{pmatrix}+o_p\left(\frac{K}{n}\right).
\end{equation}

\textbf{(3) Next, we consider the third term $III$.} By the Taylor Expansion, we can get 
\begin{equation}
\begin{split}
        & \sum_{l=1}^m \frac{K}{\mathcal{T}}
    \left[\int_{\mathcal{S}(\beta_{\tau,l}^0)}
    p_{\lambda_l}(\left|\bm{\mathcal{B}}^T(t)\bm{\theta}_{\tau,l}\right|)dt-
    \int_{\mathcal{S}(\beta_{\tau,l}^0)}p_{\lambda_l}(\left|\bm{\mathcal{B}}^T(t)\bm{\theta}_{\tau,l}^0\right|)dt
    \right]\\
    \approx & 
    \begin{pmatrix}
       \frac{K}{\mathcal{T}} \nabla \left[\int_{\mathcal{S}(\beta_{\tau,1}^0)}p_{\lambda_1}(\left|\bm{\mathcal{B}}^T(t)\bm{\theta}_{\tau,1}^0\right|)dt\right]\\
       \vdots\\
       \frac{K}{\mathcal{T}} \nabla \left[\int_{\mathcal{S}(\beta_{\tau,m}^0)}p_{\lambda_m}(\left|\bm{\mathcal{B}}^T(t)\bm{\theta}_{\tau,m}^0\right|)dt\right]\\
    \end{pmatrix}^T\begin{pmatrix}
     \bm{\theta}_{\tau}-\bm{\theta}_{\tau}^0\\
 \end{pmatrix}
 +\frac{1}{2}\begin{pmatrix}
     \bm{\theta}_{\tau}-\bm{\theta}_{\tau}^0\\
 \end{pmatrix}^T \\
 & \qquad
\begin{bmatrix}
     \frac{K}{\mathcal{T}} \nabla^2 \left[\int_{\mathcal{S}(\beta_{\tau,1}^0)}p_{\lambda_1}(\left|\bm{\mathcal{B}}^T(t)\bm{\theta}_{\tau,1}^0\right|)dt\right] & &\\
     & \ddots &\\
     & & \frac{K}{\mathcal{T}} \nabla^2 \left[\int_{\mathcal{S}(\beta_{\tau,m}^0)}p_{\lambda_m}(\left|\bm{\mathcal{B}}^T(t)\bm{\theta}_{\tau,m}^0\right|)dt\right]
\end{bmatrix}
 \begin{pmatrix}
     \bm{\theta}_{\tau}-\bm{\theta}_{\tau}^0\\
 \end{pmatrix}\\
  = & III_1+III_2
\end{split}
\end{equation}
For the bound of $\nabla \left[\int_{\mathcal{S}(\beta_{\tau,l}^0)}p_{\lambda_l}(\left|\bm{\mathcal{B}}^T(t)\bm{\theta}_{\tau,l}^0\right|)dt\right]$ and all $l=1,2,...,m$ and $j=1,2,...,K+p$, using H{\"o}lder inequality, we have
\begin{equation*}
    \begin{split}
        \left|\frac{\partial}{\partial\theta_{\tau,l,j}}\int_{\mathcal{S}(\beta_{\tau,l}^0)}p_{\lambda_l}(\left|\bm{\mathcal{B}}^T(t)\bm{\theta}_{\tau,l}^0\right|)dt\right| & =  \left|\int_{\mathcal{S}(\beta_{\tau,l}^0)}
        \frac{\partial}{\partial\theta_{\tau,l,j}^0}
        p_{\lambda_l}(\left|\bm{\mathcal{B}}^T(t)\bm{\theta}_{\tau,l}^0\right|)dt\right|\\
        & = \left|\int_{\mathcal{S}(\beta_{\tau,l}^0)}
        \dot{p}_{\lambda_l}(\left|\bm{\mathcal{B}}^T(t)\bm{\theta}_{\tau,l}^0\right|)B_j(t)dt \quad \text{sgn}(\theta_{\tau,l,j}^0)\right|\\
        & \leq \sqrt{\int_{\mathcal{S}(\beta_{\tau,l}^0)}
        \dot{p}_{\lambda_l}(|\bm{\mathcal{B}}^T(t)\bm{\theta}_{\tau,l}^0|)^2dt} \sqrt{\int_{\mathcal{S}(\beta_{\tau,l}^0)}B_j^2(t)dt}\\
        &\leq \|B_j\|_2 \sqrt{\int_{\mathcal{S}(\beta_{\tau,l}^0)}
        \dot{p}_{\lambda_l}(|\bm{\mathcal{B}}^T(t)\bm{\theta}_{\tau,l}^0|)^2dt} \\
        & = O(K^{-1/2}n^{-1/2}K^{-1}) = O(n^{-1/2}K^{-3/2}),
    \end{split}
\end{equation*}
with the condition
$\max_{l}\sqrt{\int_{\mathcal{S}(\beta_{\tau,l}^0)}
        \dot{p}_{\lambda_l}(|\bm{\mathcal{B}}^T(t)\bm{\theta}_{\tau,l}^0|)^2dt} = o(n^{-1/2}K^{-1})$ (Condition \ref{fscad}-(i)).
Then, we can also have
\begin{equation}\label{III_1}
    \begin{split}
        |III_1| & \leq \left\|\begin{pmatrix}
       \frac{K}{\mathcal{T}} \nabla \left[\int_{\mathcal{S}(\beta_{\tau,1}^0)}p_{\lambda_1}(\left|\bm{\mathcal{B}}^T(t)\bm{\theta}_{\tau,1}^0\right|)dt\right]\\
       \vdots\\
       \frac{K}{\mathcal{T}} \nabla \left[\int_{\mathcal{S}(\beta_{\tau,m}^0)}p_{\lambda_m}(\left|\bm{\mathcal{B}}^T(t)\bm{\theta}_{\tau,m}^0\right|)dt\right]
    \end{pmatrix}^T\right\|_2
    \left\|
     \bm{\theta}_{\tau}-\bm{\theta}_{\tau}^0\right\|_2\\
     & = O_p(K/n).
    \end{split}
\end{equation}

For the bound of $\nabla^2
\left[\int_{\mathcal{S}(\beta_{\tau,l}^0)}p_{\lambda_l}(\left|\bm{\mathcal{B}}^T(t)\bm{\theta}_{\tau,l}^0\right|)dt\right]$ and all $l=1,2,...,m$ and $j,k=1,2,...,K+p$, using H{\"o}lder inequality, we have 
\begin{equation*}
    \begin{split}
      \left|\frac{\partial}{\partial\theta_{\tau,l,j}\partial\theta_{\tau,l,k}}
\int_{\mathcal{S}(\beta_{\tau,l}^0)}p_{\lambda_l}(\left|\bm{\mathcal{B}}^T(t)\bm{\theta}_{\tau,l}^0\right|)dt
\right| & = 
\int_{\mathcal{S}(\beta_{\tau,l}^0)} \nabla^2 p_{\lambda_l}(\left|\bm{\mathcal{B}}^T(t)\bm{\theta}_{\tau,l}^0\right|)dt\\
& = \int_{\mathcal{S}(\beta_{\tau,l}^0)}  \ddot{p}_{\lambda_l}(\left|\bm{\mathcal{B}}^T(t)\bm{\theta}_{\tau,l}^0\right|)
\left[B_j(t)\circ \text{sgn}(\theta_{\tau,l,j}^0)\right]\left[B_k(t)\circ \text{sgn}(\theta_{\tau,l,k}^0)\right]dt\\
& \leq \sqrt{\int_{\mathcal{S}(\beta_{\tau,l}^0)}
        \ddot{p}_{\lambda_l}(|\bm{\mathcal{B}}^T(t)\bm{\theta}_{\tau,l}^0|)^2dt} 
        \sqrt{\int_{0}^{\mathcal{T}}B_j^2(t)B_k^2(t)dt}\\
& \leq \sqrt{\int_{\mathcal{S}(\beta_{\tau,l}^0)}
        \ddot{p}_{\lambda_l}(|\bm{\mathcal{B}}^T(t)\bm{\theta}_{\tau,l}^0|)^2dt} 
        \sqrt{\int_{0}^{\mathcal{T}}B_j(t)B_k(t)dt}\\
& = o(K^{-1}),
    \end{split}
\end{equation*}
with Condition \ref{fscad}-(i), 
where $\circ$ denotes the entry-wise product of two vectors and the last equation is based on the property 
$\sup_{j,k}|<B_j,B_k>|=O(K^{-1})$. Therefore,
\begin{equation}\label{III_2}
    \begin{split}
        \left|III_2\right| & \leq \frac{1}{2}\left\|
     \bm{\theta}_{\tau}-\bm{\theta}_{\tau}^0\right\|_2^2 \max_{l,j,k} \left|\frac{\partial}{\partial\theta_{\tau,l,j}\partial\theta_{\tau,l,k}}
\int_{\mathcal{S}(\beta_{\tau,l}^0)}p_{\lambda_l}(\left|\bm{\mathcal{B}}^T(t)\bm{\theta}_{\tau,l}^0\right|)dt
\right|\\
& = o_p(K/n).
    \end{split}
\end{equation}
With the above results, we have
\begin{equation}
    \label{III_com}
    III = O_p(K/n).
\end{equation}

Combining (\ref{diff}), (\ref{I_com}), (\ref{II_com}) and (\ref{III_com}), we finally have
\begin{equation}\label{diff2}
\begin{split}
   &  \mathcal{L}^*(\bm{\alpha}_{\tau},\bm{\theta}_{\tau})-
    \mathcal{L}^*(\bm{\alpha}_{\tau}^0,\bm{\theta}_{\tau}^0)\\
    \geq &   I+II+III\\
    = & O_p(K/n)+\frac{1}{2}\begin{pmatrix}
     \bm{\alpha}_{\tau}-\bm{\alpha}_{\tau}^0\\
     \bm{\theta}_{\tau}-\bm{\theta}_{\tau}^0\\
 \end{pmatrix}^TE\left\{f_{e|\bm{Z},\bm{X}}(0)
 \begin{pmatrix}
     \bm{Z}\\
     \bm{U}\\
 \end{pmatrix}^{\otimes 2}
 \right\}\begin{pmatrix}
     \bm{\alpha}_{\tau}-\bm{\alpha}_{\tau}^0\\
     \bm{\theta}_{\tau}-\bm{\theta}_{\tau}^0\\
 \end{pmatrix}\\
 > & 0,
\end{split} 
\end{equation}
where the positive term
\begin{equation*}
\begin{split}
   & \frac{1}{2}\begin{pmatrix}
     \bm{\alpha}_{\tau}-\bm{\alpha}_{\tau}^0\\
     \bm{\theta}_{\tau}-\bm{\theta}_{\tau}^0\\
 \end{pmatrix}^TE\left\{f_{e|\bm{Z},\bm{X}}(0)
 \begin{pmatrix}
     \bm{Z}\\
     \bm{U}\\
 \end{pmatrix}^{\otimes 2}
 \right\}\begin{pmatrix}
     \bm{\alpha}_{\tau}-\bm{\alpha}_{\tau}^0\\
     \bm{\theta}_{\tau}-\bm{\theta}_{\tau}^0\\
 \end{pmatrix} \\
 \geq & C_1\rho_{\min} (\bm{U}\bm{U}^T) \left\|\begin{pmatrix}
     \bm{\alpha}_{\tau}-\bm{\alpha}_{\tau}^0\\
     \bm{\theta}_{\tau}-\bm{\theta}_{\tau}^0\\
 \end{pmatrix}\right\|_2^2\\
 = & O_p(1/(nK)),
\end{split}
\end{equation*}
with Condition \ref{XZ-moment} and Lemma A.\ref{lemma2}.

Then, (\ref{diff}) holds, which means there is a local minimizer
$(\widehat{\bm{\alpha}}_{\tau},\widehat{\bm{\theta}}_{\tau})$ of 
$\mathcal{L}^*(\bm{\alpha}_{\tau},\bm{\theta}_{\tau})$
such that $\|\widehat{\bm{\alpha}}_{\tau}-\bm{\alpha}^0_{\tau}\|_2=O_p(n^{-1/2})$ and 
$\|\widehat{\bm{\theta}}_{\tau}-\bm{\theta}^0_{\tau}\|_2=O_p(n^{-1/2}K)$.
Moreover, with $\widehat{\beta}_{\tau,l}(t)=\bm{\mathcal{B}}^T(t)\widehat{\bm{\theta}}_{\tau,l}$, we have
\begin{equation}
    \begin{split}
        \left\|\widehat{\beta}_{\tau,l}-\beta_{\tau,l}\right\|_2
        \leq & \left\|\widehat{\beta}_{\tau,l}-\beta^0_{\tau,l}\right\|_2 +\left\|\beta^0_{\tau,l}-\beta_{\tau,l}\right\|_2\\
        = & \left\|(\widehat{\bm{\theta}}_{\tau,l}-\bm{\theta}^0_{\tau,l})^T \bm{\mathcal{B}}(t)\right\|_2+
        \left\|\beta^0_{\tau,l}-\beta_{\tau,l}\right\|_2\\
        = & O_p(n^{-1/2}K^{1/2})+O(K^{-r})\\
        = & O_p(n^{-1/2}K^{1/2}),    
    \end{split}
\end{equation}
with the condition $K/n^{1/(2r+1)}\rightarrow\infty$ (Condition \ref{basis-number}).

Then the proof of Theorem 1 is completed.

\subsection{Proof of Theorem 2}
\subsubsection*{Proof of (i) of Theorem 2}
To prove the first part of Theorem 2, we need to show that $\widehat{\beta}_{\tau,l}(t)=0$ for all $t\in \mathcal{N}(\beta_{\tau,l})$ with probability tending to one.

As stated in Section 3, $\bm{\mathcal{B}}_{\tau,l,1}(t)$ denotes the $K_{\tau,l}^*$ dimensional sub-vector of $\bm{\mathcal{B}}(t)$ 
such that each $B_{j}(t)$ in $\bm{\mathcal{B}}_{\tau,l,1}(t)$ has a support inside $\mathcal{S}(\beta_{\tau,l})$. We also further denote the $\bm{\mathcal{B}}_{\tau,l,2}(t)$ as the sub-vector of $\bm{\mathcal{B}}(t)$ 
such that the support of each $B_{j}(t)$ in $\bm{\mathcal{B}}_{\tau,l,2}(t)$ belongs to $\mathcal{N}(\beta_{\tau,l})$. Let $\mathcal{A}_{\tau,l,j}$ consist of indices $k$ such that $B_j(t)$ belongs to $\bm{\mathcal{B}}_{\tau,l,j}(t)$, and $F_{\tau,l,j}$ denote the union of supports of those basis functions in $\bm{\mathcal{B}}_{\tau,l,j}(t),j=1,2$.
By the local support property of B-spline basis functions, $F_{\tau,l,1}$ converges to $\mathcal{S}(\beta_{\tau,l})$ and $F_{\tau,l,2}$ converges to $\mathcal{N}(\beta_{\tau,l})$ as $K\rightarrow \infty$.


Fix a $\theta_{\tau,l,k}$ such that $k\in\mathcal{A}_{\tau,l,2}$, recall that
\begin{equation*}
    \begin{split}
        \frac{\partial \mathcal{L}^*(\bm{\alpha}_{\tau},\bm{\theta}_{\tau})}{\partial \theta_{\tau,l,k}}  = &\frac{\partial \mathcal{L}_1^*(\bm{\alpha}_{\tau},\bm{\theta}_{\tau})}{\partial \theta_{\tau,l,k}} + 
        \frac{K}{\mathcal{T}}\text{sgn} (\theta_{\tau,l,k})\int_0^\mathcal{T}\dot{p}_{\lambda_l}(\left|\bm{\mathcal{B}}^T(t)\bm{\theta}_{\tau,l}\right|)B_k(t)dt ,
    \end{split}
\end{equation*}
for the first term of RHS, we take Taylor expansion, then
\begin{equation}\label{firstd}
    \begin{split}
        \frac{\partial \mathcal{L}^*(\bm{\alpha}_{\tau},\bm{\theta}_{\tau})}{\partial \theta_{\tau,l,k}}  = &\frac{\partial \mathcal{L}_1^*(\bm{\alpha}_{\tau}^0,\bm{\theta}_{\tau}^0)}{\partial \theta_{\tau,l,k}} + \sum_{l^{'} = 1}^m\sum_{j=1}^{K+p} 
        \frac{\partial^2 \mathcal{L}_1^*(\bm{\alpha}_{\tau}^0,\bm{\theta}_{\tau}^0)}{\partial \theta_{\tau,l,k}\partial \theta_{\tau,l^{'},j}} (\theta_{\tau,l^{'},j}-\theta^0_{\tau,l^{'},j})\\
        & \qquad +
        \frac{K}{\mathcal{T}}\text{sgn} (\theta_{\tau,l,k})\int_0^\mathcal{T}\dot{p}_{\lambda_l}(\left|\bm{\mathcal{B}}^T(t)\bm{\theta}_{\tau,l}\right|)B_k(t)dt ,
    \end{split}
\end{equation}
where 
\begin{equation*}
    \begin{split}
        \frac{\partial \mathcal{L}_1^*(\bm{\alpha}_{\tau}^0,\bm{\theta}_{\tau}^0)}{\partial \theta_{\tau,l,k}} & = \frac{1}{n}\sum_{i=1}^n\left\{G_h(\bm{Z}_i^T\bm{\alpha}_{\tau}^0+
\bm{U}_i^T\bm{\theta}_{\tau}^0-Y_i)-\tau\right\} U_{i,lk}
+ 2\gamma_l
\bm{V}_{k\cdot} \bm{\theta}_{\tau,l}^0,
\end{split}
\end{equation*}
$U_{i,lk}=\int_0^{\mathcal{T}}X_{il}(t)B_{k}(t)dt$ and $\bm{V}_{k\cdot}$ denotes the $k$th row of $\bm{V}$.  

Below we will consider each term in (\ref{firstd}).

\begin{itemize}
   \item [(1)]
By Lemma A. \ref{lemma4}, $|U_{i,lk}| = O(K^{-1/2})$, $\|\bm{V}_{k\cdot}\|_{\infty}=O(K^{2q-1})$ and Conditions \ref{XZ-moment}-\ref{bandwidth}, we have 
\begin{equation}\label{firstd_1}
     \frac{\partial \mathcal{L}_1^*(\bm{\alpha}_{\tau}^0,\bm{\theta}_{\tau}^0)}{\partial \theta_{\tau,l,k}} = O_p(n^{-1/2}K^{-1/2}).
\end{equation}

\item [(2)]
Based on the Conditions \ref{XZ-moment}-\ref{bandwidth}, Lemma A.\ref{lemma5} and assumption $\left|\theta_{\tau,l^{'},j}-\theta^0_{\tau,l^{'},j}\right| = O_p(\sqrt{K/n})$, we can get 
\begin{equation}
    \label{firstd_2}
    \sum_{l^{'} = 1}^m\sum_{j=1}^{K+p} 
        \frac{\partial^2 \mathcal{L}_1^*(\bm{\alpha}_{\tau}^0,\bm{\theta}_{\tau}^0)}{\partial \theta_{\tau,l,k}\partial \theta_{\tau,l^{'},j}} (\theta_{\tau,l^{'},j}-\theta^0_{\tau,l^{'},j}) = O_p(K^{1/2}n^{-1/2}).
\end{equation}

\item [(3)]
Let $\mathcal{S}_k$ denote the support of $B_k(t)$, then, in the last term,
\begin{equation*}
    \begin{split}
       &  \left|\int_0^\mathcal{T}\dot{p}_{\lambda_l}(\left|\bm{\mathcal{B}}^T(t)\bm{\theta}_{\tau,l}\right|)dt
        - \int_{\mathcal{S}_k}\dot{p}_{\lambda_l}(\left|B_k(t)\theta_{\tau,l,k}\right|)dt\right|
        \\
       = & 
       \left|\int_{\mathcal{S}(\beta_{\tau,l})}\dot{p}_{\lambda_l}(\left|\bm{\mathcal{B}}^T(t)\bm{\theta}_{\tau,l}\right|)dt
       +\int_{\mathcal{N}(\beta_{\tau,l})}\dot{p}_{\lambda_l}(\left|\bm{\mathcal{B}}^T(t)\bm{\theta}_{\tau,l}\right|)dt
       -\int_{\mathcal{S}_k}\dot{p}_{\lambda_l}(\left|B_k(t)\theta_{\tau,l,k}\right|)dt\right|
       \\
       \leq & 
       \left|\int_{\mathcal{S}(\beta_{\tau,l})}\dot{p}_{\lambda_l}(\left|\bm{\mathcal{B}}^T(t)\bm{\theta}_{\tau,l}\right|)dt\right|
       +\left|\int_{\mathcal{N}(\beta_{\tau,l})}\dot{p}_{\lambda_l}(\left|\bm{\mathcal{B}}^T(t)\bm{\theta}_{\tau,l}\right|)dt
       -\int_{\mathcal{S}_k}\dot{p}_{\lambda_l}(\left|B_k(t)\theta_{\tau,l,k}\right|)dt
       \right|\\
       = & o(n^{-1/2}K^{-1})+O(K^{-1}) \\
       = & o(1),
    \end{split}
\end{equation*}
with the Condition $\int_{\mathcal{S}(\beta_{\tau,l})}\dot{p}_{\lambda_l}(\left|\bm{\mathcal{B}}^T(t)\bm{\theta}_{\tau,l}\right|)dt = o(n^{-1/2}K^{-1})$. Then we have 
\begin{equation}
    \label{firstd_3}
    \int_0^\mathcal{T}\dot{p}_{\lambda_l}(\left|\bm{\mathcal{B}}^T(t)\bm{\theta}_{\tau,l}\right|)dt
        = \int_{\mathcal{S}_k}\dot{p}_{\lambda_l}(\left|B_k(t)\theta_{\tau,l,k}\right|)dt +o(1).
\end{equation}
\end{itemize}

Combining (\ref{firstd}), (\ref{firstd_1}), (\ref{firstd_2}) and (\ref{firstd_3}), we finally have 
\begin{equation}
    \begin{split}
        \frac{\partial \mathcal{L}^*(\bm{\alpha}_{\tau},\bm{\theta}_{\tau})}{\partial \theta_{\tau,l,k}} 
        & = \frac{K}{\mathcal{T}}\lambda_l\text{sgn}(\theta_{\tau,l,k})\int_{\mathcal{S}_k}\dot{p}_{\lambda_l}(\left|B_k(t)\theta_{\tau,l,k}\right|)\lambda_l^{-1} B_k(t)dt   +O_p(K^{1/2}n^{-1/2})\\
        & = \lambda_l \left\{\frac{K}{\mathcal{T}}\lambda_l\text{sgn}(\theta_{\tau,l,k})\int_{\mathcal{S}_k}\dot{p}_{\lambda_l}(\left|B_k(t)\theta_{\tau,l,k}\right|)\lambda_l^{-1} B_k(t)dt   +O_p(K^{1/2}n^{-1/2}\lambda_l^{-1})\right\}.
    \end{split}
\end{equation}
If Condition \ref{fscad}-(ii) holds, $\liminf_{n\rightarrow \infty}\liminf_{\theta\rightarrow 0^+}\dot{p}_{\lambda_l}(\theta)\lambda_l^{-1}>0$, and $K^{1/2}n^{-1/2}\lambda_l^{-1}\rightarrow 0$. Besides,
$B_k(t)$ is non-negative, the sign of the derivative is completely determined by that of $\theta_{\tau,l,k}$. Now, since $(\widehat{\bm{\alpha}}_{\tau},\widehat{\bm{\theta}}_{\tau})$ minimizes 
$\mathcal{L}^*(\bm{\alpha}_{\tau},\bm{\theta}_{\tau})$, we must have $\widehat{\theta}_{\tau,l,k}=0$ for $k\in \mathcal{A}_{\tau,l,2}$.
with probability tending to one. Similar with \cite{lin2017locally}, it is easy to prove that the union $\widehat{F}_{\tau,l,2}$ of supports of basis functions associated to $\widehat{\bm{\theta}}_{\tau,l,2}$ equals to $F_{\tau,l,2}$ in probability.  Therefore $\widehat{F}_{\tau,l,2}$ converges to $\mathcal{N}(\beta_{\tau,l})$. This completes the proof of the first part of Theorem 2.

\subsubsection*{Proof of (ii) of Theorem 2}
We divide $\bm{\theta}_{\tau,l}^0$ into two parts: $\bm{\theta}_{\tau,l}^{0(1)}$ that consists of those $\theta_{\tau,l,k}$ such that $k\in \mathcal{A}_{\tau,l,1}$, and 
$\bm{\theta}_{\tau,l}^{0(2)}$ that consists of those $\theta_{\tau,l,k}$ such that $k\in \mathcal{A}_{\tau,l,2}$. Then we define $\bm{\theta}_{\tau}^{0(1)} = \left(\left(\bm{\theta}_{\tau,1}^{0(1)}\right)^T,\cdots,\left(\bm{\theta}_{\tau,m}^{0(1)}\right)^T\right)^T$ and $\bm{\theta}_{\tau}^{0(2)} = \left(\left(\bm{\theta}_{\tau,1}^{0(2)}\right)^T,\cdots,\left(\bm{\theta}_{\tau,m}^{0(2)}\right)^T\right)^T$.
Correspondingly, we also divide $\widehat{\bm{\theta}}_{\tau,l}$ into two parts, namely, $\widehat{\bm{\theta}}_{\tau,l}^{(1)}$ and $\widehat{\bm{\theta}}_{\tau,l}^{(2)}$, then define 
$\widehat{\bm{\theta}}_{\tau}^{(1)} = \left(\left(\widehat{\bm{\theta}}_{\tau,1}^{(1)}\right)^T,\cdots,\left(\widehat{\bm{\theta}}_{\tau,m}^{(1)}\right)^T\right)^T$ and $\widehat{\bm{\theta}}_{\tau}^{(2)} = \left(\left(\widehat{\bm{\theta}}_{\tau,1}^{(2)}\right)^T,\cdots,\left(\widehat{\bm{\theta}}_{\tau,m}^{(2)}\right)^T\right)^T$.
From the proof of (i), we can get that each element of $\widehat{\bm{\alpha}}_{\tau}$ and
$\widehat{\bm{\theta}}_{\tau}^{(1)}$ stay away from zero when $n$ is sufficiently large. At the same time, $\widehat{\bm{\theta}}_{\tau}^{(2)}=\bm{0}$ with probability tending to one. Thus, $\left(\widehat{\bm{\alpha}}_{\tau}^T,
\left(\widehat{\bm{\theta}}_{\tau}^{(1)}\right)^T\right)^T$ satisfies
\begin{equation*}
    \begin{split}
        \dot{\mathcal{L}}_1^*(\widehat{\bm{\alpha}}_{\tau},\widehat{\bm{\theta}}_{\tau}^{(1)})
        + \begin{pmatrix}
            \bm{0}_{d\times 1}\\ 
       \left[ \frac{K}{\mathcal{T}}\frac{\partial}{\partial\theta_{\tau,l,k}}\int_{0}^{\mathcal{T}}p_{\lambda_l}(\left|\bm{\mathcal{B}}^T(t)\widehat{\bm{\theta}}_{\tau,l}\right|)dt\right]_{k\in \mathcal{A}_{\tau,l,1}}^{l=1,\ldots,m} 
       \\
        \end{pmatrix}  = 0.  
    \end{split}
\end{equation*}
Then applying the Taylor expansion to $\frac{K}{\mathcal{T}}\frac{\partial}{\partial\theta_{\tau,l,k}}\int_{0}^{\mathcal{T}}p_{\lambda_l}(\left|\bm{\mathcal{B}}^T(t)\widehat{\bm{\theta}}_{\tau,l}\right|)dt$, we can get that
\begin{equation*}
    \begin{split}
        \frac{K}{\mathcal{T}}\frac{\partial}{\partial\theta_{\tau,l,k}}\int_{0}^{\mathcal{T}}p_{\lambda_l}\left(\left|\bm{\mathcal{B}}^T(t)\widehat{\bm{\theta}}_{\tau,l}\right|\right)dt  = & \frac{K}{\mathcal{T}}\text{sgn}(\theta_{\tau,l,k}^0)\int_{0}^{\mathcal{T}}\dot{p}_{\lambda_l}\left(\left|\bm{\mathcal{B}}^T(t)\bm{\theta}^0_{\tau,l}\right|\right)B_k(t)dt \\
        & \quad +
        \frac{K}{\mathcal{T}}\int_{0}^{\mathcal{T}}\ddot{p}_{\lambda_l}\left(\left|\bm{\mathcal{B}}^T(t)\bm{\theta}^0_{\tau,l}\right|\right)B^2_k(t)dt (\widehat{\theta}_{\tau,l,k}-\theta^0_{\tau,l,k})\\
        & = o_p(\widehat{\theta}_{\tau,l,k}-\theta^0_{\tau,l,k}),
    \end{split}
\end{equation*}
by the Condition \ref{fscad}-(i). Then, we can have 
\begin{equation*}
    \begin{split}
        \dot{\mathcal{L}}_1^*(\widehat{\bm{\alpha}}_{\tau},\widehat{\bm{\theta}}_{\tau}^{(1)})
        + \begin{pmatrix}
            \bm{0}_{d\times 1}\\ 
       o_p(\widehat{\bm{\theta}}_{\tau}^{(1)}-\bm{\theta}^{0(1)}_{\tau})
       \\
        \end{pmatrix}=0 .
    \end{split}
\end{equation*}
By Taylor expansion and the above results, we have 
\begin{equation}\label{hessian}
    \begin{split}
    \sqrt{n}
        \begin{pmatrix}
            \widehat{\bm{\alpha}}_{\tau}- \bm{\alpha}_{\tau}^0\\
            \widehat{\bm{\theta}}_{\tau}^{(1)}- \bm{\theta}_{\tau}^{0(1)}\\
        \end{pmatrix}
        & = -\left\{\ddot{\mathcal{L}}_1^*(\bm{\alpha}_{\tau}^0,\bm{\theta}_{\tau}^{0(1)})\right\}^{-1}\sqrt{n}\dot{\mathcal{L}}_1^*(\bm{\alpha}_{\tau}^0,\bm{\theta}_{\tau}^{0(1)})
        \left\{1+o_p(1)\right\}.
    \end{split}
\end{equation}

According to the proof of Lemma A.\ref{lemma4}, we have
\begin{equation*}
    \begin{split}
         \sqrt{n}\dot{\mathcal{L}}_1^*(\bm{\alpha}_{\tau}^0,\bm{\theta}_{\tau}^{0(1)})      
= & 
 \frac{1}{\sqrt{n}}\sum_{i=1}^n\left\{\mathcal{K}_h\left(
-e_i+\int_{0}^{\mathcal{T}}\bm{X}_i^T(t)\bm{\beta}^*_{\tau}(t)dt
\right) \left(\bm{U}_i^T\bm{\theta}_{\tau}^0-\int_{0}^{\mathcal{T}}\bm{X}_i^T(t)\bm{\beta}_{\tau}(t)dt\right)\right\}(\bm{Z}_i^T,\bm{U}_{\tau,i}^{*T})^T\\
& \qquad+ \frac{1}{\sqrt{n}}\sum_{i=1}^n\left\{G_h(-e_i)-\tau\right\}(\bm{Z}_i^T,\bm{U}_{\tau,i}^{*T})^T+o_p(1)\\
& = \frac{1}{\sqrt{n}}\sum_{i=1}^n\left\{G_h(-e_i)-\tau\right\}(\bm{Z}_i^T,\bm{U}_{\tau,i}^{*T})^T+o_p(1).
\end{split}
\end{equation*}

Similar with the proof of Lemma A.\ref{lemma6}, we can prove that 
 \begin{equation*}
        \frac{1}{\sqrt{n}}\sum_{i=1}^n\left\{E\left((\bm{Z}^T,\bm{U}_{\tau}^{*T})^{T}\right)^{\otimes2}\right\}^{-1/2}\left\{G_h(-e_i)-\tau\right\}(\bm{Z}_i^T,\bm{U}_{\tau,i}^{*T})^T \overset{d}{\rightarrow}  \mathbb{\bm{N}}(\bm{0},\tau(1-\tau)\bm{I}_{d+K^*_{\tau}}),
    \end{equation*}
and
    \begin{equation}\label{normal_sub}
\bm{\Sigma}_{\tau,1}^{-1/2}\sqrt{n}\dot{\mathcal{L}}_1^*(\bm{\alpha}_{\tau}^0,\bm{\theta}_{\tau}^{0(1)}) \overset{d}{\rightarrow}  \mathbb{\bm{N}}(\bm{0},\tau(1-\tau)\bm{I}_{d+K^*_{\tau}}),
    \end{equation}
where $\bm{\Sigma}_{\tau,1} = E\left\{\left((\bm{Z}^T,\bm{U}_{\tau}^{*T})^{T}\right)^{\otimes2}\right\}$.
    
Similar with the proof of Lemma A.\ref{lemma5}, we have
\begin{equation*}
    \begin{split}
        \ddot{\mathcal{L}}_0^*(\bm{\alpha}_{\tau}^0,\bm{\theta}_{\tau}^{0(1)})
        & = 
        \frac{1}{n}\sum_{i=1}^n\mathcal{K}_h(\bm{Z}_i\bm{\alpha}_{\tau}^0+\bm{U}_i^T\bm{\theta}_{\tau}^0-Y_i)\begin{pmatrix}
     \bm{Z}_{i}\\
     \bm{U}_{\tau,i}^{*}\\
 \end{pmatrix}^{\otimes 2} \\
 & = E\left\{f_{e|\bm{Z},\bm{X}}(0)\begin{pmatrix}
     \bm{Z}\\
     \bm{U}_{\tau}^{*}\\
 \end{pmatrix}^{\otimes 2}\right\} +O_p(\frac{1}{\sqrt{nh}}),
    \end{split}
\end{equation*}
and 
\begin{equation*}
    \begin{split}
        \ddot{\mathcal{L}}_1^*(\bm{\alpha}_{\tau}^0,\bm{\theta}_{\tau}^{0(1)})
 & = E\left\{f_{e|\bm{Z},\bm{X}}(0)
 \begin{pmatrix}
     \bm{Z}\\
     \bm{U}_{\tau}^{*}\\
 \end{pmatrix}^{\otimes 2}\right\} 
 + o_p\left(\frac{1}{\sqrt{nK}}\right)+O_p(\frac{1}{\sqrt{nh}})\\
 & = E\left\{f_{e|\bm{Z},\bm{X}}(0)
 \begin{pmatrix}
     \bm{Z}\\
     \bm{U}_{\tau}^{*}\\
 \end{pmatrix}^{\otimes 2}\right\} 
,
    \end{split}
\end{equation*}
then 
\begin{equation}\label{prob_sub}
    \ddot{\mathcal{L}}_1^*(\bm{\alpha}_{\tau}^0,\bm{\theta}_{\tau}^{0(1)})\overset{P}{\rightarrow} \bm{\Sigma}_{\tau,2},\qquad
 \bm{\Sigma}_{\tau,2}=
 E\left\{f_{e|\bm{Z},\bm{X}}(0)\begin{pmatrix}
     \bm{Z}\\
     \bm{U}_{\tau}^{*}\\
 \end{pmatrix}^{\otimes 2}\right\}.
\end{equation}
Thus, by (\ref{normal_sub}), (\ref{prob_sub}) and Slutsky's theorem, we can get that 
\begin{equation}\label{normal-para}
    \begin{split}
    \sqrt{n}\left(\bm{\Sigma}_{\tau,2}^{-1}\bm{\Sigma}_{\tau,1}\bm{\Sigma}_{\tau,2}^{-1}\right)^{-1/2}
        \begin{pmatrix}
            \widehat{\bm{\alpha}}_{\tau}- \bm{\alpha}_{\tau}^0\\            \widehat{\bm{\theta}}_{\tau}^{(1)}- \bm{\theta}_{\tau}^{0(1)}
        \end{pmatrix}
        \overset{d}{\rightarrow}  
        \bm{\mathbb{N}}(\bm{0},\tau(1-\tau)\bm{I})
    \end{split}
\end{equation}

Finally we prove the asymptotic normality of $\sqrt{n/K}(\widehat{\beta}_{\tau,l}(t) -\beta_{\tau,l}(t))$.
Recall that
$\sqrt{n/K}(\widehat{\beta}_{\tau,l}(t) -\beta_{\tau,l}(t))$ admits the following decomposition,
\begin{equation*}
    \widehat{\beta}_{\tau,l}(t) -\beta_{\tau,l}(t)=  \frac{1}{\sqrt{n}}\bm{\Lambda}_{\tau,l}(t)\sqrt{n}
        \begin{pmatrix}
            \widehat{\bm{\alpha}}_{\tau}- \bm{\alpha}_{\tau}^0\\
         \widehat{\bm{\theta}}_{\tau}^{(1)}- \bm{\theta}_{\tau}^{0(1)}
        \end{pmatrix}+ \beta_{\tau,l}^0(t) -\beta_{\tau,l}(t).
\end{equation*}
From Lemma A.\ref{lemma1}, for the B-spline approximation error, we have $\beta_{\tau,l}^0(t) -\beta_{\tau,l}(t) = O(K^{-r})$. Under Condition \ref{basis-number}, $nK^{-(2r+1)}\to 0$, the bias of $\widehat{\beta}_{\tau,l}(t) -\beta_{\tau,l}(t)$ is asymptotically
negligible. Then by (\ref{normal-para}) and Slutsky's theorem, we can conclude that for any given point $t$ such $\beta_{\tau,l}\neq 0$, we have 
\begin{equation}\label{normal-point}
    \sigma_{\tau,l}^{-1/2}(t)\left(\widehat{\beta}_{\tau,l}(t) -\beta(t)\right) \to N(0, 1),
\end{equation}
in distribution, where $\sigma_{\tau,l}(t)= n^{-1}\tau(1-\tau)\bm{\Lambda}_{\tau,l}(t)\bm{\Sigma}_{\tau,2}^{-1}\bm{\Sigma}_{\tau,1}\bm{\Sigma}_{\tau,2}^{-1}\bm{\Lambda}_{\tau,l}^T(t)=O(K/n)$.

(\ref{normal-point}) shows the point-wise asymptotic normality and we can construct the corresponding confidence interval. However, in many studies, it is desirable to obtain simultaneous confidence bands (SCB) for the varying coefficient functions.

Define 
\begin{equation*} 
\Theta_{\tau,l}(t) = \sigma_{\tau,l}^{-1/2}(t)n^{-1}\sqrt{\tau(1-\tau)}\bm{\Lambda}_{\tau,l}(t)\bm{\Sigma}_{\tau,2}^{-1}\bm{\Sigma}_{\tau,1}^{1/2}\sum_{\i=1}^n\bm{\zeta}_i,\qquad \bm{\zeta}_{\tau,l,i}\overset{iid}{\sim } \bm{\mathbb{N}}(\bm{0},\bm{I}_{d+K_{\tau}^*}).
\end{equation*}
It is easy to show that $\Theta_{\tau,l}(t)$ is a Gaussian process with $E(\Theta_{\tau,l}(t)) = 0$ and $Var(\Theta_{\tau,l}(t)) = 1$. And, the covariance function of the Gaussian process is 
\begin{equation*}
\begin{split}
  \mathcal{C} (t,s) 
= n^{-1}\tau(1-\tau)\sigma_{\tau,l}^{-1/2}(t)\sigma_{\tau,l}^{-1/2}(s)
\bm{\Lambda}_{\tau,l}(t)
\bm{\Sigma}_{\tau,2}^{-1}\bm{\Sigma}_{\tau,1}\bm{\Sigma}_{\tau,2}^{-1}
\bm{\Lambda}_{\tau,l}^T(s).
\end{split}
   \end{equation*}

Similar to the proofs for Theorem 3 in \citep{ma2016estimation}, and Theorem 5 in \citep{belloni2019conditional}, by the strong approximation theorem \citep{csorgo2014strong}, we can prove that
\begin{equation}\label{scb1}
    \sup_{t\in \mathcal{S}(\beta_{\tau,l})}\left|\sigma_{\tau,l}^{-1/2}(t)\left\{\widehat{\beta}_{\tau,l}(t)-\beta_{\tau,l}(t)\right\} - \Theta_{\tau,l}(t)\right| = o_p(\sqrt{K/n}).
\end{equation}
Then for $t\in \mathcal{S}(\beta_{\tau,l})$, it can be proved that 
\begin{equation}
\label{fclt}
\sigma^{-1/2}_{\tau,l}(t)(\widehat{\beta}_{\tau,l}(t)-\beta_{\tau,l}(t))\overset{d}{\rightarrow} \mathbb{G}(t),
\end{equation}
where $\mathbb{G}(t)$ is a Gaussian random process with mean $0$ defined on $\mathcal{S}(\beta_{\tau,l})$ with
the covariance function $\mathcal{C}(t,s)$.

Denote the left endpoint and right endpoint of $\mathcal{S}(\beta_{\tau,l})$ as $\mathcal{S}^L(\beta_{\tau,l})$ and  $\mathcal{S}^R(\beta_{\tau,l})$, respectively. Then we partition $\mathcal{S}(\beta_{\tau,l})$ into $\widetilde{K}_{\tau,l}+1$ equally spaced intervals with $\mathcal{S}^L(\beta_{\tau,l})=\nu_0<\nu_1<\cdots<\nu_{\widetilde{K}_{\tau,l}}<\nu_{\widetilde{K}_{\tau,l}+1}=\mathcal{S}^R(\beta_{\tau,l})$, where $\widetilde{K}_{\tau,l}\rightarrow \infty$. We construct the SCB for $\beta_{\tau,l}(t)$ over a subset of $\mathcal{S}(\beta_{\tau,l})$, namely, $\mathcal{S}_{\varepsilon}(\beta_{\tau,l}) = \left\{\nu_0,\ldots,\nu_{\tilde{K}_{\tau,l}+1}\right\}$, and $\mathcal{S}_{\varepsilon}(\beta_{\tau,l})$ becomes denser as $n\to\infty$.

And, for any $0\leq j<j^{'} \leq \tilde{K}_{\tau,l}+1$, the covariance function of the Gaussian process $\Theta_{\tau,l}(t)$ is 
\begin{equation*}
\begin{split}
  \left|Cov\left(\Theta_{\tau,l}(\nu_{j}),\Theta_{\tau,l}(\nu_{j^{'}})\right)\right|
&  = \left|n^{-1}\sigma_{\tau,l}^{-1/2}(\nu_j)\sigma_{\tau,l}^{-1/2}(\nu_{j^{'}})
\bm{\Lambda}_{\tau,l}(\nu_{j})
\bm{\Sigma}_{\tau,2}^{-1}\bm{\Sigma}_{\tau,1}\bm{\Sigma}_{\tau,2}^{-1}
\bm{\Lambda}_{\tau,l}^T(\nu_{j^{'}})\right|\\
& \propto K^{-1}\left|
\bm{\Lambda}_{\tau,l}(\nu_{j})
\bm{\Sigma}_{\tau,2}^{-1}\bm{\Sigma}_{\tau,1}\bm{\Sigma}_{\tau,2}^{-1}
\bm{\Lambda}_{\tau,l}^T(\nu_{j^{'}})\right|\\
& \propto \left|
\bm{\Lambda}_{\tau,l}(\nu_{j})
\bm{\Lambda}_{\tau,l}^T(\nu_{j^{'}})\right|\\
& = \left|
\bm{\mathcal{B}}_{\tau,l}^T(\nu_j)\bm{\mathcal{B}}_{\tau,l}(\nu_{j^{'}})\right|\\
& = \left|\sum_{k=1}^{K+p}B_{k}(\nu_j)B_{k}(\nu_{j^{'}})\right|.
\end{split}
   \end{equation*}
Since the B-spline basis function has a local support property, that is, each B-spline basis function is nonzero over no more than $p+1$ consecutive sub-intervals \citep{deboor01}, we can have $ \left|Cov\left(\Theta_{\tau,l}(\nu_{j}),\Theta_{\tau,l}(\nu_{j^{'}})\right)\right| \leq C_B^{-|j-j^{'}|}$, with some positive constant $C_B$. Now combined with Lemma 1 in \cite{ma2011jump}, we can conclude that for any $a \in (0,1)$,
\begin{equation*}
\lim_{n\to\infty}P\left\{\sup_{t\in \mathcal{S}_{\varepsilon}(\beta_{\tau,l})}\left|
\sigma_{\tau,l}^{-1/2}(t)\left\{\widehat{\beta}_{\tau,l}(t)-\beta_{\tau,l}(t)\right\}\right|\leq \mathcal{Q}_{\tau,l}(a) \right\} = 1-a,
\end{equation*}
where 
\begin{equation*}
\begin{split}
    \mathcal{Q}_{\tau,l}(a) & = (2\log |\mathcal{S}_{\varepsilon}(\beta_{\tau,l})|)^{1/2} \\
    & - (2\log |\mathcal{S}_{\varepsilon}(\beta_{\tau,l})|)^{-1/2}\left\{\log(-0.5\log(1-a))
+0.5\left[\log(\log |\mathcal{S}_{\varepsilon}(\beta_{\tau,l})|) + \log (4\pi)\right]\right\}.
\end{split}
\end{equation*}
Then an asymptotic $100(1-a)\%$ simultaneous confidence band (SCB)
for $\beta_{\tau,l}(t)$ over $\mathcal{S}_{\varepsilon}(\beta_{\tau,l})$ is given by $\widehat{\beta}_{\tau,l}(t)\pm\sigma_{\tau,l}^{1/2}(t)\mathcal{Q}_{\tau,l}(a)$.
The proof of (ii) of Theorem 2 is completed.

\subsection{Proof of Theorem 3}
We want to prove the asymptotic normality of $\sqrt{n}( \widehat{\bm{\alpha}}_{\tau}- \bm{\alpha}_{\tau}^0 )$, and $\sqrt{n}( \widehat{\bm{\alpha}}_{\tau}- \bm{\alpha}_{\tau}^0 )$ can be expressed as 
\begin{equation*}
    \sqrt{n}( \widehat{\bm{\alpha}}_{\tau}- \bm{\alpha}_{\tau}^0 ) = \bm{\mathcal{I}}_{\tau}
    \sqrt{n}
        \begin{pmatrix}
            \widehat{\bm{\alpha}}_{\tau}- \bm{\alpha}_{\tau}^0\\            \widehat{\bm{\theta}}_{\tau}^{(1)}- \bm{\theta}_{\tau}^{0(1)}
        \end{pmatrix},
\end{equation*}
then by (\ref{normal-para}), we can have
\begin{equation}\label{normal-alpha}
    \begin{split}
    \sqrt{n}
        \begin{pmatrix}
            \widehat{\bm{\alpha}}_{\tau}- \bm{\alpha}_{\tau}^0\\  
        \end{pmatrix}
        \overset{d}{\rightarrow}  
        \bm{\mathbb{N}}(\bm{0},\tau(1-\tau)\bm{\mathcal{I}}_{\tau}\bm{\Sigma}_{\tau,2}^{-1}\bm{\Sigma}_{\tau,1}\bm{\Sigma}_{\tau,2}^{-1}\bm{\mathcal{I}}_{\tau}^T),
    \end{split}
\end{equation}
where $\left\|Cov(\widehat{\bm{\alpha}}_{\tau}- \bm{\alpha}_{\tau}^0)\right\|_{\infty} = O(1/n)$.
Then we complete the proof of Theorem 3.

\subsection{Proof of Theorem 4}
We choose to randomly split the original dataset into two parts $Data_{I}$ and $Data_{II}$, where $Data_{I} = \left\{(Y_i,\bm{Z}_i,\bm{X}_i(t),t\in[0,\mathcal{T}),n_1+1\leq i\leq n\right\}$, $Data_{II} = \left\{(Y_i,\bm{Z}_i,\bm{X}_i(t),t\in[0,\mathcal{T}),1\leq i\leq n_1\right\}$,
$n_1 = \lfloor n/2 \rfloor$ and $\lfloor v \rfloor$ denotes the 
\textcolor{blue}{greatest integer less than or equal to $v$.}
Denote the estimators based on $Data_I$ as $\widetilde{\bm{\alpha}}_{\tau,I}$ and $\widetilde{\bm{\theta}}_{\tau,I}$, which is the solution of the following objective function 
\begin{equation}\label{dataII}
\begin{split}
     (\widetilde{\bm{\alpha}}_{\tau,I},
    \widetilde{\bm{\theta}}_{\tau,I}) 
=  &  \arg\min_{\bm{\alpha}_{\tau},\bm{\theta}_{\tau}}
  \frac{1}{n_2}\sum_{i=n_1+1}^{n}\left(\rho_{\tau}*\mathcal{K}_h\right)(Y_i-\bm{Z}_i^T\bm{\alpha}_{\tau}-\bm{U}_i^T\bm{\theta}_{\tau})\\
   & +\bm{\theta}^T_{\tau}(\bm{\Gamma}\otimes\bm{V})\bm{\theta}_{\tau}
+\sum_{l=1}^m\sum_{j=1}^Kp_{\lambda_l}\left(\sqrt{\frac{K}{\mathcal T}\bm{\theta}_{\tau,l}^T \bm{W}_j\bm{\theta}_{\tau,l}}\right),
\end{split}
\end{equation}
where $n_2=n-n_1$.

Define the set $\widehat{S}_{\bm{\theta},I} = \{j: \widetilde{\theta}_{\tau,I,j}\neq 0, 1\leq j\leq m(K+p)\}$ and $\widehat{\bm{T}}_{\bm{\theta},I} = \{\bm{v}\in \mathbb{R}^{m(K+p)}: v_j= 0, \forall j\in \widehat{S}_{\bm{\theta},I}^c\}$, where 
$\widehat{S}_{\bm{\theta},I}^c$ is the complement of the set $\widehat{S}_{\bm{\theta},I}$. To make the notation consistent, we denote the estimators based on the $Data_{II}$ as $\widehat{\bm{\alpha}}_{\tau,II}$ and $\widetilde{\bm{\theta}}_{\tau,II}$, which is from
\begin{equation}\label{dataI}
\begin{split}
     (\widehat{\bm{\alpha}}_{\tau,II},
    \widetilde{\bm{\theta}}_{\tau,II}) 
=  &  \arg\min_{\bm{\alpha}_{\tau},\bm{\theta}_{\tau}\in \widehat{\bm{T}}_{\bm{\theta},I} }
  \frac{1}{n_1}\sum_{i=1}^{n_1}\left(\rho_{\tau}*\mathcal{K}_h\right)(Y_i-\bm{Z}_i^T\bm{\alpha}_{\tau}-\bm{U}_i^T\bm{\theta}_{\tau})\\
   & +\bm{\theta}^T_{\tau}(\bm{\Gamma}\otimes\bm{V})\bm{\theta}_{\tau},
\end{split}
\end{equation}
then $\widehat{\bm{\theta}}_{\tau,II}$ includes those zero terms obtained from (\ref{dataII}). Similar to the proof in \cite{feng2011wild} and \cite{cheng2022regularized}, it is easy to have 
$\widetilde{\bm{\alpha}}_{\tau,I} \to \bm{\alpha}_{\tau}^0$, $\widehat{\bm{\alpha}}_{\tau,II} \to \bm{\alpha}_{\tau}^0$,
$\widetilde{\bm{\theta}}_{\tau,I} \to \bm{\theta}_{\tau}^0$, and
$\widetilde{\bm{\theta}}_{\tau,II} \to \bm{\theta}_{\tau}^0$ in probability.

Recall that $\widehat{\bm{\alpha}}_{\tau}^{(b)}$ $\widehat{\bm{\theta}}_{\tau}^{(b)}$ are the estimates which satisfy:
\begin{equation}\label{bootobj}
\begin{split}
     (\widehat{\bm{\alpha}}_{\tau}^{(b)},
    \widehat{\bm{\theta}}_{\tau}^{(b)}) 
=  &  \arg\min_{\bm{\alpha}_{\tau},\bm{\theta}_{\tau}}
  \frac{1}{n_1}\sum_{i=1}^{n_1}\left(\rho_{\tau}*\mathcal{K}_h\right)(Y_i^{(b)}-\bm{Z}_i^T\bm{\alpha}_{\tau}-\bm{U}_i^T\bm{\theta}_{\tau})\\
   & +\bm{\theta}^T_{\tau}(\bm{\Gamma}\otimes\bm{V})\bm{\theta}_{\tau}.
\end{split}
\end{equation}
Based on the bootstrapped sample, denote the following derivative function
\begin{equation}
\begin{split}
    \bm{\mathcal{G}}(Y_i^{(b)},\bm{\alpha}_{\tau},
    \bm{\theta}_{\tau})= & \left\{G_h(\bm{Z}_i^T\bm{\alpha}_{\tau}+
\bm{U}_i^T\bm{\theta}_{\tau}-Y_i^{(b)})-\tau\right\}(\bm{Z}_i^T,\bm{U}_i^T)^T\\
&+2 
    \begin{pmatrix}
        \bm{0}_{d\times d} & \\
        & \bm{\Gamma}\otimes\bm{V}
    \end{pmatrix}
   \begin{pmatrix}
       \bm{\alpha}_{\tau}\\
       \bm{\theta}_{\tau}
   \end{pmatrix}\\
\end{split}
\end{equation}
and the estimator $(\widehat{\bm{\alpha}}_{\tau}^{(b)},\widehat{\bm{\theta}}_{\tau}^{(b)})$ based on the bootstrapped dataset is the solution to the equation:
\begin{equation}
    \frac{1}{n_1}\sum_{i=1}^{n_1} \bm{\mathcal{G}}(Y_i^{(b)},\bm{\alpha}_{\tau},
    \bm{\theta}_{\tau}) =0.
\end{equation}

Consider the following identity
\begin{equation}\label{identy}
    \begin{split}
        & E\left\{\bm{\mathcal{G}}(Y_i^{(b)},\widehat{\bm{\alpha}}_{\tau}^{(b)},
    \widehat{\bm{\theta}}_{\tau}^{(b)})-\bm{\mathcal{G}}(Y_i^{(b)},\widehat{\bm{\alpha}}_{\tau,II},
    \widehat{\bm{\theta}}_{\tau,II})\right\}\\
      =& -\frac{1}{n_1}\sum_{i=1}^{n_1}\left\{\bm{\mathcal{G}}(Y_i^{(b)},\widehat{\bm{\alpha}}_{\tau,II},
    \widehat{\bm{\theta}}_{\tau,II})-\bm{\mathcal{G}}(Y_i^{(b)},\widehat{\bm{\alpha}}_{\tau}^{(b)},
    \widehat{\bm{\theta}}_{\tau}^{(b)})\right\}\\
    & +\frac{1}{n_1}\sum_{i=1}^{n_1}\left\{\bm{\mathcal{G}}(Y_i^{(b)},\widehat{\bm{\alpha}}_{\tau,II},
    \widehat{\bm{\theta}}_{\tau,II})-\bm{\mathcal{G}}(Y_i^{(b)},\widehat{\bm{\alpha}}_{\tau}^{(b)},
    \widehat{\bm{\theta}}_{\tau}^{(b)})\right\}
    - E\left\{\bm{\mathcal{G}}(Y_i^{(b)},\widehat{\bm{\alpha}}_{\tau,II},
    \widehat{\bm{\theta}}_{\tau,II})-\bm{\mathcal{G}}(Y_i^{(b)},\widehat{\bm{\alpha}}_{\tau}^{(b)},
    \widehat{\bm{\theta}}_{\tau}^{(b)})\right\}\\
    =& -\frac{1}{n_1}\sum_{i=1}^{n_1}\bm{\mathcal{G}}(Y_i^{(b)},\widehat{\bm{\alpha}}_{\tau,II},
    \widehat{\bm{\theta}}_{\tau,II})\\
    & +\frac{1}{n_1}\sum_{i=1}^{n_1}\left\{\bm{\mathcal{G}}(Y_i^{(b)},\widehat{\bm{\alpha}}_{\tau,II},
    \widehat{\bm{\theta}}_{\tau,II})-\bm{\mathcal{G}}(Y_i^{(b)},\widehat{\bm{\alpha}}_{\tau}^{(b)},
    \widehat{\bm{\theta}}_{\tau}^{(b)})\right\} 
    - E\left\{\bm{\mathcal{G}}(Y_i^{(b)},\widehat{\bm{\alpha}}_{\tau,II},
    \widehat{\bm{\theta}}_{\tau,II})-\bm{\mathcal{G}}(Y_i^{(b)},\widehat{\bm{\alpha}}_{\tau}^{(b)},
    \widehat{\bm{\theta}}_{\tau}^{(b)})\right\}.
    \end{split}
\end{equation}

\textbf{(1) First, we want to show that }
\begin{equation}\label{consist}
    \frac{1}{n_1}\sum_{i=1}^{n_1}\left\{\bm{\mathcal{G}}(Y_i^{(b)},\widehat{\bm{\alpha}}_{\tau,II},
    \widehat{\bm{\theta}}_{\tau,II})-\bm{\mathcal{G}}(Y_i^{(b)},\widehat{\bm{\alpha}}_{\tau}^{(b)},
    \widehat{\bm{\theta}}_{\tau}^{(b)})\right\} - E\left\{\bm{\mathcal{G}}(Y_i^{(b)},\widehat{\bm{\alpha}}_{\tau,II},
    \widehat{\bm{\theta}}_{\tau,II})-\bm{\mathcal{G}}(Y_i^{(b)},\widehat{\bm{\alpha}}_{\tau}^{(b)},
    \widehat{\bm{\theta}}_{\tau}^{(b)})\right\} = o_p(1),
\end{equation}
it suffices by Chebyshev's inequality to show that
\begin{equation}\label{cheby}
    \frac{1}{n_1}Cov\left\{\bm{\mathcal{G}}(Y_i^{(b)},\widehat{\bm{\alpha}}_{\tau,II},
    \widehat{\bm{\theta}}_{\tau,II})-\bm{\mathcal{G}}(Y_i^{(b)},\widehat{\bm{\alpha}}_{\tau}^{(b)},
    \widehat{\bm{\theta}}_{\tau}^{(b)})\right\}\to 0.
\end{equation}
We have 
\begin{equation}\label{cov}
    \begin{split}
        & Cov\left\{\bm{\mathcal{G}}(Y_i^{(b)},\widehat{\bm{\alpha}}_{\tau,II},
    \widehat{\bm{\theta}}_{\tau,II})-\bm{\mathcal{G}}(Y_i^{(b)},\widehat{\bm{\alpha}}_{\tau}^{(b)},
    \widehat{\bm{\theta}}_{\tau}^{(b)})\right\}\\
    = & Var\left\{G_h(\bm{Z}_i^T\widehat{\bm{\alpha}}_{\tau,II}+
\bm{U}_i^T\widehat{\bm{\theta}}_{\tau,II}-Y_i^{(b)})-G_h(\bm{Z}_i^T\widehat{\bm{\alpha}}_{\tau}^{(b)}+
\bm{U}_i^T\widehat{\bm{\theta}}_{\tau}^{(b)}-Y_i^{(b)})\right\}(\bm{Z}_i^T,\bm{U}_i^T)^T(\bm{Z}_i^T,\bm{U}_i^T)\\
=& E\left[G_h(\bm{Z}_i^T\widehat{\bm{\alpha}}_{\tau,II}+
\bm{U}_i^T\widehat{\bm{\theta}}_{\tau,II}-Y_i^{(b)})-G_h(\bm{Z}_i^T\widehat{\bm{\alpha}}_{\tau}^{(b)}+
\bm{U}_i^T\widehat{\bm{\theta}}_{\tau}^{(b)}-Y_i^{(b)})\right]^2 (\bm{Z}_i^T,\bm{U}_i^T)^T(\bm{Z}_i^T,\bm{U}_i^T)\\
& -\left\{E\left[G_h(\bm{Z}_i^T\widehat{\bm{\alpha}}_{\tau,II}+
\bm{U}_i^T\widehat{\bm{\theta}}_{\tau,II}-Y_i^{(b)})-G_h(\bm{Z}_i^T\widehat{\bm{\alpha}}_{\tau}^{(b)}+
\bm{U}_i^T\widehat{\bm{\theta}}_{\tau}^{(b)}-Y_i^{(b)})\right]\right\}^2(\bm{Z}_i^T,\bm{U}_i^T)^T(\bm{Z}_i^T,\bm{U}_i^T).\\
    \end{split}
\end{equation}
Let $\Delta = \bm{Z}_i^T(\bm{\alpha}_{\tau}-\widehat{\bm{\alpha}}_{\tau,II})+\bm{U}_i^T(\bm{\theta}_{\tau}-\widehat{\bm{\theta}}_{\tau,II})$ and 
$\widehat{\Delta}^{(b)} = \bm{Z}_i^T(\widehat{\bm{\alpha}}_{\tau,II}-\widehat{\bm{\alpha}}_{\tau}^{(b)})+\bm{U}_i^T(\widehat{\bm{\theta}}_{\tau,II}-\widehat{\bm{\theta}}_{\tau}^{(b)})$. Note that 
\begin{equation}
\begin{split}
      & \frac{\partial}{\partial e_i}\left[G_h(-w_i|\widehat{e}_i|)-G_h(-w_i|\widehat{e}_i|-\widehat{\Delta}^{(b)})\right] \\
      =&   \frac{\partial}{\partial e_i}\left[G_h(-w_i|e_i+\Delta|)-G_h(-w_i|e_i+\Delta|-\widehat{\Delta}^{(b)})\right]\\
      =&  -w_i \text{sgn}(e_i+\Delta) \left[\mathcal{K}_h(-w_i|e_i+\Delta|)-\mathcal{K}_h(-w_i|e_i+\Delta|-\widehat{\Delta}^{(b)})\right]\\
\end{split} 
\end{equation}
then when covariates are given, by integration by parts, we have 
\begin{equation}\label{1moment}
    \begin{split}
&E\left[G_h(\bm{Z}_i^T\widehat{\bm{\alpha}}_{\tau,II}+
\bm{U}_i^T\widehat{\bm{\theta}}_{\tau,II}-Y_i^{(b)})-G_h(\bm{Z}_i^T\widehat{\bm{\alpha}}_{\tau}^{(b)}+
\bm{U}_i^T\widehat{\bm{\theta}}_{\tau}^{(b)}-Y_i^{(b)})\right]\\
= & E\left[G_h(-w_i|\widehat{e}_i|)-G_h(-w_i|\widehat{e}_i|-\widehat{\Delta}^{(b)})\right]\\
=& E\left\{w_i\int \text{sgn}(\varepsilon+\Delta)F_{e|\bm{Z},\bm{X}}(\varepsilon)\left[\mathcal{K}_h(-w_i|e_i+\Delta|)-\mathcal{K}_h(-w_i|e_i+\Delta|- \widehat{\Delta}^{(b)})\right]d\varepsilon\right\}\\
=& E\left\{w_i\int \text{sgn}(\varepsilon+\Delta)F_{e|\bm{Z},\bm{X}}(\varepsilon)\mathcal{K}_h(-w_i|e_i+\Delta|)d\varepsilon\right\} \\
& -E\left\{w_i\int \text{sgn}(\varepsilon+\Delta)F_{e|\bm{Z},\bm{X}}(\varepsilon)\mathcal{K}_h(-w_i|e_i+\Delta|- \widehat{\Delta}^{(b)})d\varepsilon\right\}.\\
  \end{split}
\end{equation}
When $w_i>0$, using a change of variable, we have
    \begin{align*}
        & w_i\int \text{sgn}(\varepsilon+\Delta)F_{e|\bm{Z},\bm{X}}(\varepsilon)\mathcal{K}_h(-w_i|e_i+\Delta|)d\varepsilon\\
        = & \int_{\Delta}^{+\infty} w_i\int F_{e|\bm{Z},\bm{X}}(\varepsilon)\mathcal{K}_h(-w_i(e_i+\Delta))d\varepsilon\\
        &
        - \int_{-\infty}^{\Delta} w_i\int F_{e|\bm{Z},\bm{X}}(\varepsilon)\mathcal{K}_h(w_i(e_i+\Delta))d\varepsilon\\
        =& \int_{0}^{-\infty} - F_{e|\bm{Z},\bm{X}}(-hv/w_i-\Delta)\mathcal{K}(v)dv -
        \int_{-\infty}^0F_{e|\bm{Z},\bm{X}}(hv/w_i-\Delta)\mathcal{K}(v)dv\\
        =& \int_{-\infty}^0 F_{e|\bm{Z},\bm{X}}(-hv/w_i-\Delta)\mathcal{K}(v)dv -
        \int_{-\infty}^0F_{e|\bm{Z},\bm{X}}(hv/w_i-\Delta)\mathcal{K}(v)dv\\
        \approx & \int_{-\infty}^0 \left[F_{e|\bm{Z},\bm{X}}(0)+f_{e|\bm{Z},\bm{X}}(0)(-hv/w_i-\Delta)\right]\mathcal{K}(v)dv \\
        &
        -
        \int_{-\infty}^0\left[F_{e|\bm{Z},\bm{X}}(0)+f_{e|\bm{Z},\bm{X}}(0)(hv/w_i-\Delta)\right]\mathcal{K}(v)dv\\
        = &-2h/w_if_{e|\bm{Z},\bm{X}}(0) \int_{-\infty}^0v\mathcal{K}(v)dv+o_p(1),
    \end{align*}
and 
    \begin{align*}
        & w_i\int \text{sgn}(\varepsilon+\Delta)F_{e|\bm{Z},\bm{X}}(\varepsilon)\mathcal{K}_h(-w_i|e_i+\Delta|- \widehat{\Delta}^{(b)})d\varepsilon\\
        = & \int_{\Delta}^{+\infty} w_i\int F_{e|\bm{Z},\bm{X}}(\varepsilon)\mathcal{K}_h(-w_i(e_i+\Delta)- \widehat{\Delta}^{(b)})d\varepsilon\\
        &
        - \int_{-\infty}^{\Delta} w_i\int F_{e|\bm{Z},\bm{X}}(\varepsilon)\mathcal{K}_h(w_i(e_i+\Delta)- \widehat{\Delta}^{(b)})d\varepsilon\\
        =& \int_{0}^{-\infty} - F_{e|\bm{Z},\bm{X}}(-(hv+ \widehat{\Delta}^{(b)})/w_i-\Delta)\mathcal{K}(v)dv -
        \int_{-\infty}^0F_{e|\bm{Z},\bm{X}}((hv+ \widehat{\Delta}^{(b)})/w_i-\Delta)\mathcal{K}(v)dv\\
        =& \int_{-\infty}^0 F_{e|\bm{Z},\bm{X}}(-(hv+ \widehat{\Delta}^{(b)})/w_i-\Delta)\mathcal{K}(v)dv -
        \int_{-\infty}^0F_{e|\bm{Z},\bm{X}}((hv+ \widehat{\Delta}^{(b)})/w_i-\Delta)\mathcal{K}(v)dv\\
        \approx & \int_{-\infty}^0 \left[F_{e|\bm{Z},\bm{X}}(0)+f_{e|\bm{Z},\bm{X}}(0)(-(hv+ \widehat{\Delta}^{(b)})/w_i-\Delta)\right]\mathcal{K}(v)dv \\
        & -
        \int_{-\infty}^0\left[F_{e|\bm{Z},\bm{X}}(0)+f_{e|\bm{Z},\bm{X}}(0)((hv+ \widehat{\Delta}^{(b)})/w_i-\Delta)\right]\mathcal{K}(v)dv\\
        = &-2h/w_if_{e|\bm{Z},\bm{X}}(0) \int_{-\infty}^0v\mathcal{K}(v)dv
        - 2\widehat{\Delta}^{(b)}w_i^{-1} f_{e|\bm{Z},\bm{X}}(0) \int_{-\infty}^0\mathcal{K}(v)dv
        +o_p(1)\\
        = &-2h/w_if_{e|\bm{Z},\bm{X}}(0) \int_{-\infty}^0v\mathcal{K}(v)dv
        - \widehat{\Delta}^{(b)}w_i^{-1} f_{e|\bm{Z},\bm{X}}(0)
        +o_p(1)\\
    \end{align*}
Similarly, when $w_i<0$, we have 
\begin{equation*}
    \begin{split}
        & w_i\int \text{sgn}(\varepsilon+\Delta)F_{e|\bm{Z},\bm{X}}(\varepsilon)\mathcal{K}_h(-w_i|e_i+\Delta|)d\varepsilon\\
        = &-2h/w_if_{e|\bm{Z},\bm{X}}(0) \int_{-\infty}^0v\mathcal{K}(v)dv+o_p(1),
    \end{split}
\end{equation*}
and 
\begin{equation*}
    \begin{split}
        & w_i\int \text{sgn}(\varepsilon+\Delta)F_{e|\bm{Z},\bm{X}}(\varepsilon)\mathcal{K}_h(-w_i|e_i+\Delta|- \widehat{\Delta}^{(b)})d\varepsilon\\ 
         = &-2h/w_if_{e|\bm{Z},\bm{X}}(0) \int_{-\infty}^0v\mathcal{K}(v)dv
        + \widehat{\Delta}^{(b)}w_i^{-1} f_{e|\bm{Z},\bm{X}}(0)
        +o_p(1).\\
    \end{split}
\end{equation*}
Combing these results and (\ref{1moment}), we have
\begin{equation}\label{1momentnew}
    \begin{split}
&E\left[G_h(\bm{Z}_i^T\widehat{\bm{\alpha}}_{\tau,II}+
\bm{U}_i^T\widehat{\bm{\theta}}_{\tau,II}-Y_i^{(b)})-G_h(\bm{Z}_i^T\widehat{\bm{\alpha}}_{\tau}^{(b)}+
\bm{U}_i^T\widehat{\bm{\theta}}_{\tau}^{(b)}-Y_i^{(b)})\right]=
\widehat{\Delta}^{(b)} f_{e|\bm{Z},\bm{X}}(0)
        +o_p(\widehat{\Delta}^{(b)}).\\
  \end{split}
\end{equation}
Let $w_i|e_i+\Delta^*|$ lies between $ w_i|\widehat{e}_i|$ and $w_i|\widehat{e}_i|-\widehat{\Delta}^{(b)}$, then by Lagrange mean value theorem and Condition 4, we can get
\begin{equation}\label{2monment}
    \begin{split}
&E\left[G_h(\bm{Z}_i^T\widehat{\bm{\alpha}}_{\tau,II}+
\bm{U}_i^T\widehat{\bm{\theta}}_{\tau,II}-Y_i^{(b)})-G_h(\bm{Z}_i^T\widehat{\bm{\alpha}}_{\tau}^{(b)}+
\bm{U}_i^T\widehat{\bm{\theta}}_{\tau}^{(b)}-Y_i^{(b)})\right]^2\\
= & E\left[G_h(-w_i|\widehat{e}_i|)-G_h(-w_i|\widehat{e}_i|-\widehat{\Delta}^{(b)})\right]^2\\
=& E\left[-w_i\text{sgn}(e_i+\Delta^*)\mathcal{K}_h(-w_i|e_i+\Delta^*|)\widehat{\Delta}^{(b)}\right]^2\\
=& o_p(1/h).
    \end{split}
\end{equation}
By Conditions 1, 3, 4, and 8, 
combined (\ref{cov}), (\ref{1moment}) and (\ref{2monment}), (\ref{cheby}) holds. So we can show that $$n_1^{-1}\sum_{i=1}^{n_1}\left\{\bm{\mathcal{G}}(Y_i^{(b)},\widehat{\bm{\alpha}}_{\tau,II},
    \widehat{\bm{\theta}}_{\tau,II})-\bm{\mathcal{G}}(Y_i^{(b)},\widehat{\bm{\alpha}}_{\tau}^{(b)},
    \widehat{\bm{\theta}}_{\tau}^{(b)})\right\} - E\left\{\bm{\mathcal{G}}(Y_i^{(b)},\widehat{\bm{\alpha}}_{\tau,II},
    \widehat{\bm{\theta}}_{\tau,II})-\bm{\mathcal{G}}(Y_i^{(b)},\widehat{\bm{\alpha}}_{\tau}^{(b)},
    \widehat{\bm{\theta}}_{\tau}^{(b)})\right\} = o_p(1).$$

\textbf{(2) Second, we want to prove
}
\begin{equation*}
    \begin{split}
    \sqrt{n_1}
        \begin{pmatrix}
        \widehat{\bm{\alpha}}_{\tau,II}-\widehat{\bm{\alpha}}_{\tau}^{(b)}\\
        \widehat{\bm{\theta}}_{\tau,II}-\widehat{\bm{\theta}}_{\tau}^{(b)}
    \end{pmatrix} 
    & = E\left\{f_{e\mid \bm{Z},\bm{X}}(0)
 \begin{pmatrix}
     \bm{Z}\\
     \bm{U}\\
 \end{pmatrix}^{\otimes 2}\right\}^{-1}\frac{1}{\sqrt{n_1}}\sum_{i=1}^{n_1}\bm{\mathcal{G}}(Y_i^{(b)},\widehat{\bm{\alpha}}_{\tau,II},
    \widehat{\bm{\theta}}_{\tau,II})+o_p(n^{-1/2}).
    \end{split}
\end{equation*}

When covariates are given and Conditions 6-9 hold, we have
\begin{equation}\label{secondpart}
    \begin{split}
        & E\left\{\bm{\mathcal{G}}(Y_i^{(b)},\widehat{\bm{\alpha}}_{\tau}^{(b)},
    \widehat{\bm{\theta}}_{\tau}^{(b)})-\bm{\mathcal{G}}(Y_i^{(b)},\widehat{\bm{\alpha}}_{\tau,II},
    \widehat{\bm{\theta}}_{\tau,II})\right\}\\
= &     E\left[G_h(\bm{Z}_i^T\widehat{\bm{\alpha}}_{\tau}^{(b)}+
\bm{U}_i^T\widehat{\bm{\theta}}_{\tau}^{(b)}-Y_i^{(b)})-G_h(\bm{Z}_i^T\widehat{\bm{\alpha}}_{\tau,II}+
\bm{U}_i^T\widehat{\bm{\theta}}_{\tau,II}-Y_i^{(b)})\right](\bm{Z}_i^T,\bm{U}_i^T)^T+o_p(1)\\
=& E \left[G_h(-w_i|\widehat{e}_i|-\widehat{\Delta}^{(b)})-G_h(-w_i|\widehat{e}_i|)\right](\bm{Z}_i^T,\bm{U}_i^T)^T+o_p(1).\\
    \end{split}
\end{equation}

For any vector $\bm{u}_1$, $\bm{u}_2\in\mathbb{R}^d$ and $\bm{v}_1,\bm{v}_2\in\mathbb{R}^{m\times (K+p)}$, similar with the proof of (\ref{1moment}), we have 
\begin{equation}\label{gh}
    \begin{split}
      & E \left[G_h(-w_i|e_i-\bm{Z}_i^T\bm{u}_1-\bm{U}_i^T\bm{v}_1|-\bm{Z}_i^T\bm{u}_2-\bm{U}_i^T\bm{v}_2)-G_h(-w_i|e_i-\bm{Z}_i^T\bm{u}_1-\bm{U}_i^T\bm{v}_1|)\right] \\
     =  & -(\bm{Z}_i^T\bm{u}_2+\bm{U}_i^T\bm{v}_2)E \left[ \mathcal{K}_h(-w_i|e_i-\bm{Z}_i^T\bm{u}_1-\bm{U}_i^T\bm{v}_1|)\right]\\
     = & -(\bm{Z}_i^T\bm{u}_2+\bm{U}_i^T\bm{v}_2)
     \left[f_{e|\bm{Z},\bm{X}}(0)+O(\bm{Z}_i^T\bm{u}_1+\bm{U}_i^T\bm{v}_1)\right].
    \end{split}
\end{equation}
When $\|\bm{u}\|_2 = O_p(n^{-1/2})$ and $\|\bm{u}\|_2 = O_p(n^{-1/2}K)$, we have
\begin{equation}\label{uv}
    \begin{split}
      & E \left[G_h(-w_i|e_i-\bm{Z}_i^T\bm{u}_1-\bm{U}_i^T\bm{v}_1|-\bm{Z}_i^T\bm{u}_2-\bm{U}_i^T\bm{v}_2)-G_h(-w_i|e_i-\bm{Z}_i^T\bm{u}_1-\bm{U}_i^T\bm{v}_1|)\right] \\
     = & -(\bm{Z}_i^T\bm{u}_2+\bm{U}_i^T\bm{v}_2)
     f_{e|\bm{Z},\bm{X}}(0)+o_p(n^{-1/2}).
    \end{split}
\end{equation}
By (\ref{secondpart}), (\ref{gh}) and (\ref{uv}), we have 
\begin{equation}\label{expect}
    \begin{split}
         & E\left\{\bm{\mathcal{G}}(Y_i^{(b)},\widehat{\bm{\alpha}}_{\tau}^{(b)},
    \widehat{\bm{\theta}}_{\tau}^{(b)})-\bm{\mathcal{G}}(Y_i^{(b)},\widehat{\bm{\alpha}}_{\tau,II},
    \widehat{\bm{\theta}}_{\tau,II})\right\}\\
    =&- E\left\{f_{e|\bm{Z},\bm{X}}(0)(\bm{Z}_i^T,\bm{U}_i^T)^T(\bm{Z}_i^T,\bm{U}_i^T)\right\}
    \begin{pmatrix}
        \widehat{\bm{\alpha}}_{\tau,II}-\widehat{\bm{\alpha}}_{\tau}^{(b)}\\
        \widehat{\bm{\theta}}_{\tau,II}-\widehat{\bm{\theta}}_{\tau}^{(b)}
    \end{pmatrix}
    +o_p(n_1^{-1/2}).
    \end{split}
\end{equation}
Inserting (\ref{expect}) to (\ref{identy}) and by (\ref{consist}), we further have
\begin{equation}\label{identy}
    \begin{split}
        & E\left\{\bm{\mathcal{G}}(Y_i^{(b)},\widehat{\bm{\alpha}}_{\tau}^{(b)},
    \widehat{\bm{\theta}}_{\tau}^{(b)})-\bm{\mathcal{G}}(Y_i^{(b)},\widehat{\bm{\alpha}}_{\tau,II},
    \widehat{\bm{\theta}}_{\tau,II})\right\}+o_p(1)\\
    = &  - E\left\{f_{e|\bm{Z},\bm{X}}(0)(\bm{Z}_i^T,\bm{U}_i^T)^T(\bm{Z}_i^T,\bm{U}_i^T)\right\}\begin{pmatrix}
        \widehat{\bm{\alpha}}_{\tau,II}-\widehat{\bm{\alpha}}_{\tau}^{(b)}\\
        \widehat{\bm{\theta}}_{\tau,II}-\widehat{\bm{\theta}}_{\tau}^{(b)}
    \end{pmatrix}\\
    =& -\frac{1}{n_1}\sum_{i=1}^{n_1}\bm{\mathcal{G}}(Y_i^{(b)},\widehat{\bm{\alpha}}_{\tau,II},
    \widehat{\bm{\theta}}_{\tau,II})+o_p(1).
    \end{split}
\end{equation}
Then, we have
\begin{equation}
    \begin{split}
    \sqrt{n_1}
        \begin{pmatrix}
        \widehat{\bm{\alpha}}_{\tau,II}-\widehat{\bm{\alpha}}_{\tau}^{(b)}\\
        \widehat{\bm{\theta}}_{\tau,II}-\widehat{\bm{\theta}}_{\tau}^{(b)}
    \end{pmatrix} 
    & = E\left\{f_{e\mid \bm{Z},\bm{X}}(0)
 \begin{pmatrix}
     \bm{Z}\\
     \bm{U}\\
 \end{pmatrix}^{\otimes 2}\right\}^{-1}\frac{1}{\sqrt{n}}\sum_{i=1}^{n_1}\bm{\mathcal{G}}(Y_i^{(b)},\widehat{\bm{\alpha}}_{\tau,II},
    \widehat{\bm{\theta}}_{\tau,II})+o_p(n_1^{-1/2}),
    \end{split}
\end{equation}
where 
\begin{equation}
    \begin{split}
    \frac{1}{\sqrt{n_1}}\sum_{i=1}^{n_1}\bm{\mathcal{G}}(Y_i^{(b)},\widehat{\bm{\alpha}}_{\tau,II},
    \widehat{\bm{\theta}}_{\tau,II}) = \frac{1}{\sqrt{n_1}}\sum_{i=1}^{n_1}\left\{G_h(-w_i|\widehat{e}_i|)-\tau\right\}(\bm{Z}_i^T,\bm{U}_i^T)^T+o_p(1).
    \end{split}
\end{equation}
For $G_h(-w_i|\widehat{e}_i|)$, when covariates are given, we have 
\begin{equation*}
    E(G_h(-w_i|\widehat{e}_i|)) = \tau-E(w_i^{-1})2hf_{e|\bm{Z},\bm{X}}(0) \int_{-\infty}^0v\mathcal{K}(v)dv+o_p(1)=\tau+o_p(1),
\end{equation*}
and 
\begin{equation*}
    \begin{split}
        E(G_h^2(-w_i|\widehat{e}_i|)) = \tau+o_p(1),
    \end{split}
\end{equation*}
then asymptotic mean and variance of $\left\{G_h(-w_i|\widehat{e}_i|)-\tau\right\}$ are $0$ and $\tau(1-\tau)$, respectively.

Finally, we can have 
\begin{equation}
    \begin{split}
    \sqrt{n_1}\left(\bm{\Sigma}_{2}^{-1}\bm{\Sigma}_1\bm{\Sigma}_{2}^{-1}\right)^{-1/2}
        \begin{pmatrix}
        \widehat{\bm{\alpha}}_{\tau}^{(b)}-\widehat{\bm{\alpha}}_{\tau,II}\\
        \widehat{\bm{\theta}}_{\tau}^{(b)}-\widehat{\bm{\theta}}_{\tau,II}
    \end{pmatrix} 
    & \overset{d}{\rightarrow}  \mathbb{N} (\bm{0},\tau(1-\tau)\bm{I}),
   \end{split}
\end{equation}
where 
$$
\bm{\Sigma}_1 = E\left\{\left((\bm{Z}^T,\bm{U}^T)^T\right)^{\otimes2}\right\},\quad 
\bm{\Sigma}_2 = E\left\{f_{e\mid \bm{Z},\bm{X}}(0)
 \begin{pmatrix}
     \bm{Z}\\
     \bm{U}\\
 \end{pmatrix}^{\otimes 2}\right\}.
$$
Correspondingly, we have 
\begin{equation}\label{normal-boot}
    \begin{split}
    \sqrt{n_1}\left(\bm{\Sigma}_{\tau,2}^{-1}\bm{\Sigma}_{\tau,1}\bm{\Sigma}_{\tau,2}^{-1}\right)^{-1/2}
        \begin{pmatrix}
            \widehat{\bm{\alpha}}_{\tau}^{(b)}- \widehat{\bm{\alpha}}_{\tau,II}\\          \widehat{\bm{\theta}}_{\tau}^{(b),(1)}- \widehat{\bm{\theta}}_{\tau,II}^{(1)}
        \end{pmatrix}
        \overset{d}{\rightarrow}  
        \bm{\mathbb{N}}(\bm{0},\tau(1-\tau)\bm{I}).
    \end{split}
\end{equation}
This completes the proof of Theorem 4 by Theorem 2-3 and the asymptotic result above.

\bibliographystyle{apalike}      
\bibliography{bibfile}   